\documentclass[aps,prb,twocolumn,floatfix,superscriptaddress]{revtex4-2}
\usepackage{graphicx}
\usepackage{multirow}
\usepackage{color}
\usepackage{amsmath}
\usepackage{amsfonts}
\usepackage{amssymb}
\usepackage{graphicx}
\graphicspath{{Figures/}}
\usepackage{epstopdf}
\usepackage{empheq}
\usepackage{float}
\usepackage{cases}
\usepackage{mathtools}
\usepackage{dcolumn}
\usepackage{bbold}
\usepackage{xfrac}
\usepackage{bm}

\usepackage[dvipsnames]{xcolor}
\usepackage[colorlinks]{hyperref}
\hypersetup{
colorlinks=True,
linkbordercolor=White,
linkcolor=BrickRed,
citecolor=Blue,	 
}

\DeclarePairedDelimiter\bra{\langle}{\rvert}
\DeclarePairedDelimiter\ket{\lvert}{\rangle}
\DeclarePairedDelimiter\braket{\langle}{\rangle}

\renewcommand{\vec}[1]{\mathbf{#1}}

\newcommand\YMN[1]{#1}

\begin{document}

\title{Linear-in-momentum spin orbit interactions in planar Ge/GeSi heterostructures and spin qubits}

\author{Esteban A. Rodr\'iguez-Mena}
\affiliation{Univ. Grenoble Alpes, CEA, IRIG-MEM-L\_Sim, Grenoble, France.}
\author{Jos\'e Carlos Abadillo-Uriel}
\affiliation{Univ. Grenoble Alpes, CEA, IRIG-MEM-L\_Sim, Grenoble, France.}
\affiliation{Instituto de Ciencia de Materiales de Madrid (ICMM), Consejo Superior de Investigaciones Cientificas (CSIC), Sor Juana Ines de la Cruz 3, 28049 Madrid, Spain}%
\author{Gaëtan Veste}
\affiliation{Univ. Grenoble Alpes, CEA, LETI, Grenoble, France.}%
\author{Biel Martinez}
\affiliation{Univ. Grenoble Alpes, CEA, IRIG-MEM-L\_Sim, Grenoble, France.}%
\affiliation{Univ. Grenoble Alpes, CEA, LETI, Grenoble, France.}%
\author{Jing Li}
\affiliation{Univ. Grenoble Alpes, CEA, LETI, Grenoble, France.}%
\author{Benoît Sklénard}
\affiliation{Univ. Grenoble Alpes, CEA, LETI, Grenoble, France.}%
\author{Yann-Michel Niquet}
\email{yniquet@cea.fr}
\affiliation{Univ. Grenoble Alpes, CEA, IRIG-MEM-L\_Sim, Grenoble, France.}%

\date{\today}

\begin{abstract}
We investigate the existence of linear-in-momentum spin-orbit interactions in the valence band of Ge/GeSi heterostructures using an atomistic tight-binding method. We show that symmetry breaking at the Ge/GeSi interfaces gives rise to a linear Dresselhaus-type interaction for heavy-holes. This interaction results from the heavy-hole/light-hole mixings induced by the interfaces and can be captured by a suitable correction to the minimal Luttinger-Kohn, four bands $\vec{k}\cdot\vec{p}$ Hamiltonian. It is dependent on the steepness of the Ge/GeSi interfaces, and is suppressed if interdiffusion is strong enough. Besides the Dresselhaus interaction, the Ge/GeSi interfaces also make a contribution to the in-plane gyromagnetic $g$-factors of the holes. The tight-binding calculations also highlight the existence of a small linear Rashba interaction resulting from the couplings between the heavy-hole/light-hole manifold and the conduction band enabled by the low structural symmetry of Ge/GeSi heterostructures. These interactions can be leveraged to drive the hole spin. The linear Dresselhaus interaction may, in particular, dominate the physics of the devices for out-of-plane magnetic fields. When the magnetic field lies in-plane, it is, however, usually far less efficient than the $g$-tensor modulation mechanisms arising from the motion of the dot in non-separable, inhomogeneous electric fields and strains. 
\end{abstract}

\maketitle

\section{Introduction}

Hole spins in semiconductor quantum dots \cite{Burkard23,Fang23} can be efficiently manipulated with electric fields thanks to the strong spin-orbit interaction (SOI) in the valence bands of these materials \cite{Winkler03,Kloeffel11,Kloeffel18}. Fast electrically-driven spin Rabi oscillations have thus been reported in Si/SiO$_2$ \cite{Maurand16,Crippa18,Camenzind22} and Ge/GeSi \cite{Watzinger18,Hendrickx20b,Froning21,wang2022ultrafast} heterostructures. Silicon and Germanium can, in particular, be purified isotopically in order to get rid of the detrimental hyperfine interactions with the nuclear spins \cite{fischer2008spin,Mazzocchi19,moutanabbir23}. Although SOI also couples the spin to charge noise, there has been theoretical and experimental demonstrations of the existence of operational ``sweet spots'' where the hole is resilient to dephasing \cite{Froning21,Bosco21,Wang21,Piot22,Hendrickx23}. At these sweet spots, the Rabi frequency can actually be maximal owing to ``reciprocal sweetness'' relations with the dephasing rate \cite{michal2022tunable}, which allows for fast manipulation with lifetimes comparable to electrons in an artificial SOI (micro-magnets) \cite{Piot22,Yoneda18}. Holes also hold promises for strong spin-photon interactions, enabling long-range coupling between spin quantum bits (qubits) using circuit quantum electrodynamics \cite{Kloeffel13,Bosco22,yu2022strong,michal2022tunable}.

Ge/GeSi hole spin qubits \cite{Scappucci20} have made remarkable progress in the past few years \cite{Hendrickx20b,Hendrickx20}, with the demonstration of a four qubits processor \cite{Hendrickx21} and the achievement of charge control in a sixteen dots array \cite{Borsoi22}. Epitaxial Ge/GeSi interfaces are indeed much cleaner than Si/SiO$_2$ interfaces, which reduces in principle the level of charge noise and disorder near the qubits \cite{Martinez2022}. Also, holes are much lighter in Ge than in Si, which further mitigates their sensitivity to disorder and relaxes the constraints on dot size and gate pitch.

The landscape of spin-orbit interactions in Ge/GeSi nanostructures is very rich. The bulk valence band Bloch functions of Ge and Si are degenerate at $\Gamma$ and give rise to ``heavy-hole'' (HH) and ``light-hole'' (LH) bands with different masses \cite{Winkler03,Luttinger56,KP09}. The admixture of HH and LH components by confinement in low-dimensional structures leads to various kinds of couplings between the hole spin and its momentum. In planar (2D) heterostructures grown along $z=[001]$, the resulting Rashba-type SOI is cubic in the in-plane wave vector components $k_x$ and $k_y$ at lowest order in perturbation \cite{Rashba88,Winkler03,Marcellina17,Wang21,Terrazos21}. In nanowire (1D) structures, this ``direct'' Rashba SOI becomes typically linear-in-momentum \cite{Kloeffel11,Kloeffel18}. In quantum dot (0D) structures, the relevant SOI depends on the symmetry of the device. The cubic Rashba interaction dominates the physics of disk-shaped (2D-like) dots, while the linear Rashba interaction can prevail in squeezed (1D-like) dots \cite{Bosco21b}. 

The interplay between the kinetic and Zeeman Hamiltonian of holes also shapes the gyromagnetic $g$-factors of the dots. These corrections arise at the same order in the HH-LH band gap than the Rashba SOI \cite{Ares13,Michal21,Uriel22}, and are, therefore, independent manifestations of spin-orbit coupling in the valence bands. Both the Rashba SOI and the $g$-factor modulations can be leveraged to drive the spin in an AC electric field resonant with the Zeeman splitting: the motion of the dot as a whole couples to the spin through the Rashba spin/momentum interaction on the one hand \cite{Rashba03,Golovach06,Rashba08}, while the deformations of the dot give rise to ``$g$-tensor modulation resonance'' ($g$-TMR) on the other hand \cite{Kato03}. The Rabi oscillations generally result from a combination of these two mechanisms \cite{Crippa18,Venitucci18}.

Ge dots in Ge/GeSi heterostructures are usually manipulated under in-plane magnetic fields as this best decouples the hole spin from the nuclear spins \cite{fischer2008spin}. The actual mechanisms at play in these structures are still an open question. Indeed, the cubic Rashba SOI is little efficient \cite{Uriel22}, while significant direct linear Rashba SOI calls for heavily squeezed dots \cite{Bosco21b}. It has been shown in Refs.~\cite{Uriel22} and \cite{martinez2022hole} that a variety of $g$-TMR mechanisms can actually give rise to Rabi oscillations for in-plane magnetic fields \cite{Hendrickx23}. They involve the non-separability of the confinement potential (coupling between the in-plane and vertical motions of the driven dot), the inhomogeneity of the AC drive field (that squeezes the dot dynamically), and the inhomogeneous strain fields \cite{Corley2023} induced by the thermal contraction of the metal gates upon cool-down (that modulate the $g$-tensor of the driven dot). 

It is yet unclear whether the simplest Luttinger-Kohn (LK) Hamiltonian \cite{Winkler03,Luttinger56,KP09} used to model the HH/LH manifold \cite{martinez2022hole,Uriel22,Salfi22,Malkoc22,Wang22,sarkar2023} catches all linear-in-momentum SOIs that could help to drive spin qubits under in-plane magnetic fields. Indeed, this Hamiltonian is usually more symmetric in nanostructures than the atomic lattice and may, therefore, miss some of the emerging interactions. Also, the HH/LH manifold can mix with remote bands, which, while much farther, may bring sizable corrections to the effective SOIs. In particular, Refs.~\cite{Xiong21,Xiong22b} have highlighted with atomistic pseudo-potential calculations the existence of a linear-in-momentum SOI (primarily interpreted as a Rashba-type interaction) in 2D heterostructures. This linear SOI is not captured by the LK model but is much stronger than the cubic Rashba SOI in the range $k\sim\pi/d$ relevant for quantum dots with diameters $d\gtrsim 50$\,nm \cite{Liu22}. These calculations were, however, performed on heavily strained Ge/Si superlattices instead of Ge/GeSi superlattices.

In this work, we perform atomistic tight-binding (TB) calculations on the Ge/Ge$_{0.8}$Si$_{0.2}$ superlattices used in the recent experiments on Ge spin qubits \cite{Hendrickx20b,Hendrickx20,Hendrickx21,Hendrickx23}. We indeed evidence the existence of a significant linear-in-momentum SOI in these 2D heterostructures. We show, however, that it is actually dominated by a Dresselhaus-type interaction resulting from symmetry breaking by the Ge/GeSi interfaces (interface inversion asymmetry or IIA \cite{Vervoort97,Vervoort99,Winkler03,Luo10}). This Dresselhaus SOI is highly dependent on the quality of these interfaces and disappears once they get interdiffused over a few monolayers. It results from peculiar HH-LH mixings and can thus be captured by a suitable correction to the LK Hamiltonian \cite{Winkler03,Aleiner92,Ivchenko96}. This correction also interferes with the Zeeman Hamiltonian of the hole and thus slightly shifts the in-plane $g$-factors of the quantum dots. The Dresselhaus SOI goes along with a much smaller linear Rashba SOI that is independent on the status of the interfaces and results from mixings between the HH/LH manifold and the remote conduction bands allowed by the global structural inversion asymmetry (SIA) of the heterostructure \cite{Winkler03}. 

We review spin-orbit interactions in a 2D heavy-hole gas in section \ref{sec:SO}, then discuss the TB calculations in section \ref{sec:TB}, the nature of the SOIs and their description in the LK Hamiltonian in section \ref{sec:KP}, and finally draw conclusions for Ge/SiGe spin qubits in section \ref{sec:qubits}.

\section{Spin-orbit interactions in a 2D heavy-hole gas}
\label{sec:SO}

In bulk Germanium, the Bloch functions of the six topmost valence bands are, in the simplest TB approximation, bonding combinations of the atomic $4p$ orbitals with angular momentum $\ell=1$ and spin $s=\frac{1}{2}$ \cite{Winkler03,Luttinger56,KP09}. These Bloch functions are mixed by the atomic spin-orbit interaction $H_\mathrm{so}\propto\vec{L}\cdot\vec{S}$, where $\vec{L}$ and $\vec{S}$ are the orbital and spin angular momenta, respectively. They are consequently split into a quadruplet and a doublet, which can be mapped respectively onto the $J=\tfrac{3}{2}$ and $J=\tfrac{1}{2}$ eigenstates of the total angular momentum $\vec{J}=\vec{L}+\vec{S}$. Choosing $z=[001]$ as the quantization axis, the $\ket{J=\tfrac{3}{2},J_z=\pm\tfrac{3}{2}}$ Bloch functions give rise to ``heavy-hole'' (HH) bands along $z$, while the $\ket{J=\tfrac{3}{2},J_z=\pm\tfrac{1}{2}}$ Bloch functions give rise to ``light-hole'' (LH) bands.

The electronic properties of a Ge/GeSi quantum well grown along $z=[001]$ can be characterized by its subband structure $E_n(\vec{k})$, where $\vec{k}\equiv(k_x,k_y)$ is the wave vector in the ($x=[100]$, $y=[010]$) plane. The two topmost valence subbands at $\vec{k}=\vec{0}$ are pure $\ket{J=\tfrac{3}{2},J_z=\pm\tfrac{3}{2}}$ states due to the heavier HH mass. This remains so in homogeneous, compressive biaxial strains $\varepsilon_{xx}=\varepsilon_{yy}=\varepsilon_\parallel<0$, $\varepsilon_{zz}=\varepsilon_\perp>0$ that further promote heavy holes. These subbands are twofold degenerate at $\Gamma$ owing to time reversal symmetry (Kramers' degeneracy). However, they may be non-degenerate at finite $\vec{k}$ as a result of spin-orbit coupling. 

We label $\ket{0,\pm\tfrac{3}{2}}$ the confined Bloch functions of the topmost HH subbands at $\vec{k}=\vec{0}$. These subbands can be generally described by an effective Hamiltonian acting in the $\{\ket{0,+\tfrac{3}{2}},\ket{0,-\tfrac{3}{2}}\}$ subspace \cite{Winkler03}:
\begin{equation}
H_\mathrm{eff}(\vec{k})=\varepsilon(\vec{k})I+\eta_1(\vec{k})\sigma_1+\eta_2(\vec{k})\sigma_2+\eta_3(\vec{k})\sigma_3\,,
\label{eq:Heff}
\end{equation}
where $\sigma_i$ are the Pauli matrices. The eigenenergies $E_\pm(\vec{k})$ of this Hamiltonian are:
\begin{equation}
E_\pm(\vec{k})=\varepsilon(\vec{k})\pm\sqrt{\eta_1^2(\vec{k})+\eta_2^2(\vec{k})+\eta_3^2(\vec{k})}\,,
\label{eq:E12}
\end{equation}
so that the spin splitting is:
\begin{equation}
\Delta E(\vec{k})=2\sqrt{\eta_1^2(\vec{k})+\eta_2^2(\vec{k})+\eta_3^2(\vec{k})}\,.
\end{equation}
The functions $\eta_i(\vec{k})$ can be computed numerically from, e.g., a TB calculation (see section \ref{sec:TB}) or analytically (to some order in $\vec{k}$) from perturbation theory (see section \ref{sec:KP}). Since time-reversal symmetry transforms $\vec{k}$ into $-\vec{k}$, and $\sigma_i$ into $-\sigma_i$, $\varepsilon(\vec{k})$ must be an even function of $\vec{k}$ while the $\eta_i(\vec{k})$'s must be odd. They can thus be expanded in powers of the wave vector as:
\begin{equation}
\eta_i(\vec{k})=\sum_{u\in\{x,y\}} \alpha_{i,u} k_u+\sum_{u,v,w\in\{x,y\}} \beta_{i,uvw} k_u k_v k_w+...
\label{eq:etak}
\end{equation}
The patterns of non-zero coefficients $\alpha_{i,u}$ and $\beta_{i,uvw}$ are ruled by the spatial symmetries that leave the Hamiltonian invariant.

\begin{table}
\begin{tabular}{l|rrrrrrr}
\toprule
& $M_{x^\prime}$ & $M_{y^\prime}$ & $C_{2z}$ & $C_{4z}$ & $C_{2x}$ & $C_{2y}$ & $I$ \\
\hline
$\vec{x}^\prime$ & $-\vec{x}^\prime$ &  $\vec{x}^\prime$  & $-\vec{x}^\prime$ &  $\vec{y}^\prime$ & $-\vec{y}^\prime$ & $\vec{y}^\prime$ & $-\vec{x}^\prime$ \\
$\vec{y}^\prime$ &  $\vec{y}^\prime$ &  $-\vec{y}^\prime$ & $-\vec{y}^\prime$ & $-\vec{x}^\prime$ & $-\vec{x}^\prime$ & $\vec{x}^\prime$ & $-\vec{y}^\prime$ \\
\hline
$\sigma_1$ & $\sigma_1$  & $-\sigma_1$ & $-\sigma_1$ & $-\sigma_2$ &  $\sigma_2$ & $-\sigma_2$ & $\sigma_1$  \\
$\sigma_2$ & $-\sigma_2$ & $\sigma_2$  & $-\sigma_2$ & $\sigma_1$  &  $\sigma_1$ & $-\sigma_1$ & $\sigma_2$  \\
$\sigma_3$ & $-\sigma_3$ & $-\sigma_3$ & $\sigma_3$  & $\sigma_3$  & $-\sigma_3$ & $-\sigma_3$ & $\sigma_3$ \\
\hline
\botrule
\end{tabular}
\caption{Effects of different symmetry operations on the real and reciprocal space vectors $\vec{x}^\prime\parallel[110]$ and $\vec{y}^\prime\parallel[\overline{1}10]$, and on the Pauli matrices $\sigma_i$ of $\frac{3}{2}$ spins: the mirrors $M_{x^\prime}$ and $M_{y^\prime}$ orthogonal to $\vec{x}^\prime$ and $\vec{y}^\prime$, the $\pi$ ($C_{2z}$) and $\pi/2$ ($C_{4z}$) rotations around the $z$ axis, the $\pi$ rotations $C_{2x}$ and $C_{2y}$ around $x$ and $y$, and the inversion $I$. We emphasize that this table only holds for an appropriate choice of phase for the $\ket{0,\pm\tfrac{3}{2}}$ Bloch functions; other choices lead to different forms for Eq.~\eqref{eq:Heff} [see, e.g., Eqs.~\eqref{eq:H1_cubic} and \eqref{eq:H3_cubic}], but to the same physics anyhow, as the resulting Hamiltonians differ by an unitary transform.}
\label{tab:symmetries}
\end{table}

We shall forget in a first place about the disordered nature of the GeSi alloy and treat it as a homogeneous material. The point group of a heterostructure with an even number of Ge monolayers (MLs) and atomically flat interfaces is $D_{2h}$, which is centrosymmetric \cite{Golub04,Nestoklon08}. The point group of a heterostructure with an odd number of Ge MLs is $D_{2d}$; under a homogeneous vertical electric field $E_z$, the symmetry is lowered to $C_{2v}$ in both cases: the system is only invariant by a $\pi$ rotation $C_{2z}$ around the $z$ axis, and by the mirrors $M_{x^\prime}$ and $M_{y^\prime}$ orthogonal to $x^\prime=[110]$ and $y^\prime=[\overline{1}10]$. The phase of the $\ket{0,\pm\tfrac{3}{2}}$ basis functions can then be chosen so that the $\sigma_i$ matrices transform as in Table~\ref{tab:symmetries} under the different symmetry operations (also see Appendix \ref{app:symmetries}). We first emphasize that no symmetry-invariant Hamiltonian like Eq.~\eqref{eq:etak} can be built in a centrosymmetric group as $\vec{k}\to-\vec{k}$ but $\sigma_i\to\sigma_i$ under inversion. Therefore, there is no spin splitting in the $D_{2h}$ group (even number of Ge MLs without vertical electric field) \cite{Winkler03}. In the $C_{2v}$ group, the following linear-in-$k$ Hamiltonian is invariant under the $C_{2z}$, $M_{x^\prime}$ and $M_{y^\prime}$ operations:
\begin{equation}
H_\mathrm{eff}^{(1)}(\vec{k})=\alpha_{1,y^\prime}k_{y^\prime}\sigma_1^\prime+\alpha_{2,x^\prime}k_{x^\prime}\sigma_2^\prime\,,
\label{eq:H1}
\end{equation}
as well as the following cubic-in-$k$ Hamiltonian:
\begin{align}
H_\mathrm{eff}^{(3)}(\vec{k})&=\beta_{1,y^\prime y^\prime y^\prime}k_{y^\prime}^3\sigma_1^\prime+\beta_{2,x^\prime x^\prime x^\prime}k_{x^\prime}^3\sigma_2^\prime \nonumber \\ 
&+\beta_{1,x^\prime y^\prime x^\prime}k_{x^\prime}k_{y^\prime}k_{x^\prime}\sigma_1^\prime \nonumber \\ 
&+\beta_{2,y^\prime x^\prime y^\prime}k_{y^\prime}k_{x^\prime}k_{y^\prime}\sigma_2^\prime \,.
\label{eq:H3}
\end{align}
We have primed all Pauli matrices to highlight the particular choice of $\ket{0,\pm\tfrac{3}{2}}$ basis functions. We can alternatively write the above invariants as a function of $k_x$ and $k_y$:
\begin{equation}
H_\mathrm{eff}^{(1)}(\vec{k})=\alpha_\mathrm{D}(k_x\sigma_1+k_y\sigma_2)+\alpha_\mathrm{R}(k_x\sigma_2+k_y\sigma_1)\,,
\label{eq:H1_cubic}
\end{equation}
where:
\begin{subequations}
\begin{align}
\alpha_\mathrm{D}&= \frac{1}{2}\left(\alpha_{1,y^\prime}-\alpha_{2,x^\prime}\right) \\
\alpha_\mathrm{R}&=-\frac{1}{2}\left(\alpha_{1,y^\prime}+\alpha_{2,x^\prime}\right) \,. 
\end{align}
\end{subequations}
The first $\propto\alpha_\mathrm{D}$ term is the Dresselhaus-type interaction for $\frac{3}{2}$ spins, and the second $\propto\alpha_\mathrm{R}$ one is the Rashba-type interaction \footnote{The Rashba and Dresselhaus Hamiltonians for holes slightly differ from those for electrons, $H_\mathrm{R}\propto k_x\sigma_2-k_y\sigma_1$ and $H_\mathrm{D}\propto k_x\sigma_1-k_y\sigma_2$, because spins $\tfrac{3}{2}$ do not transform as spins $\tfrac{1}{2}$ under the symmetry operations of Table~\ref{tab:symmetries} (also see Appendix \ref{app:symmetries}).}. We emphasize that the $\ket{0,\pm\tfrac{3}{2}}$ basis functions have also been rotated by $-\pi/4$ around $z$ when going from Eq.~\eqref{eq:H1} to Eq.~\eqref{eq:H1_cubic} (which amounts to a different phase choice, see Appendix \ref{app:symmetries}). The $\sigma_{1,2}$ and $\sigma_{1,2}^\prime$ matrices act, therefore, on different basis sets of the same subspace. We can likewise transform the cubic Hamiltonian:
\begin{align}
H_\mathrm{eff}^{(3)}(\vec{k})&=\beta_\mathrm{D}^\prime(k_x^3\sigma_1+k_y^3\sigma_2)+\beta_\mathrm{R}^\prime(k_x^3\sigma_2+k_y^3\sigma_1) \nonumber \\ 
&+\beta_\mathrm{D}(k_yk_xk_y\sigma_1+k_xk_yk_x\sigma_2) \nonumber \\ 
&+\beta_\mathrm{R}(k_yk_xk_y\sigma_2+k_xk_yk_x\sigma_1)\,,
\label{eq:H3_cubic}
\end{align}
with:
\begin{subequations}
\begin{align} 
\beta_\mathrm{D}^\prime&=\frac{1}{4}\left(\beta_{1,y^\prime y^\prime y^\prime}-\beta_{2,x^\prime x^\prime x^\prime}+\beta_{1,x^\prime y^\prime x^\prime}-\beta_{2,y^\prime x^\prime y^\prime}\right) \\
\beta_\mathrm{R}^\prime&=-\frac{1}{4}\left(\beta_{1,y^\prime y^\prime y^\prime}+\beta_{2,x^\prime x^\prime x^\prime}+\beta_{1,x^\prime y^\prime x^\prime}+\beta_{2,y^\prime x^\prime y^\prime}\right) \\
\beta_\mathrm{D}&=\frac{1}{4}\left(3\beta_{1,y^\prime y^\prime y^\prime}-3\beta_{2,x^\prime x^\prime x^\prime}-\beta_{1,x^\prime y^\prime x^\prime}+\beta_{2,y^\prime x^\prime y^\prime}\right) \\
\beta_\mathrm{R}&=-\frac{1}{4}\left(3\beta_{1,y^\prime y^\prime y^\prime}+3\beta_{2,x^\prime x^\prime x^\prime}-\beta_{1,x^\prime y^\prime x^\prime}-\beta_{2,y^\prime x^\prime y^\prime}\right)\,,
\end{align}
\end{subequations}
the corresponding cubic Dresselhaus and Rashba coefficients. We would like to emphasize that there are additional symmetry operations ($C_{2x}$ and $C_{2y}$ axes in the $D_{2d}$ group) that impose $\alpha_\mathrm{R}=\beta_\mathrm{R}=\beta_\mathrm{R}^\prime=0$ when $E_z=0$. We conclude from Eqs.~\eqref{eq:H1} and \eqref{eq:H3} that the effective spin-orbit Hamiltonian of the topmost HH subbands can be completely characterized up to third-order in $\vec{k}$ by the subband structure along $x^\prime=[110]$, $y^\prime=[\overline{1}10]$, and $x=[100]$. 

In the simplest, minimal $\vec{k}\cdot\vec{p}$ theory, the heavy and light holes are described by the four bands Luttinger-Kohn Hamiltonian (see section \ref{sec:KP}) \cite{Winkler03,Luttinger56,KP09}. The latter gives rise to a ``cubic Rashba'' contribution to $H_\mathrm{eff}^{(3)}(\vec{k})$ as a result of HH/LH mixings at $\vec{k}\ne\vec{0}$ \cite{Rashba88,Marcellina17,Wang21,Terrazos21}. It does not, however, bring forth a linear-in-$k$ spin-orbit interaction $H_\mathrm{eff}^{(1)}(\vec{k})$ in planar heterostructures. We evidence below with TB calculations that this interaction exists nonetheless and further explore its nature in section \ref{sec:KP}. We discuss its role in Ge spin-orbit qubits in section \ref{sec:qubits}.

Beforehand, we would like to discuss the role of alloy disorder in this problem. Strictly speaking, a GeSi alloy is a disordered material with no atomistic symmetry. Nonetheless, the alloy has {\it on average} (and on the macroscopic scale relevant when $|\vec{k}|\to 0$) the same diamond-like symmetry as pure silicon and germanium. Therefore, the above symmetry analysis holds for the average $\eta_i(\vec{k})$'s computed in large ``supercells'' with side $L\gg a$ in the $(xy)$ plane (with $a=5.658$\,\AA\ the lattice parameter of Ge). We will address below how the $\eta_i(\vec{k})$'s converge with increasing $L$. 

\section{Tight-binding calculations in {Ge/GeSi} heterostructures}
\label{sec:TB}

\subsection{Methodology}

\begin{figure}
\includegraphics[width=0.8\columnwidth]{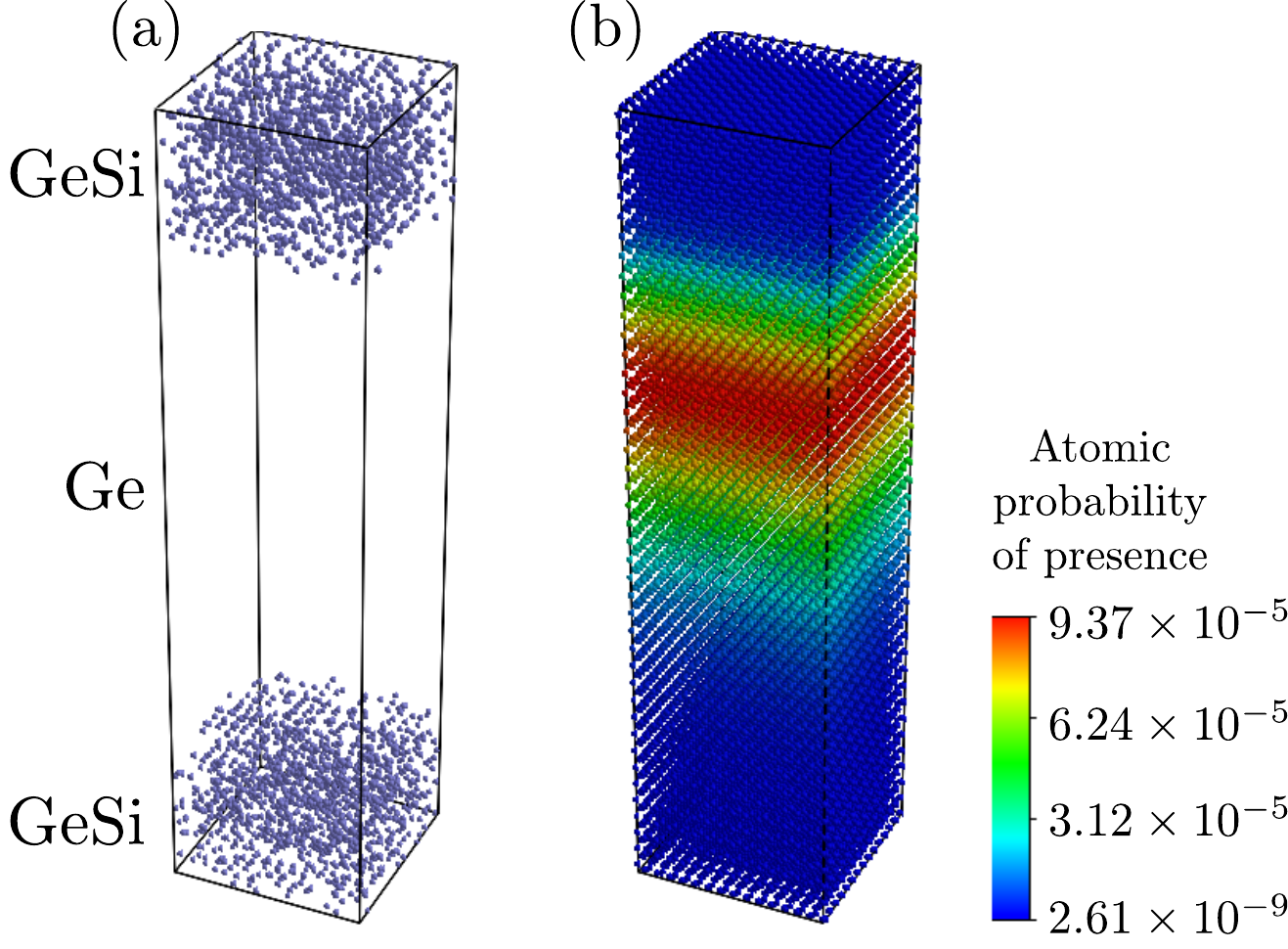}
\caption{(a) Supercell of a (Ge)$_{112}$/(Ge$_{0.8}$Si$_{0.2}$)$_{56}$ superlattice with 112 MLs ($\approx 15.8$\,nm) thick Ge wells separated by 56 MLs ($\approx 7.9$\,nm) thick Ge$_{0.8}$Si$_{0.2}$ barriers. Only the silicon atoms are shown for clarity. The structure is periodic in all directions. The in-plane side of the unit cell is $L=15L_0=6$\,nm. (b) Probability of presence of the hole on each atom, at vertical electric field $E_z=3$\,mV/nm.}
\label{fig:structure}
\end{figure}

The band structure of Ge/Ge$_{0.8}$Si$_{0.2}$ superlattices is computed with a nearest-neighbor $sp^3d^5s^*$ TB model \cite{DiCarlo03} that accounts for strains and reproduces the valence and conduction bands of Ge and Si over the whole first Brillouin zone \cite{Niquet09}. The superlattices comprise $N_\mathrm{Ge}=112$ MLs ($\approx 15.8$ nm) thick Ge wells separated by 56 MLs ($\approx 7.9$ nm) thick Ge$_{0.8}$Si$_{0.2}$ barriers. The Ge$_{0.8}$Si$_{0.2}$ alloy is either modeled as a random distribution (RD) of Ge and Si atoms (Fig.~\ref{fig:structure}) or as a virtual crystal (VC). This VC is a diamond-like material with a single kind of atom whose TB parameters are the appropriate averages of those of Si and Ge \footnote{In a Ge$_{1-x}$Si$_x$ alloy, the atomic energies $E_\alpha$ of orbital $\alpha\in\{s,p,d,s^*\}$ (as well as the other on-site parameters) are averaged according to the probability of finding a Si or Ge atom in the alloy:
\begin{equation*}
E_\alpha(x)=xE_\alpha(\mathrm{Si})+(1-x)E_\alpha(\mathrm{Ge})\,.
\end{equation*}
The nearest-neighbor parameters $V_{\alpha\beta}$ are averaged according to the probability of finding a Si$-$Si, Ge$-$Ge, Si$-$Ge or Ge$-$Si bond:
\begin{align*}
V_{\alpha\beta}(x)&=x^2V_{\alpha\beta}(\mathrm{Si}-\mathrm{Si})+(1-x)^2V_{\alpha\beta}(\mathrm{Ge}-\mathrm{Ge}) \\
&+x(1-x)[V_{\alpha\beta}(\mathrm{Si}-\mathrm{Ge})+V_{\alpha\beta}(\mathrm{Ge}-\mathrm{Si})]\,,
\end{align*}
where $V_{\alpha\beta}(\mathrm{A}-\mathrm{B})$ is the matrix element between orbital $\alpha$ on atom A and orbital $\beta$ on atom B.}. As such a VC does not break translational symmetry in the ($xy$) plane, the band structure can be directly computed in the minimal, primitive unit cell with unstrained side $L_0=a/\sqrt{2}$ along $x^\prime=[110]$ and $y^\prime=[\overline{1}10]$. Conversely, RD calculations are run in much larger unit cells with sides up to $L=41L_0$ and must be averaged over a few tens of realizations of the alloy. The VC unit cell hence contains $168$ atoms, while the largest RD unit cell contains $282\,408$ atoms.

In line with Ref. \cite{Sammak19,martinez2022hole,Uriel22}, we assume that the superlattice is grown on a thick Ge$_{0.8}$Si$_{0.2}$ buffer with a residual tensile in-plane strain $\varepsilon_{xx}=\varepsilon_{yy}=0.26$\%. The average in-plane strain in the Ge film is therefore $\varepsilon_{xx}=\varepsilon_{yy}=\varepsilon_\parallel=-0.63$\% \cite{martinez2022hole}. The atomic positions in the whole superlattice are relaxed with Keating's valence force field \cite{Keating66,Niquet09}. A sawtooth electric potential can be applied to the superlattice, characterized by a vertical electric field $E_z$ in the Ge well and $-2E_z$ in the Ge$_{0.8}$Si$_{0.2}$ barrier.

We compute the superlattice band structure on a path from $k_{y^\prime}=0.01$\,\AA$^{-1}$ to $\Gamma$ then to $k_{x^\prime}=0.01$\,\AA$^{-1}$, and on a path from $\Gamma$ to $k_x=0.01$\,\AA$^{-1}$, and monitor the splitting $\Delta E(\vec{k})=E_+(\vec{k})-E_-(\vec{k})$ between the two topmost valence bands. This splitting is not, however, sufficient to de-embed the different $\eta_i$'s in Eq.~\eqref{eq:E12}. Therefore, we reconstruct the effective Hamiltonian, Eq.~\eqref{eq:Heff}, from the projections of the TB wave functions $\Psi_+(\vec{k})$ and $\Psi_-(\vec{k})$ on the $\ket{0,+\tfrac{3}{2}}$ and $\ket{0,-\tfrac{3}{2}}$ Bloch functions at $\Gamma$. For that purpose, we couple a small magnetic field $B_z$ to the physical spin in order to split (and identify) the $\ket{0,+\tfrac{3}{2}}$ and $\ket{0,-\tfrac{3}{2}}$ states at $\vec{k}=\vec{0}$ \footnote{More precisely, the $\ket{0,+\tfrac{3}{2}}$ and $\ket{0,-\tfrac{3}{2}}$ states are first computed at zero magnetic field, then the matrix $B_zS_z$ (with $S_z$ the physical spin) is diagonalized in the subspace $\{\ket{0,+\tfrac{3}{2}}$,$\ket{0,-\tfrac{3}{2}}\}$. The amplitude of $B_z$ is, therefore, irrelevant.}. We next choose the phase of the calculated $\ket{0,+\tfrac{3}{2}}$ and $\ket{0,-\tfrac{3}{2}}$ Bloch functions so that the Pauli matrices transform as close as possible to Table~\ref{tab:symmetries} under the symmetry operations of the $C_{2v}$ group (given that a perfect match is not possible in a RD alloy as the latter has an average macroscopic, but not an atomistic $C_{2v}$ symmetry). We finally introduce the projection matrix
\begin{equation}
P(\vec{k})=
\begin{pmatrix}
\big\langle 0,+\tfrac{3}{2}\big|\Psi_+(\vec{k})\big\rangle & \big\langle 0, +\tfrac{3}{2}\big|\Psi_-(\vec{k})\big\rangle \\
\big\langle 0,-\tfrac{3}{2}\big|\Psi_+(\vec{k})\big\rangle & \big\langle 0,-\tfrac{3}{2}\big|\Psi_-(\vec{k})\big\rangle 
\end{pmatrix}\,.
\label{eq:PTB}
\end{equation}
The effective Hamiltonian in the $\{\ket{0,+\tfrac{3}{2}}, \ket{0,-\tfrac{3}{2}}\}$ basis set is then
\begin{equation}
H_\mathrm{eff}(\vec{k})=P(\vec{k})
\begin{pmatrix}
E_+(\vec{k}) & 0 \\ 0 & E_-(\vec{k})
\end{pmatrix}    
P(\vec{k})^\dagger\,,
\label{eq:HeffTB}
\end{equation}
which can be uniquely decomposed as Eq.~\eqref{eq:Heff}. We emphasize that the $P(\vec{k})$ matrix is not strictly unitary as $\{\Psi_+(\vec{k}), \Psi_-(\vec{k})\}$ admix remote band contributions when $k=|\vec{k}|$ increases; however, $|P^\dagger(\vec{k})P(\vec{k})-I|$ remains typically smaller than 0.025 in the investigated $\vec{k}$ vector range (and so is the relative error on the extracted $\eta_i$'s). 

\subsection{Results}

\begin{figure}
\includegraphics[width=0.9\columnwidth]{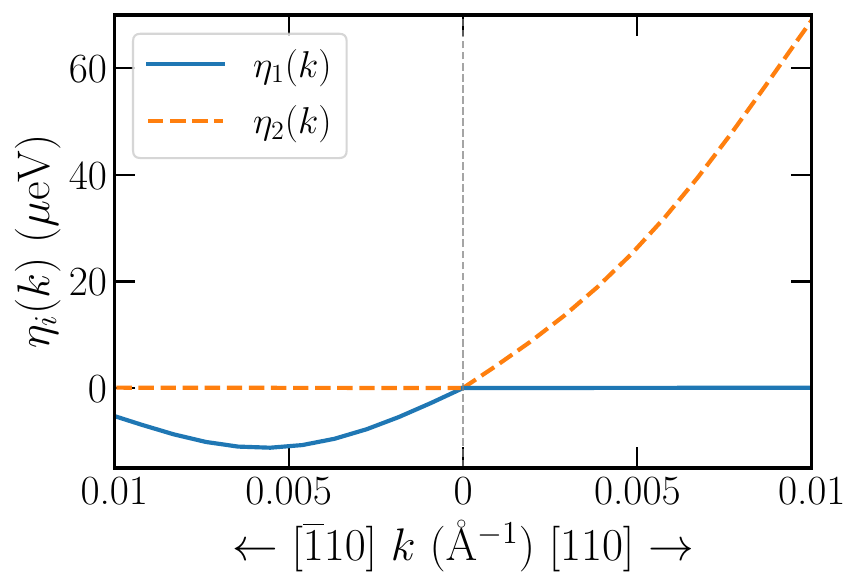}
\caption{The functions $\eta_{1,2}(\vec{k})$ [Eq.~\eqref{eq:Heff}] computed with the TB model in the (Ge)$_{112}$/(Ge$_{0.8}$Si$_{0.2}$)$_{56}$ superlattice, along a path from $k_{y^\prime}=0.1$\,\AA$^{-1}$ to $\Gamma$ then to $k_{x^\prime}=0.1$\,\AA$^{-1}$. The unit cell size is $L=32L_0$ and the vertical electric field is $E_z=3$\,mV/nm. The $\eta_i(\vec{k})$'s are averaged over 70 realizations of the RD alloy.}
\label{fig:etas}
\end{figure}

\begin{figure}
\includegraphics[width=0.9\columnwidth]{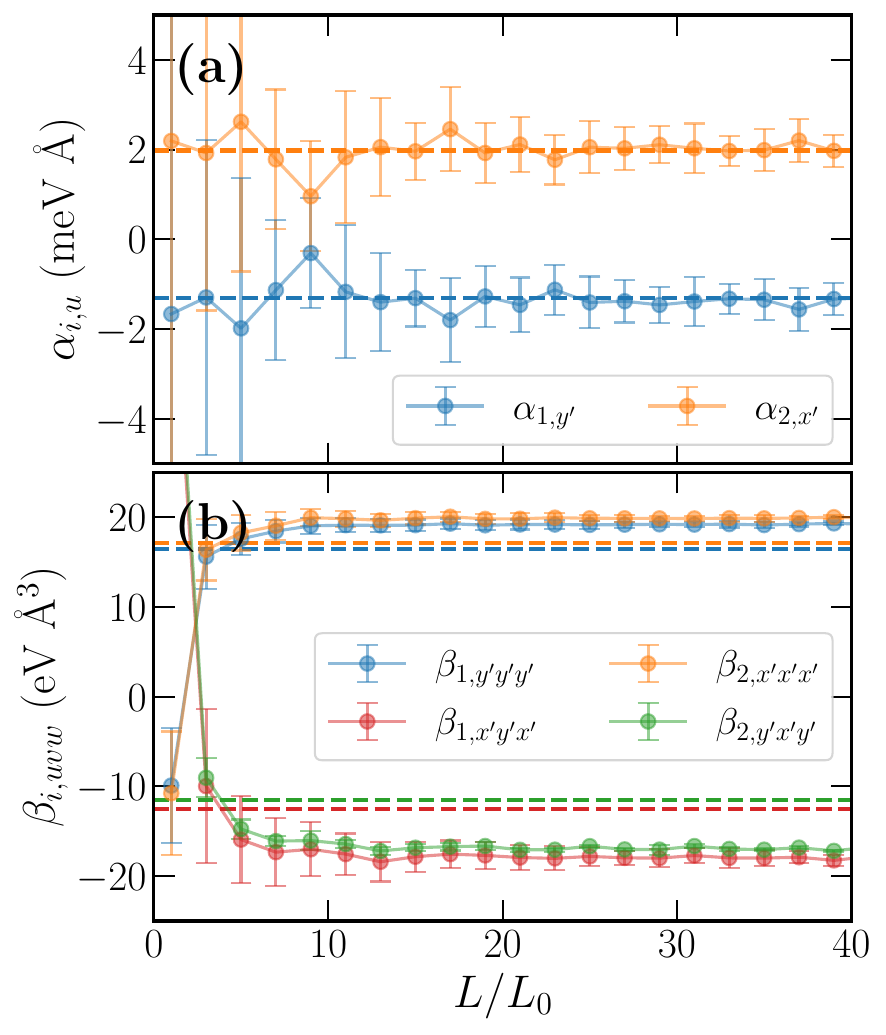}
\caption{(a) The linear coefficients $\alpha_{1,y^\prime}$ and $\alpha_{2,x^\prime}$ [Eq.~\eqref{eq:H1}] computed in the (Ge)$_{112}$/(Ge$_{0.8}$Si$_{0.2}$)$_{56}$ superlattices as a function of the supercell side $L$ ($E_z=3$\,mV/nm). The dots are the $\alpha$'s fitted to the band structure averaged over 70 realizations of the RD alloy. The error bars are the standard deviation of the $\alpha$'s fitted to each single realization. The horizontal dashed lines are the $\alpha$'s computed in the VC approximation. (b) Same for the cubic coefficients $\beta_{1,y^\prime y^\prime y^\prime}$, $\beta_{2,x^\prime x^\prime x^\prime}$, $\beta_{1,x^\prime y^\prime x^\prime}$,  and $\beta_{2,y^\prime x^\prime y^\prime}$. The latter two are extracted on the path from $\Gamma$ to $k_x=0.01$\,\AA$^{-1}$.}
\label{fig:convergence}
\end{figure}

As an example, we plot in Fig.~\ref{fig:etas} the $\eta_i$'s computed along the path $k_{y^\prime}=0.01$\,\AA$^{-1}\to\Gamma\to k_{x^\prime}=0.01$\,\AA$^{-1}$. The unit cell size is $L=32L_0$ and the vertical electric field is $E_z=3$\,mV/nm. The $\eta_i$'s are averaged over 70 realizations of the RD alloy. As expected from section \ref{sec:SO}, only $\eta_1$ is sizable along $[\overline{1}10]$, and only $\eta_2$ is sizable along $[110]$ ($\eta_3$, not displayed on Fig.~\ref{fig:etas}, is negligible in both directions). The effective spin Hamiltonian reads therefore:
\begin{equation}
H_\mathrm{eff}(\vec{k})=\eta_1(k_{y^\prime})\sigma_1^\prime+\eta_2(k_{x^\prime})\sigma_2^\prime\,.
\end{equation}
We next fit $\eta_1$ and $\eta_2$ with a fifth-order polynomial accounting for linear, cubic, and residual higher-order in $k$ spin-orbit interactions:
\begin{subequations}
\label{eq:alphabetas}
\begin{align}
\eta_1(k_{y^\prime})&=\alpha_{1,y^\prime} k_{y^\prime}+\beta_{1,y^\prime y^\prime y^\prime} k_{y^\prime}^3+\gamma_{1} k_{y^\prime}^5 \\
\eta_2(k_{x^\prime})&=\alpha_{2,x^\prime} k_{x^\prime}+\beta_{2,x^\prime x^\prime x^\prime} k_{x^\prime}^3+\gamma_{2} k_{x^\prime}^5\,.
\end{align}
\end{subequations}
The $\gamma$'s make sizable contributions only at large $k$ and are therefore little relevant in large quantum dots. The values of the linear coefficients $\alpha_{1,y^\prime}$ and $\alpha_{2,x^\prime}$ are plotted as a function of the unit cell side $L$ in Fig.~\ref{fig:convergence}a. The fits on the average $\eta_i$'s (dots) are reported along with the standard deviation on each single realization of the RD alloy (error bars). The horizontal dashed lines are the $\alpha$'s obtained in the VC approximation (that are independent on $L$). Although the different realizations of the RD alloy are still scattered at $L=41L_0=16.4$ nm, the average $\alpha$'s are reasonably well converged and match the VC data. Similar convergence is reached for the cubic coefficients $\beta$ (Fig.~\ref{fig:convergence}b), but the VC approximation performs slightly worse \YMN{(possibly because cubic interactions probe larger $k$/shorter wavelengths and are thus more sensitive to alloy disorder in the barriers).} We conclude from these plots that alloy disorder may significantly scatter the spin-orbit parameters of small quantum dots with diameters $d\lesssim 20$\,nm; yet for practical quantum dots with diameters $d\gtrsim 50$ nm, the VC approximation provides a reliable description of spin-orbit splittings, especially at first-order in $k$.

\begin{figure}
\includegraphics[width=0.9\columnwidth]{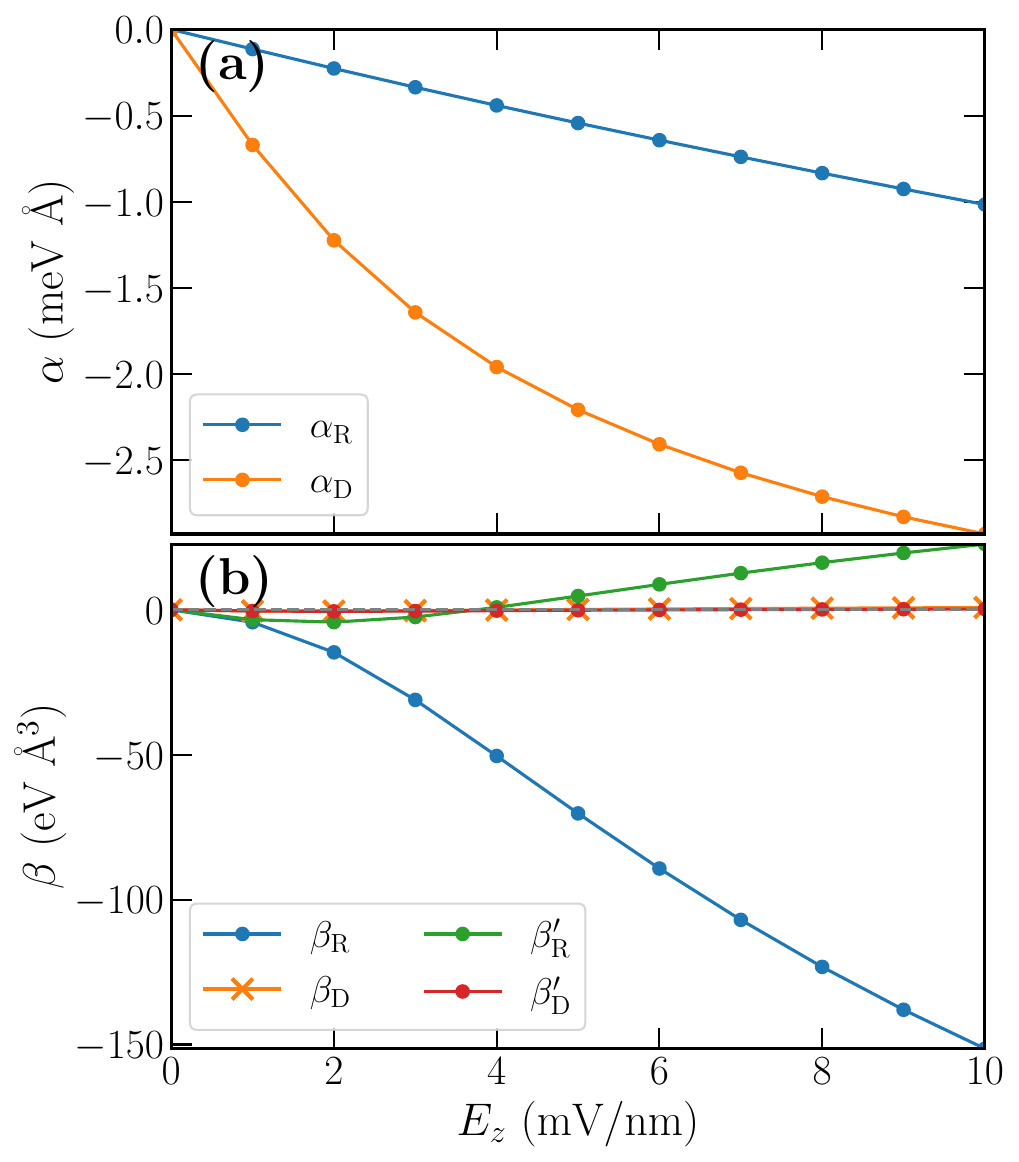}
\caption{(a) Linear [Eq.~\eqref{eq:H1_cubic}] and (b) cubic [Eq.~\eqref{eq:H3_cubic}] Rashba and Dresselhaus coefficients as a function of the vertical electric field $E_z$, computed in the VC approximation in a  (Ge)$_{112}$/(Ge$_{0.8}$Si$_{0.2}$)$_{56}$ superlattice.}
\label{fig:alphabetaEz}
\end{figure}

\begin{figure}
\includegraphics[width=0.9\columnwidth]{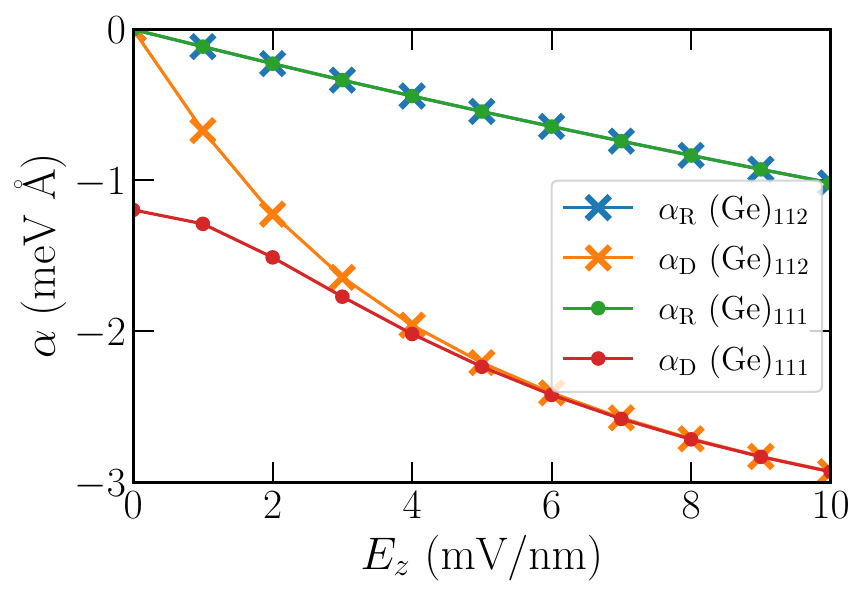}
\caption{Linear Rashba and Dresselhaus coefficients [Eq.~\eqref{eq:H1_cubic}] as a function of the vertical electric field $E_z$, computed in the VC approximation in (Ge)$_{112}$/(Ge$_{0.8}$Si$_{0.2}$)$_{56}$ and (Ge)$_{111}$/(Ge$_{0.8}$Si$_{0.2}$)$_{56}$ superlattices.}
\label{fig:monolayer}
\end{figure}

The linear Rashba and Dresselhaus coefficients $\alpha_\mathrm{R}$ and $\alpha_\mathrm{D}$ calculated in the VC approximation are plotted as a function the vertical electric field $E_z$ in Fig.~\ref{fig:alphabetaEz}a, and the $\beta$'s are plotted in Fig.~\ref{fig:alphabetaEz}b. The $\alpha$'s first increase linearly with $E_z$, then $\alpha_\mathrm{D}$ bends. Indeed, they must both tend to zero when $E_z\to 0$ as the inversion symmetry is restored (on average in the alloy). The inflection of $\alpha_\mathrm{D}$ at large $E_z$ results from an increase of the HH/LH band gap discussed in more detail in section \ref{sec:KP}. The $\beta$'s also tend to zero when $E_z\to 0$ but show a more complex behavior. They highlight the prevalence of cubic Rashba interactions (the Dresselhaus components being non-zero but negligible on the scale of Fig.~\ref{fig:alphabetaEz}b). This is qualitatively consistent with the known existence of a cubic Rashba SOI due to HH/LH mixings by vertical confinement in $[001]$ heterostructures \cite{Marcellina17,Wang21,Terrazos21}. These interactions will be discussed more quantitatively in section \ref{sec:KP}. We also plot on Fig.~\ref{fig:monolayer} the $\alpha$'s computed in a superlattice with $N_\mathrm{Ge}=111$ instead of $N_\mathrm{Ge}=112$ Ge MLs. For such an odd number of MLs, the system lacks inversion symmetry down to $E_z=0$ so that $\alpha_\mathrm{D}$ does not vanish. It matches however the 112 MLs data at large $E_z$ where electrical confinement prevails over structural confinement (the hole getting squeezed at the top interface, see Fig.~\ref{fig:structure}). The linear Rashba coefficient $\alpha_\mathrm{R}$ is, on the other hand, almost insensitive to the parity of $N_\mathrm{Ge}$ (and is still zero when $E_z=0$, as expected from section \ref{sec:SO}). The cubic Rashba coefficients (not shown) are, likewise, little dependent on the parity of $N_\mathrm{Ge}$, while for odd $N_\mathrm{Ge}$ the cubic Dresselhaus coefficients remain non-zero (but negligible) when $E_z\to 0$. 

The TB calculations therefore highlight the existence of a linear-in-$k$ spin-orbit interaction in the Ge films, as already evidenced in Refs.~\cite{Xiong21,Xiong22b}. There are, nonetheless, two major differences with these works. First, we account for realistic GeSi alloys (either as RDs or as VCs), whereas Refs. \cite{Xiong21,Xiong22b} considered pure Ge/Si superlattices where the spin-orbit interactions are expected to be different (since the confinement is sharper). Second, we find that the spin-orbit coefficients $\alpha_{1,y^\prime}$ (along $[\overline{1}10]$) and $\alpha_{2,x^\prime}$ (along $[110]$) differ in both magnitude and sign, as allowed by the symmetry analysis of section \ref{sec:SO}. According to Eqs.~\eqref{eq:H1_cubic}, such a SOI can be interpreted in the cubic axes as a dominant Dresselhaus-type interaction with coefficient $\alpha_\mathrm{D}=(\alpha_{1,y^\prime}-\alpha_{2,x^\prime})/2$, along with a much smaller Rashba-type interaction with coefficient $\alpha_\mathrm{R}=-(\alpha_{1,y^\prime}+\alpha_{2,x^\prime})/2$. The spin splittings calculated in Ref.~\cite{Xiong21} were, on the opposite, primarily interpreted as the fingerprints of a Rashba interaction. However, as discussed in section \ref{sec:TB}, the spin splittings are not necessarily sufficient to determine the nature of the SOI and shall be seconded with an analysis of the wave functions along the lines of Eqs.~\eqref{eq:PTB}--\eqref{eq:HeffTB}. For comparison, we model the (Ge)$_{40}$/(Si)$_{20}$ superlattice of Ref.~\cite{Xiong21} in Appendix \ref{app:GeSi} and recover the same qualitative results (but a stronger linear-in-$k$ SOI) that we can unambiguously assign to a Dresselhaus-type interaction. We have also benchmarked TB against {\it ab initio} calculations, achieving reasonable agreement on the linear Dresselhaus coefficient (see Appendix \ref{app:abinitio}).

\subsection{Role of interfaces and interdiffusion}
\label{sec:interdiffusion}

The strong asymmetry between the $[110]$ and $[\overline{1}10]$ axes underlying the Dresselhaus interaction can be related to symmetry breaking by the Ge/GeSi interfaces (IIA) \cite{Ivchenko96,Vervoort99, Winkler03}. Indeed, at such an interface, the in-plane projection of all bonds from the Ge well to the GeSi barrier are oriented either along $[110]$ or along $[\overline{1}10]$. The orientation changes each time a ML of Ge is added at the interface. The coefficients $\alpha_{1,y^\prime}$ and $\alpha_{2,x^\prime}$ actually swap (and thus $\alpha_\mathrm{D}$ changes sign) when the whole Ge well is shifted up or down by one ML. In Ge wells with an even (odd) number of MLs, the orientation of the bonds is the same (is different) at the two interfaces, whose effects thus add up (cancel each other).

\begin{figure}
\includegraphics[width=0.9\columnwidth]{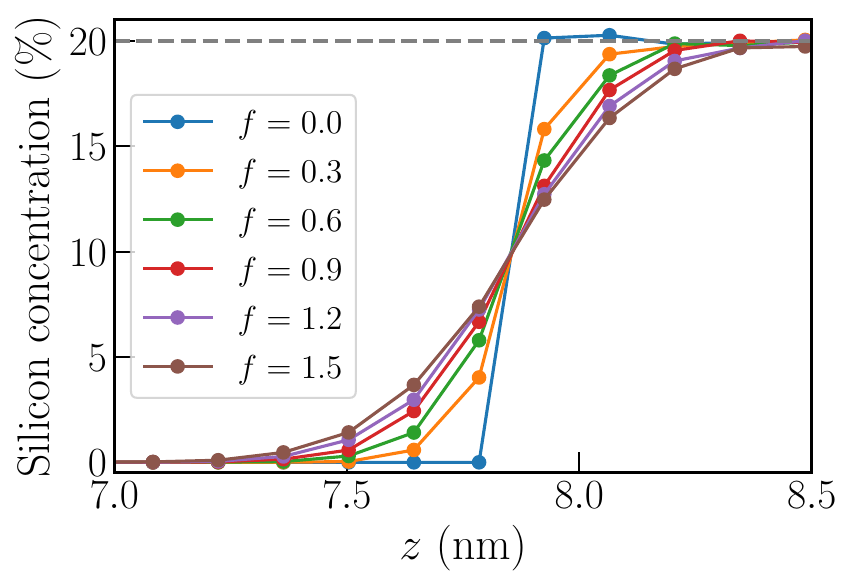}
\caption{Silicon concentration profiles near the top interface of interdiffused (Ge)$_{112}$/(Ge$_{0.8}$Si$_{0.2}$)$_{56}$ superlattices for different fractions $f$ of swapped nearest-neighbor pairs. The center of the Ge well is at $z=0$.}
\label{fig:Siconc}
\end{figure}

The difference between $\alpha_{1,y^\prime}$ and $\alpha_{2,x^\prime}$ shall thus average out if the interfaces between Ge and GeSi are strongly enough interdiffused. Actually, the symmetry of the heterostructure is then promoted to $C_{4v}$, which contains a fourfold rotation axis around $z$. This additional symmetry operation imposes $\alpha_{1,y^\prime}=\alpha_{2,x^\prime}$ according to Table~\ref{tab:symmetries}. The resulting SOI shall therefore appear as a pure Rashba-type interaction in the cubic axis set.

\begin{figure}
\includegraphics[width=0.9\columnwidth]{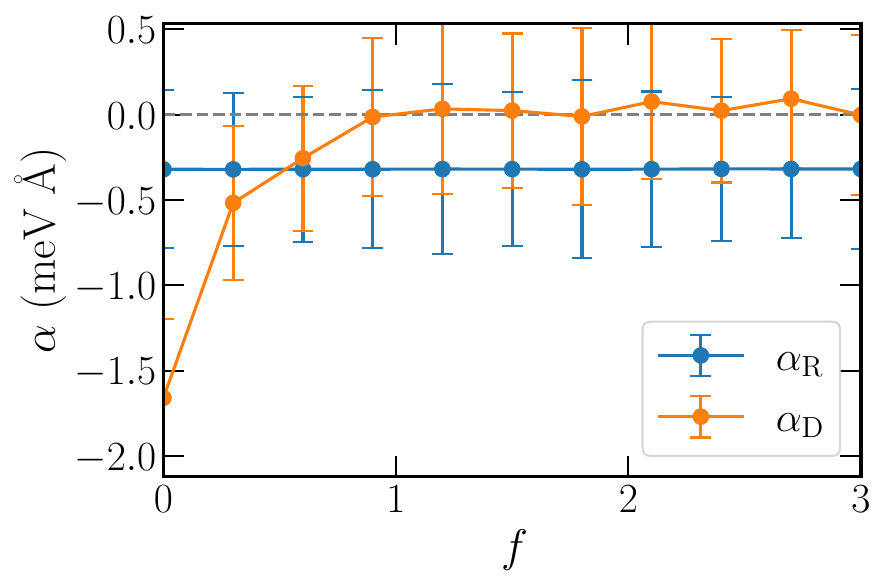}
\caption{Linear Rashba and Dresselhaus coefficients [Eq.~\eqref{eq:H1_cubic}] computed in (Ge)$_{112}$/(Ge$_{0.8}$Si$_{0.2}$)$_{56}$ superlattices as a function of the interdiffusion strength $f$, at vertical electric field $E_z=3$ mV/nm. The data are computed in supercells with side $L=32L_0$ and are averaged over $70$ realizations of the RD alloy.}
\label{fig:alphaf}
\end{figure}

\begin{figure}
\includegraphics[width=0.9\columnwidth]{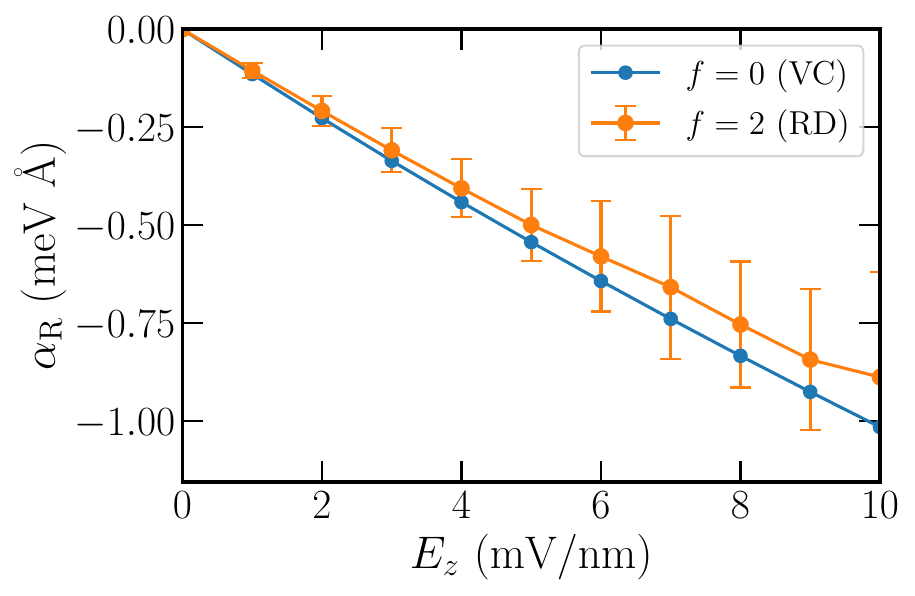}
\caption{The Rashba coefficient $\alpha_\mathrm{R}(f)$ in (Ge)$_{112}$/(Ge$_{0.8}$Si$_{0.2}$)$_{56}$ superlattices as a function of the vertical electric field $E_z$ at $f=0$ and $f=2$. The data at $f=2$ are averaged over 20 realizations of the RD alloy, while the data at $f=0$ are computed in the VC.}
\label{fig:rashbaEz}
\end{figure}

To highlight this trend, we have computed the TB band structures of interdiffused RD alloys. For that purpose, we start from the previous structures with sharp Ge/GeSi interfaces, then randomly swap $M$ pairs of nearest-neighbor atoms in the supercell. The strength of the interdiffusion is hence controlled by the ratio $f=M/N$ between $M$ and the total number of atoms $N$ in the supercell. The resulting Si concentration profiles near the top interface, averaged over 70 realizations of the alloy, are plotted in Fig.~\ref{fig:Siconc} for various $f$'s. The sharp step for $f=0$ is smoothed over $\approx 6$ MLs for $f=1$. The coefficients $\alpha_\mathrm{D}$ and $\alpha_\mathrm{R}$ computed with the TB model are plotted as a function of $f$ in Fig.~\ref{fig:alphaf} (at vertical electric field $E_z=3$\,mV/nm). As expected, $\alpha_{1,y^\prime}\to\alpha_{2,x^\prime}\approx0.31$\,meV\,\AA\ as soon as $f\gtrsim 1$ so that $\alpha_\mathrm{D}\to 0$. Moreover, the linear Rashba coefficient $\alpha_\mathrm{R}$ is almost independent on $f$. This is further emphasized in Fig.~\ref{fig:rashbaEz}, which compares $\alpha_\mathrm{R}(f=2)$ with $\alpha_\mathrm{R}(f=0)$ as a function of the vertical electric field $E_z$. Therefore, the interdiffusion only suppresses the linear Dresselhaus SOI, while the linear Rashba SOI appears unrelated to the interfaces. We will further discuss its nature in the next section. The same conclusions hold for the cubic Dresselhaus and Rashba components, which are respectively suppressed and (almost) independent on the interdiffusion strength.

\section{Nature of the spin-orbit interactions and $\vec{k}\cdot\vec{p}$ models}
\label{sec:KP}

In this section, we further discuss the nature of the different spin-orbit interactions evidenced in the previous section and their description in the $\vec{k}\cdot\vec{p}$ approximation. We first introduce the four bands Luttinger-Kohn Hamiltonian, then discuss the linear Dresselhaus interaction, and finally the Rashba interactions.

\subsection{The Luttinger-Kohn Hamiltonian}

The Luttinger-Kohn $\vec{k}\cdot\vec{p}$ Hamiltonian provides the simplest, minimal description of the HH/LH manifold of diamond-like materials \cite{Winkler03,Luttinger56,KP09}. It reads in the basis set of $J_z=\{+\tfrac{3}{2},+\tfrac{1}{2},-\tfrac{1}{2},-\tfrac{3}{2}\}$ bulk Bloch functions at $\Gamma$:
\begin{equation}
H_\mathrm{LK}=-\begin{pmatrix}
P+Q & -S & R & 0 \\
-S^\dagger & P-Q & 0 & R \\
R^\dagger & 0 & P-Q & S \\
0 & R^\dagger & S^\dagger & P+Q
\end{pmatrix}
\label{eq:LK}\,,
\end{equation}
where: 
\begin{subequations}
\begin{align}
P&=\frac{\hbar^2}{2m_0}\gamma_1(k_x^2+k_y^2+k_z^2) \\
Q&=\frac{\hbar^2}{2m_0}\gamma_2(k_x^2+k_y^2-2k_z^2) \\
R&=\frac{\hbar^2}{2m_0}\sqrt{3}\left[-\gamma_2(k_x^2-k_y^2)+2i\gamma_3k_xk_y\right] \\
S&=\frac{\hbar^2}{2m_0}2\sqrt{3}\gamma_3(k_x-ik_y)k_z\,,
\end{align}
\end{subequations}
with $\gamma_1$, $\gamma_2$ and $\gamma_3$ the Luttinger parameters that characterize the mass of the holes ($\gamma_1=13.38$, $\gamma_2=4.24$ and $\gamma_3=5.69$ in Ge). In biaxial strains $\varepsilon_{xx}=\varepsilon_{yy}=\varepsilon_\parallel$, $\varepsilon_{zz}=\varepsilon_\perp$, the HH are further split from the LH Bloch functions by the transformation:
\begin{equation}
Q\to Q-b_v(\varepsilon_\parallel-\varepsilon_\perp)\,,
\end{equation}
where $b_v$ is the uniaxial deformation potential of the valence bands ($b_v=-2.16$\,eV in Ge). In the present calculations, $\varepsilon_\parallel=-0.63\%$ and $\varepsilon_\perp=0.47\%$ in the Ge well \footnote{\YMN{In Ge$_{0.8}$Si$_{0.2}$, $\gamma_1=11.56$, $\gamma_2=3.46$, $\gamma_3=4.84$, $\kappa=2.64$, $q=0.05$, and $b_v=-2.19$\,eV. The unstrained valence band offset is $\Delta_\mathrm{VBO}=0.138$\,eV, and the hydrostatic valence band deformation potentials are $a_v=2$\,eV in Ge and $a_v=2.02$\,eV in  Ge$_{0.8}$Si$_{0.2}$}}.

In Ge/GeSi planar heterostructures, the confined hole wave functions are described by a set of four HH/LH envelopes that fulfill the set of differential equations obtained by substituting $k_z\to-i\tfrac{\partial}{\partial z}$ in Eq.~\eqref{eq:LK} \cite{Luttinger55,KP09}. There are, therefore, no explicit relations to the atomic lattice in this envelope functions approximation, so that the LK Hamiltonian is usually more symmetric than expected. As shown below, the main features evidenced in section \ref{sec:TB} can nonetheless be captured with simple corrections to the LK Hamiltonian.

In a finite magnetic field $\vec{B}$, the $J_z=\{+\tfrac{3}{2},+\tfrac{1}{2},-\tfrac{1}{2},-\tfrac{3}{2}\}$ Bloch functions are also split and mixed by the Zeeman Hamiltonian:
\begin{equation}
H_\mathrm{Z}=-2\mu_B(\kappa\mathbf{B}\cdot\mathbf{J}+q\mathbf{B}\cdot\mathbf{J}^3)\,,
\label{eq:Hz}
\end{equation}
with $\mathbf{J}$ the spin $\tfrac{3}{2}$ operator, $\mathbf{J}^3\equiv(J_x^3,J_y^3,J_z^3)$, $\mu_B$ the Bohr magneton, and $\kappa$, $q$ the isotropic and cubic Zeeman parameters ($\kappa=3.41$ and $q=0.06$ in Ge). The action of $\mathbf{B}$ on the envelopes of the hole is accounted for by the substitution $\mathbf{k}\to-i\boldsymbol{\nabla}+e\mathbf{A}/\hbar$ in $H_\mathrm{LK}$, with $\mathbf{A}=\frac{1}{2}\mathbf{B}\times\mathbf{r}$ the magnetic vector potential.

\subsection{The linear Dresselhaus interaction}

As discussed in section \ref{sec:interdiffusion}, the linear Dresselhaus interaction is strongly dependent on the steepness of the Ge/GeSi interfaces. It arises primarily from HH/LH mixings induced by the change of Bloch functions at these interfaces. 

The effects of an abrupt interface at $z=z_0$ on the $J=\tfrac{3}{2}$ manifold can actually be described by the following correction to the LK Hamiltonian \cite{Winkler03,Aleiner92,Ivchenko96}:
\begin{equation}
H_\mathrm{int}=s_\mathrm{int}\frac{ic_\mathrm{int}}{2\sqrt{3}}\delta(z-z_0) 
\begin{pmatrix}
0 & 0 & -1 & 0 \\
0 & 0 & 0 & -1 \\
1 & 0 & 0 & 0 \\
0 & 1 & 0 & 0
\end{pmatrix}\,,
\label{eq:Hint}
\end{equation}
where $c_\mathrm{int}$ is the coupling strength (in eV\,\AA) and $s_\mathrm{int}=\pm 1$ changes sign every time the interface is shifted by one ML. This interface Hamiltonian lowers the symmetry of the original LK Hamiltonian.

The effect of $H_\mathrm{int}$ on the band structure of the film can be qualitatively captured by a Schrieffer-Wolff (SW) transformation. Let us introduce the pure HH subband states $\ket{n,\pm\tfrac{3}{2}}$ at $\vec{k}=\vec{0}$, with energies $E_n^\mathrm{HH}$, and the pure LH subband states $\ket{n,\pm\tfrac{1}{2}}$, with energies $E_n^\mathrm{LH}$. The $R$ and $S$ terms of the LK Hamiltonian as well as $H_\mathrm{int}$ mix these pure HH and LH states at finite $\vec{k}$; the effective Hamiltonian in the $\{\ket{0,+\tfrac{3}{2}},\,\ket{0,-\tfrac{3}{2}}\}$ subspace reads to second order in these perturbations:
\begin{equation}
{\cal H}_{hh^\prime}=\sum_{l=\pm\tfrac{1}{2},n} \frac{\bra{0,h}H_\mathrm{c}\ket{n,l}\bra{n,l}H_\mathrm{c}^\prime\ket{0,h^\prime}}{E_0^\mathrm{HH}-E_n^\mathrm{LH}}\,,
\label{eq:SW}
\end{equation}
where $h,h^\prime=\pm\tfrac{3}{2}$ and $H_\mathrm{c},\,H_\mathrm{c}^\prime\in\{R,\,S,\,H_\mathrm{int}\}$. For the sake of demonstration, we shall assume for now that the Luttinger parameters are the same on both sides of the interface, and that $E_0^\mathrm{HH}-E_n^\mathrm{LH}\approx E_0^\mathrm{HH}-E_0^\mathrm{LH}=\Delta_\mathrm{LH}$ whatever $n\ge 0$. Setting $H_\mathrm{c}=H_\mathrm{int}$ and $H_\mathrm{c}^\prime=S$ (or vice-versa) then yields, thanks to the closure relation $\sum_n\braket{z|n,l}\braket{n,l|z^\prime}=\delta(z-z^\prime)$:
\begin{align}
{\cal H}(\vec{k})&=s_\mathrm{int}\frac{\gamma_3 c_\mathrm{int}}{\Delta_\mathrm{LH}}\frac{\hbar^2}{2m_0}\left.\frac{d}{dz}|\psi_0(z)|^2\right|_{z=z_0} \nonumber \\
&\times(k_x\sigma_1+k_y\sigma_2)\,,
\end{align}
where $\psi_0(z)=\braket{z|0,\pm\tfrac{3}{2}}$ is the ground-state HH envelope. $H_\mathrm{int}$ thus gives rise to a linear Dresselhaus interaction whose strength is proportional to $c_\mathrm{int}$ and $\gamma_3$, and to the gradient of the probability of presence of the hole at the interface. It is also inversely proportional to the fundamental HH-LH band gap $\Delta_\mathrm{LH}$. In a Ge well with thickness $L_\mathrm{W}$ at zero vertical electric field \cite{Michal21},
\begin{equation}
\Delta_\mathrm{LH}\approx\frac{2\pi^2\hbar^2\gamma_2}{m_0L_\mathrm{W}^2}+2b_v(\varepsilon_\parallel-\varepsilon_\perp)\,,
\label{eq:deltaLH}
\end{equation}
where the first term accounts for structural confinement and the second one for strains. The interactions at the two interfaces of the well add up if their $s_\mathrm{int}$ are different (odd number of Ge MLs), and cancel each other if their $s_\mathrm{int}$ are the same (even number of Ge MLs), as the gradient of $|\psi_0(z)|^2$ is opposite at the top and bottom interfaces when $E_z=0$.

\begin{figure}
\includegraphics[width=0.9\columnwidth]{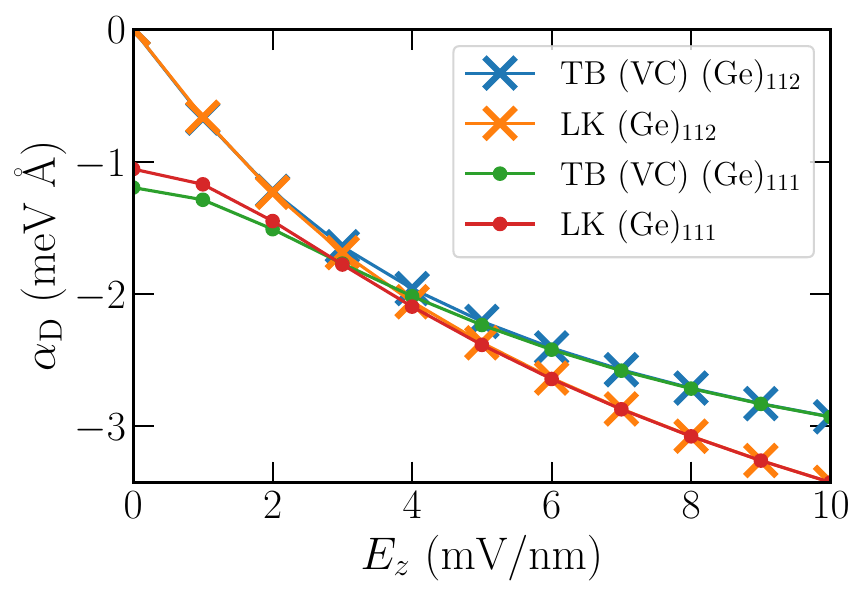}
\caption{Comparison between the TB and LK Dresselhaus coefficients $\alpha_\mathrm{D}$ computed as a function of the vertical electric field $E_z$ in $\approx 15.9$ nm thick Ge films with even/odd number of MLs embedded in Ge$_{0.8}$Si$_{0.2}$ barriers.}
\label{fig:DresselhausKPEz}
\end{figure}

\begin{figure}
\includegraphics[width=0.9\columnwidth]{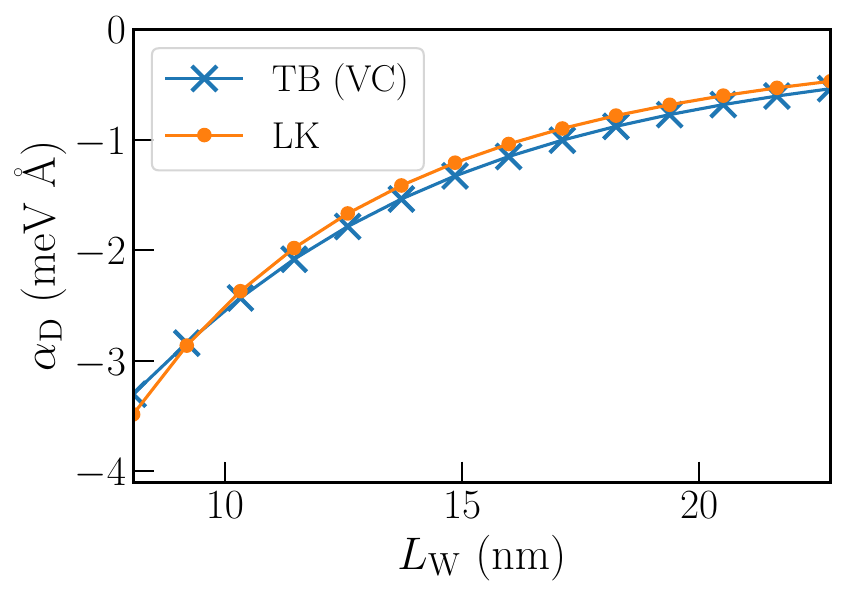}
\caption{Comparison between the TB and LK Dresselhaus coefficients $\alpha_\mathrm{D}$ computed as a function of the thickness $L_\mathrm{W}$ of Ge films with odd number of MLs embedded in Ge$_{0.8}$Si$_{0.2}$ barriers. The vertical electric field is zero.}
\label{fig:DresselhausKPLw}
\end{figure}

\begin{figure}
\includegraphics[width=0.9\columnwidth]{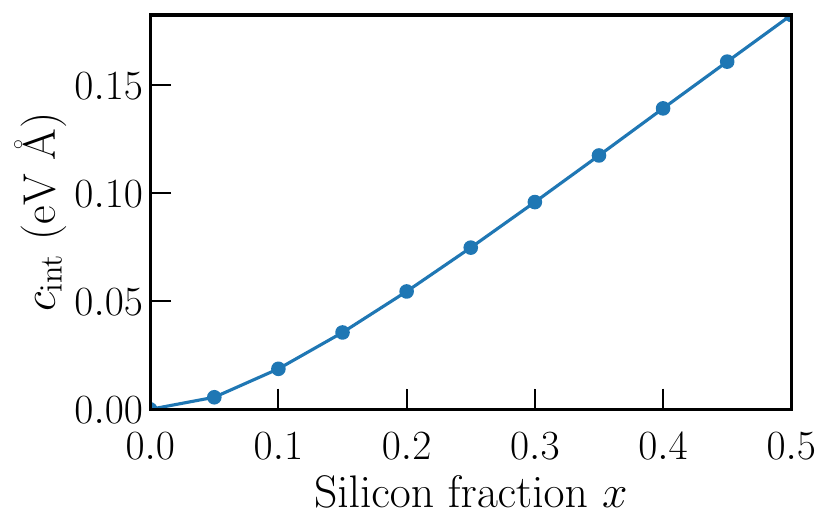}
\caption{The coupling strength $c_\mathrm{int}$ [Eq.~\eqref{eq:Hint}] extracted from TB VC calculations as a function of the concentration $x$ of silicon in the Ge$_{1-x}$Si$_x$ alloy. We assume that the heterostructures are grown on a GeSi buffer with the same concentration and no residual in-plane strain.}
\label{fig:cintconc}
\end{figure}

The value of $c_\mathrm{int}$ can be fitted to the TB linear Dresselhaus coefficient $\alpha_\mathrm{D}$. For that purpose, we have modeled the same superlattice structures as in TB with a numerical implementation of the four bands LK Hamiltonian including Eq.~\eqref{eq:Hint} \footnote{\YMN{In Figs.~\ref{fig:DresselhausKPEz} to \ref{fig:alpha3beta3Ez}, we use the Luttinger parameters, deformation potentials and strains of the TB and valence force field models \cite{Niquet09}, in order to achieve a meaningful comparison between the TB and LK data. In section \ref{sec:qubits}, we use the parameters given in the main text for consistency with our previous works \cite{martinez2022hole,Uriel22}. There are no qualitative differences between the two sets of parameters, though.}}. \YMN{We use finite differences with symmetric ordering for the derivatives at the Ge/GeSi interface (e.g., $\gamma_2 k_z^2\to-\tfrac{\partial}{\partial z}\gamma_2\tfrac{\partial}{\partial z}$) and discuss the alternative Burt-Foreman ordering \cite{Burt92,Foreman93,Foreman97,Veprek07} in Appendix \ref{app:BF}.} The resulting $\vec{k}\cdot\vec{p}$ band structures and spin splittings are non perturbative and include, therefore, all orders beyond Eq.~\eqref{eq:SW}. We reach good agreement at small electric fields $E_z$ using $c_\mathrm{int}=55.9$\,meV\,\AA\ (see Fig.~\ref{fig:DresselhausKPEz}). The parameter $s_\mathrm{int}$ is actually $+1$ (resp. $-1$) when the bonds from Ge to GeSi project onto $[110]$ (resp $[\overline{1}10]$). Note, again, that $\alpha_\mathrm{D}$ thus changes sign each time the whole Ge well is shifted up or down by one ML. The linear Dresselhaus coefficient $\alpha_\mathrm{D}$ bends upwards at large $E_z$ because the increase of the gradient of $|\psi_0(z_0)|^2$ is partly compensated by the increase of the HH-LH band gap $\Delta_\mathrm{LH}$ (the effective, electric width \cite{Michal21} of the well $\ell_\mathrm{E}\propto E_z^{-1/3}$ becoming much smaller that its structural thickness $L_\mathrm{W}$ \footnote{In the limit $\ell_\mathrm{E}\ll L_\mathrm{W}$, the potential in Ge can be approximated by a triangular well. It can be shown from the eigensolutions of this triangular well (the Airy function) or from a simple dimensional analysis that both $\Delta_\mathrm{LH}\propto 1/\ell_\mathrm{E}^2\propto E_z^{2/3}$ and $d|\Psi(z)|^2/dz\propto 1/\ell_\mathrm{E}^2\propto E_z^{2/3}$ so that $\alpha_\mathrm{D}$ is expected to saturate. This saturation is however delayed by the biaxial strain contribution to $\Delta_\mathrm{LH}$ [second term of Eq.~\eqref{eq:deltaLH}] and by the finite band offset between Ge and GeSi.}). There is a small discrepancy between the $\vec{k}\cdot\vec{p}$ and TB $\alpha_\mathrm{D}$'s at large $E_z$, which results from slightly different responses to the electric field.

We also plot in Fig.~\ref{fig:DresselhausKPLw} the $\alpha_\mathrm{D}$'s computed at zero electric field in Ge films with various odd numbers of MLs. The TB trends are again very well reproduced by the $\vec{k}\cdot\vec{p}$ calculations, which supports the relevance and versatility of Eq.~\eqref{eq:Hint}. We also emphasize that our model is different from Ref.~\cite{Xiong21}. Indeed, Ref.~\cite{Xiong21} assumes some interface-induced HH/LH mixing coefficients $a_i$ and $b_i$ (that shall in principle depend on $L_\mathrm{W}$ and $E_z$), whereas we start from the Hamiltonian, Eq.~\eqref{eq:Hint}, that gives rise to these mixings (with an unique parameter $c_\mathrm{int}$ that is independent on $L_\mathrm{W}$ and $E_z$).

We finally plot in Fig.~\ref{fig:cintconc} the value of $c_\mathrm{int}$ computed for different concentrations $x$ of silicon in the Ge$_{1-x}$Si$_x$ alloy. We assume that the heterostructures are grown on a GeSi buffer with the same concentration and no residual in-plane strain (at variance with the former calculations). Therefore, only the Ge well is now strained. The extracted $c_\mathrm{int}(x=0.2)=55.5$\,meV\,\AA\ remains nonetheless very close to the value $c_\mathrm{int}=55.9$\,meV\,\AA\ obtained previously with a residual in-plane strain $\varepsilon_\parallel=0.26$\% in the buffer. The interface Hamiltonian $H_\mathrm{int}$ is, therefore, weakly dependent on strains, contrary to the HH-LH band gap $\Delta_\mathrm{LH}$. As expected, the coupling strength $c_\mathrm{int}$ increases almost linearly with the silicon fraction except for small $x$ for reasons likely similar to interdiffusion (blurred interfaces).

When the interface gets interdiffused, Eq.~\eqref{eq:Hint} can still be used with a rescaled $c_\mathrm{int}$. According to the TB calculations, $c_\mathrm{int}\approx 0$ once the interface is interdiffused over more than 5 MLs.

\subsection{The linear Rashba interaction}

The linear Rashba interaction is independent on the status of the Ge/GeSi interfaces and shows no saturation with increasing $E_z$. Therefore, it arises most likely from mixings between the valence and conduction bands enabled by the structural inversion asymmetry (SIA) at finite electric field. As a matter of fact, $\alpha_\mathrm{R}$ is little dependent on strains, which rules out a direct HH/LH mixing mechanism (whose strength would be $\propto 1/\Delta_\mathrm{LH}$, hence be primarily controlled by strains). In extended $\vec{k}\cdot\vec{p}$ models, there are actually linear-in-$k$ terms coupling the $\Gamma_{8\mathrm{v}}/\Gamma_{7\mathrm{v}}$ valence bands and the lowest $\Gamma_{6\mathrm{c}}$ conduction bands; they do not, however, give rise to a linear Rashba interaction in the HH manifold of a planar heterostructure \cite{Winkler03}. On the other hand, couplings between the $\Gamma_{8\mathrm{v}}/\Gamma_{7\mathrm{v}}$ valence bands and the higher-lying $\Gamma_{8\mathrm{c}}/\Gamma_{7\mathrm{c}}$ conduction bands with the same symmetry do so at third-order in perturbation; the resulting linear Rashba coefficient reads \cite{Winkler03}:
\begin{align}
\alpha_\mathrm{R}&\approx-\frac{eQ^2}{3}\left(\frac{1}{E_0^{\prime2}}-\frac{1}{(E_0^{\prime}+\Delta_0^\prime)^2}\right)E_z \nonumber \\
&\approx(-0.5\,\mathrm{\AA}^2)eE_z\,,
\end{align}
where $E_0^{\prime}$ is the gap between the $\Gamma_{8\mathrm{v}}$ valence bands and the $\Gamma_{7\mathrm{c}}$ conduction bands, $\Delta_0^\prime$ is the splitting between the $\Gamma_{8\mathrm{c}}$ and $\Gamma_{7\mathrm{c}}$ conduction bands, and $Q$ is an interband momentum matrix element between the $\Gamma_{8\mathrm{v}}/\Gamma_{7\mathrm{v}}$ and $\Gamma_{8\mathrm{c}}/\Gamma_{7\mathrm{c}}$ manifolds \cite{Winkler03,Richard04,Durnev14,Malkoc22}. This simple estimate is half the slope of Fig.~\ref{fig:alphabetaEz}, but the correct order of magnitude. Due to this remote nature, the linear Rashba SOI can only be introduced in LK Hamiltonian as an {\it ad-hoc} correction in the $J_z=\{+\tfrac{3}{2},-\tfrac{3}{2}\}$ subspace:
\begin{equation}
H_\mathrm{R}^{(1)}=\alpha_\mathrm{R}(k_x\sigma_2+k_y\sigma_1)\,,
\end{equation}
where $\alpha_\mathrm{R}$ must be tabulated as a function of the well thickness and electric field with the TB calculations (see, e.g., Fig.~\ref{fig:rashbaEz}).

\subsection{The cubic Rashba interaction}

\begin{figure}
\includegraphics[width=0.9\columnwidth]{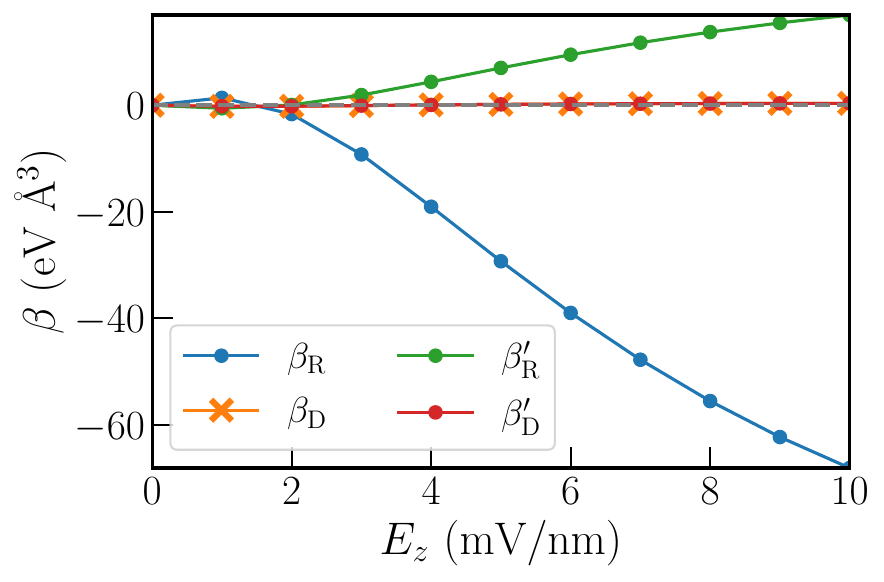}
\caption{Cubic [Eq.~\eqref{eq:H3_cubic}] Rashba and Dresselhaus coefficients as a function of the vertical electric field $E_z$, computed with the LK Hamiltonian in a (Ge)$_{112}$/(Ge$_{0.8}$Si$_{0.2}$)$_{56}$ superlattice.}
\label{fig:alpha3beta3Ez}
\end{figure}

The admixture of HH and LH components by vertical confinement is known to give rise to a cubic Rashba interaction whose strength is, in a first approximation, proportional to $1/\Delta_\mathrm{LH}$ \cite{Winkler03,Marcellina17,Wang21,Terrazos21}. This contribution to the cubic Rashba SOI is, therefore, captured by the LK model. \YMN{The cubic Rashba and Dresselhaus coefficients of the LK model ($H_\mathrm{int}$ included) are plotted as a function of the vertical electric field $E_z$ in Fig.~\ref{fig:alpha3beta3Ez}. They are computed non-perturbatively by fitting the $\eta_i(\vec{k})$'s obtained from the diagonalization of the LK Hamiltonian of the superlattice (same procedure as for the TB coefficients).} They are in qualitative agreement with the TB data (Fig.~\ref{fig:alphabetaEz}): the cubic Dresselhaus interactions (which result from higher order contributions of $H_\mathrm{int}$ to the effective Hamiltonian) are negligible with respect to the cubic Rashba interactions. Yet the TB $\beta_\mathrm{R}$ is $\approx 2\times$ larger than the LK coefficient at high vertical electric field. The LK $\beta_\mathrm{R}$ also shows a ``sweet spot'' (zero) at small field that does not exist in the TB data. This is particularly relevant because the $\propto\beta_\mathrm{R}$ term of Eq.~\eqref{eq:H3_cubic} is the only cubic interaction contributing to the Rabi oscillations of disk-shaped dots in the linear response regime \cite{Wang21,Terrazos21}. We also assign these discrepancies to the couplings between the HH/LH subspace and remote conduction and valence bands \cite{Winkler03,Richard04,Durnev14,Malkoc22}. The inclusion of the split-off, $J=\tfrac{1}{2}$ valence bands only (6 bands, Dresselhaus-Kipp-Kittel $\vec{k}\cdot\vec{p}$ model \cite{Dresselhaus55}) however lowers the cubic Rashba coefficients at high field and further degrades the comparison with TB data. 

\section{Application to {Ge} hole spin qubits}
\label{sec:qubits}

We now address the effects of the linear Rashba and Dresselhaus interactions on the physics of Germanium devices. The linear Dresselhaus interaction is likely difficult to detect in magneto-transport experiments that probe the devices over long length scales \cite{Moriya14,Morisson16,Mizokuchi17}, as it will be averaged out by interface steps (or superseded by the cubic Rashba at large Fermi wave vector \cite{Xiong22}). It may, however, survive at the scale of a quantum dot if interdiffusion is locally limited. The interdiffusion lengths reported in the literature are in fact pretty long \cite{Sammak19,Pena23}, but it remains unclear how homogeneous and symmetric (top/bottom interfaces) they are. We analyze, therefore, the impact of the linear Rashba and Dresselhaus interactions on the manipulation of Ge hole spin qubits in the ``best case'' scenario (no interdiffusion) in order to assess their maximal impact. We first remind the analytical expressions for the Rabi frequency in a simple model with harmonic in-plane confinement potential. We compare, in particular, the contributions of these interactions with those of the $g$-TMR mechanisms discussed in Refs.~\cite{Uriel22} and \cite{martinez2022hole}. We then highlight that the effective interface Hamiltonian, Eq.~\eqref{eq:Hint}, also gives rise to $g$-factor corrections. We finally support this analysis with numerical simulations on a realistic structure.

\subsection{Analytical estimates of the Rabi frequencies}
\label{sec:analytical}

We first consider a simple analytical model for a quantum dot confined in a Ge/GeSi heterostructure by the electric field from accumulation or depletion gates. We assume a Ge well with thickness $L_\mathrm{W}$, a homogeneous vertical electric field $E_z$, and a harmonic in-plane confinement potential
\begin{equation}
V(x,y)=-\frac{\hbar^2}{2m_\parallel r_\parallel^4}(x^2+y^2)
\end{equation}
with $m_\parallel$ the in-plane mass of the heavy holes. If the vertical confinement is much stronger than the in-plane confinement ($r_\parallel\gg L_\mathrm{W})$, the ground-state is a HH pseudo-spin doublet 
\begin{equation}
\{\ket{\Uparrow}\approx\ket{\varphi_0}\ket{0,+\tfrac{3}{2}},\ket{\Downarrow}\approx\ket{\varphi_0}\ket{0,-\tfrac{3}{2}}\}\,,
\label{eq:GS}
\end{equation}
where $\varphi_0$ is the eigenfunction of the 2D harmonic oscillator:
\begin{equation}
\varphi_0(x, y)=\frac{1}{\sqrt{\pi}r_\parallel}\exp\left(-\frac{x^2+y^2}{2r_\parallel^2}\right)\,.
\end{equation}
This doublet is split by a homogeneous magnetic field $\mathbf{B}$ whose action in the $\{\ket{\Uparrow},\ket{\Downarrow}\}$ subspace can be described by the effective Zeeman Hamiltonian
\begin{equation}
\tilde{H}_\mathrm{Z}=-\frac{1}{2}\mu_B(g_xB_x\sigma_1+g_yB_y\sigma_2+g_zB_z\sigma_3)
\label{eq:HZeff}
\end{equation}
with $\mu_B$ the Bohr magneton and $g_u$ the gyromagnetic factors. For a pure HH in bulk Ge, $g_z=g_\perp=6\kappa+27q/2=21.3$ and $g_x=-g_y=g_\parallel=3q=0.18$ from Eq.~\eqref{eq:Hz}. Vertical, lateral and magnetic confinements however admix a small LH component into Eq.~\eqref{eq:GS} and decrease $g_\perp$ down to $\approx 13.5$ and $g_\parallel$ down to $\approx 0.15$ ($L_\mathrm{W}=16$ nm, $r_\parallel=30$ nm) \cite{Michal21,martinez2022hole,Wang22}. The Larmor frequency of the doublet is thus $f_\mathrm{L}=\omega_\mathrm{L}/(2\pi)=\mu_B \sqrt{g_\parallel B_\parallel^2+g_\perp B_z^2}/h$ with $B_\parallel=\sqrt{B_x^2+B_y^2}$.

A hole in the $\{\ket{0,+\tfrac{3}{2}}, \ket{0,-\tfrac{3}{2}}\}$ subbands is moreover subject to Rashba and Dresselhaus SOIs. Leaving out cubic interactions for now, the spin-orbit Hamiltonian
\begin{equation}
H_\mathrm{so}=\frac{\alpha_\mathrm{D}}{\hbar}(p_x\sigma_1+p_y\sigma_2)+\frac{\alpha_\mathrm{R}}{\hbar}(p_x\sigma_2+p_y\sigma_1)
\label{eq:Hso}
\end{equation}
couples the spin of the hole to its velocity; shaking the dot as a whole with a homogeneous AC electric field $E_x=E_\mathrm{ac}\cos(\omega_\mathrm{L}t)$ resonant with the Larmor frequency can thus give rise to Rabi oscillations. 

We can find the Rabi frequency with a simplified semi-classical treatment of the momentum operators in Eq.~\eqref{eq:Hso}. When driven, the dot moves by $\delta x(t)=\delta x_\mathrm{ac}\cos(\omega_\mathrm{L}t)$, where $\delta x_\mathrm{ac}=eE_\mathrm{ac}m_\parallel r_\parallel^4/\hbar^2$, and acquires a classical momentum:
\begin{equation}
p_x(t)=m_\parallel v_x(t)=-m_\parallel\delta x_\mathrm{ac}\omega_\mathrm{L}\sin(\omega_\mathrm{L}t)\,.
\end{equation}
The effective time-dependent Hamiltonian for the $\{\ket{\Uparrow}, \ket{\Downarrow}\}$ doublet reads therefore:
\begin{align}
\tilde{H}_\mathrm{s}=&-\frac{1}{2}\mu_B(g_\parallel B_x\sigma_1-g_\parallel B_y\sigma_2+g_\perp B_z\sigma_3)\nonumber \\
&-\frac{1}{\hbar}m_\parallel\omega_\mathrm{L}(\alpha_\mathrm{D}\sigma_1+\alpha_\mathrm{R}\sigma_2)\delta x_\mathrm{ac}\sin(\omega_\mathrm{L}t)\,.
\end{align}
We next introduce
\begin{equation}
\boldsymbol{\Omega}^\prime=\frac{2}{\hbar^2}m_\parallel\omega_\mathrm{L}\delta x_\mathrm{ac}
\begin{pmatrix}
\alpha_\mathrm{D} \\
\alpha_\mathrm{R} \\
0
\end{pmatrix}
\end{equation}
as well as the unit vectors $\boldsymbol{\omega}^\prime=\boldsymbol{\Omega}^\prime/|\boldsymbol{\Omega}^\prime|$ and $\boldsymbol{\omega}=\mu_B(g_\parallel B_x,-g_\parallel B_y, g_\perp B_z)/(\hbar\omega_\mathrm{L})$, so that
\begin{equation}
\tilde{H}_\mathrm{s}=-\frac{1}{2}\hbar\omega_\mathrm{L}\sigma_{\boldsymbol{\omega}}-\frac{1}{2}\hbar|\boldsymbol{\Omega}^\prime|\sin(\omega_\mathrm{L}t)\sigma_{\boldsymbol{\omega}^\prime}
\end{equation}
with $\sigma_\vec{u}=u_x\sigma_1+u_y\sigma_2+u_z\sigma_3$. We finally split $\boldsymbol{\Omega}^\prime=\Omega_\parallel^\prime\boldsymbol{\omega}+\boldsymbol{\Omega}_\perp^\prime$ into components parallel and perpendicular to $\boldsymbol{\omega}$, and get:
\begin{align}
\tilde{H}_\mathrm{s}=&-\frac{1}{2}\hbar\left[\omega_\mathrm{L}+\Omega_\parallel^\prime\sin(\omega_\mathrm{L}t)\right]\sigma_{\boldsymbol{\omega}} \nonumber \\
&-\frac{1}{2}\hbar|\boldsymbol{\Omega}_\perp^\prime|\sin(\omega_\mathrm{L}t)\sigma_{\boldsymbol{\omega}_\perp^\prime}\,,
\end{align}
where $\boldsymbol{\omega}_\perp^\prime$ is the unit vector along $\boldsymbol{\Omega}_\perp^\prime$. In the rotating wave approximation, the Rabi frequency at resonance is then simply 
\begin{align}
f_\mathrm{R}&=\frac{1}{4\pi}\left|\boldsymbol{\Omega}_\perp^\prime\right|=\frac{1}{4\pi}\left|\boldsymbol{\omega}\times\boldsymbol{\Omega}^\prime\right| \nonumber \\
&=\frac{1}{2\pi\hbar^3}m_\parallel\delta x_\mathrm{ac}\mu_B
\big[g_\parallel^2(\alpha_\mathrm{R}B_x+\alpha_\mathrm{D}B_y)^2  \nonumber \\
&+g_\perp^2B_z^2(\alpha_\mathrm{R}^2+\alpha_\mathrm{D}^2) \big]^{1/2}\,.
\label{eq:fRabi}
\end{align}
This is the same result as the full quantum mechanical treatment of Refs.~\cite{Golovach06,Rashba08,Bosco21b}. Since $g_\perp\gg g_\parallel$, the Rabi frequency at constant magnetic field strength $B=|\vec{B}|$ has, in a first approximation, a $\propto B_z$ envelope (this argument also holds for cubic Rashba SOI, whose contribution to the Rabi frequency also shows an approximate $\propto B_z$ envelope \cite{martinez2022hole}). At constant Larmor frequency $f_\mathrm{L}$ on the other hand, the Rabi frequency for an in-plane magnetic field oriented along the unit vector $\mathbf{b}=(b_x,b_y)$ is:
\begin{equation}
f_\mathrm{R}(b_x,b_y)=f_\mathrm{L}\frac{m_\parallel}{\hbar^2} |\alpha_\mathrm{R}b_x+\alpha_\mathrm{D}b_y|\delta x_\mathrm{ac}\,,
\label{eq:fRabifL}
\end{equation}
while for a magnetic field along $z$:
\begin{equation}
f_\mathrm{R}(\vec{B}\parallel\vec{z})=f_\mathrm{L}\frac{m_\parallel}{\hbar^2}\sqrt{\alpha_\mathrm{R}^2+\alpha_\mathrm{D}^2}\delta x_\mathrm{ac}\,.
\label{eq:fRabifLz}
\end{equation}
The Dresselhaus interaction does not contribute when $\vec{B}\parallel\vec{x}$, while the Rashba interaction does not contribute when $\vec{B}\parallel\vec{y}$ (since $\tilde{H}_\mathrm{Z}$ and $H_\mathrm{so}$ then share the same pseudo-spin eigenvectors). 

\begin{figure}
\includegraphics[width=.85\columnwidth]{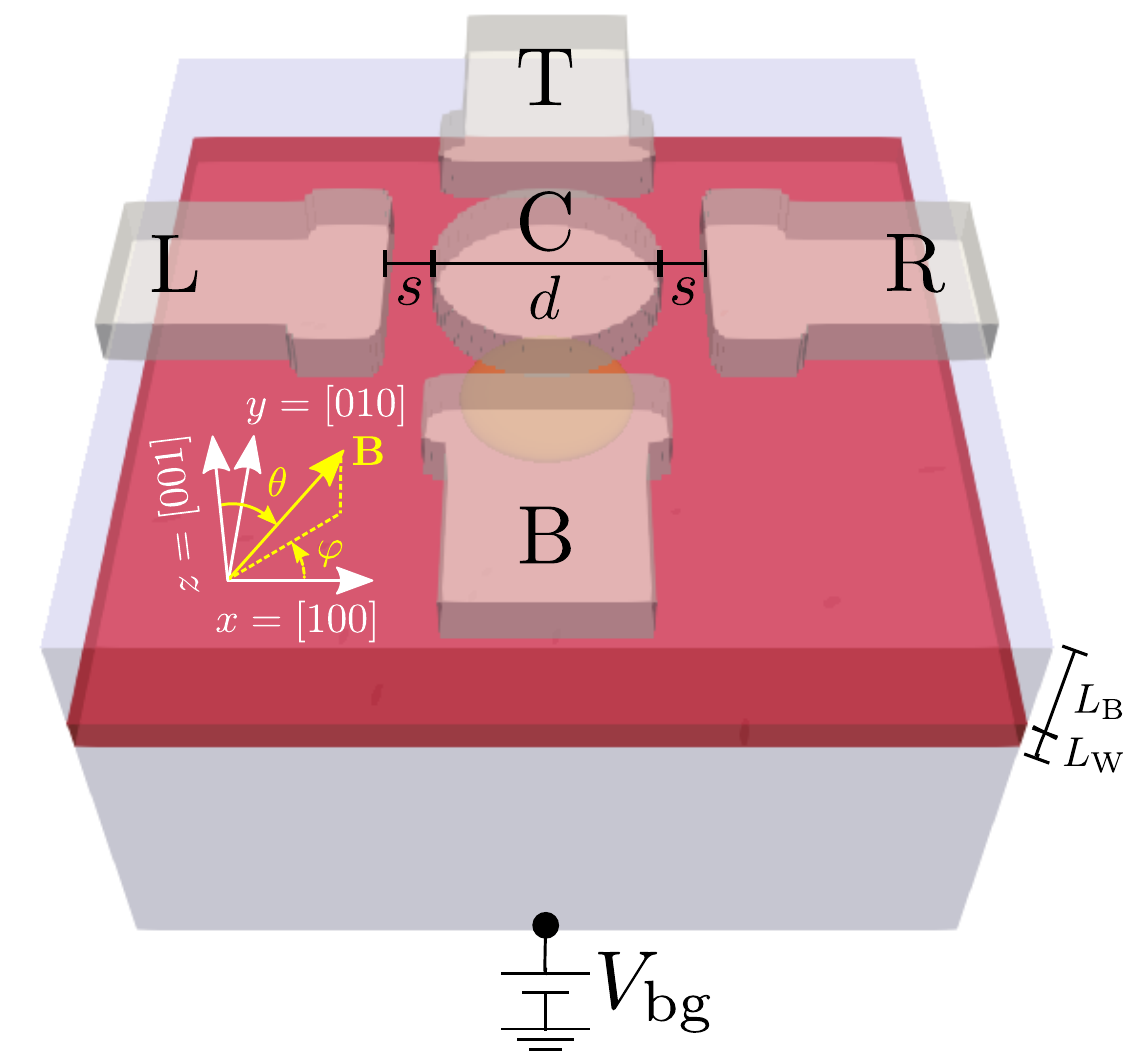}
\caption{The hole spin qubit device. The Ge well (red) is $L_\mathrm{W}=16$\,nm thick, the upper  Ge$_{0.8}$Si$_{0.2}$ barrier (light blue) is $L_\mathrm{B}=50$\,nm thick, the diameter of the C gate is $d=100$ nm and the gap with the side gates is $s=20$ nm. All gates are embedded in Al$_2$O$_3$, and are insulated from the heterostructure by 5\,nm of this material. The substrate below the 150 nm thick lower barrier acts as an effective back gate, which can be used to tune independently the depth of the quantum dot and the vertical electric field, but is grounded in the present work. We assume, as in Ref.~\cite{Sammak19}, that the Ge$_{0.8}$Si$_{0.2}$ barriers are not fully relaxed, and experience residual in-plane strain $\varepsilon_{xx}=\varepsilon_{yy}=\varepsilon_\parallel=0.26$\% and out-of-plane strain $\varepsilon_{zz}=\varepsilon_\perp=-0.19$\%. Consequently, the strains in the Ge well are $\varepsilon_\parallel=-0.63\%$ and $\varepsilon_\perp=0.47\%$. The yellow contour is the isodensity surface that encloses 90\% of the ground-state hole charge at $V_\mathrm{C}=-40$\,mV.}
\label{fig:device}
\end{figure}

We can give estimates of $f_\mathrm{R}$ for the devices modeled in Refs.~\cite{Uriel22,martinez2022hole} and reproduced in Fig.~\ref{fig:device}. We focus on in-plane magnetic fields ($B_z=0$), which best decouple the heavy holes from hyperfine dephasing noise. The quantum dot is shaped by the potential $V_\mathrm{C}$ applied to the central C gate with all side gates grounded. The hole is driven by opposite modulations $\delta V_\mathrm{L}=-\delta V_\mathrm{R}=(V_\mathrm{ac}/2)\cos\omega_\mathrm{L}t$ on the L and R gates. For a small bias $V_\mathrm{C}=-40$ mV, the vertical electric field is $E_z\approx 0.25$\,mV/nm, the radius of the dot is $r_\parallel\approx 27$\,nm, and $\delta x_\mathrm{ac}/V_\mathrm{ac}\approx 0.85$\,nm/mV. From Figs.~\ref{fig:rashbaEz} and \ref{fig:DresselhausKPEz}, we estimate $\alpha_\mathrm{D}=-1.20$\,meV\,\AA\ and $\alpha_\mathrm{R}=-0.029$\,meV\,\AA\ (assuming, as a best case scenario, an odd number of Ge MLs with no interdiffusion). The mass of the holes, inferred from the band structure, is $m_{\parallel}=0.077\,m_0$ (with $m_0$ the bare electron mass). At $f_\mathrm{L}=5$\,GHz, the Rabi frequencies computed from Eq.~\eqref{eq:fRabifL} are thus
\begin{subequations}
\label{eq:fRestimates}
\begin{align}
f_\mathrm{R}(\vec{B}\parallel\vec{x})&=0.012\,\text{MHz/mV} \\
f_\mathrm{R}(\vec{B}\parallel\vec{y})&=0.52\,\text{MHz/mV}\,.
\end{align}
\end{subequations}
They are normalized for a drive amplitude $V_\mathrm{ac}=1$\,mV. These frequencies are rather small. In particular, the Rashba interaction competes with the non-separability (NS) mechanism discussed in Ref.~\cite{martinez2022hole} [$f_\mathrm{R}(\vec{B}\parallel\vec{x})\approx 4.5$\,MHz/mV], and more so with the strain-induced $g$-TMR introduced in Ref.~\cite{Uriel22} [$f_\mathrm{R}(\vec{B}\parallel\vec{x})\approx 60$\,MHz/mV]. It is, therefore negligible with respect to these two mechanisms. The Dresselhaus interaction, if not washed out by interdiffusion, enables Rabi oscillations for $\vec{B}\parallel\vec{y}$, where both the NS and strain-induced $g$-TMR are forbidden (at least in the LK model). As $\delta x_\mathrm{ac}\propto 1/\omega_\parallel^2\propto r_\parallel^4$, the Rashba and Dresselhaus contributions to the Rabi frequency scale as $r_\parallel^4$, while the NS contribution shows a weak dependence on $r_\parallel$ and may thus be overcome in large dots. The strain-induced $g$-TMR does, however, also scale close to $r_\parallel^4$, and can therefore hardly be outweighed by the linear Rashba and Dresselhaus interactions.

At larger vertical electric field, $\alpha_\mathrm{D}$ and $\alpha_\mathrm{R}$ can increase significantly. At the maximum electric field $E_z=3$\,mV/nm, above which the hole is pulled out from the well to the surface of the heterostructure \cite{Su17}, $\alpha_\mathrm{D}=-1.77$\,meV\,\AA\ and $\alpha_\mathrm{R}=-0.34$\,meV\,\AA, so that $f_\mathrm{R}(\vec{B}\parallel\vec{x})=0.14$\,MHz/mV and $f_\mathrm{R}(\vec{B}\parallel\vec{y})=0.76$\,MHz/mV ($f_\mathrm{L}=5$\,GHz). They remain nonetheless negligible with respect to strain-induced $g$-TMR.

To conclude this discussion, we would like to remind that there exists other linear Rashba interactions not captured by the present atomistic calculations in planar systems. The first one is the direct linear Rashba interaction specific to 1D systems such as nanowires \cite{Kloeffel11,Kloeffel18}. It arises from HH/LH mixings by the $R$ and $S$ terms of the LK Hamiltonian, and is therefore the 1D counterpart of the cubic Rashba interaction in 2D heterostructures. It is irrelevant in symmetric (disk-shaped) quantum dots, but can be significant in elongated (``squeezed'') ones that look more one-dimensional \cite{Bosco21b}. This direct Rashba interaction is, however, sizable only when the small in-plane axis of the dot is comparable to its thickness $\ell_\mathrm{E}$. This condition is very stringent and difficult to achieve in buried heterostructures. The dot can, in principle, be squeezed by applying positive gate voltages on the L and R (or B and T) gates of Fig.~\ref{fig:device}, but the hole gets pulled out from the well and trapped at the GeSi/Al$_2$O$_3$ interface \cite{Su17} long before direct Rashba prevails over the NS mechanism in all cases we have investigated up to now. Inhomogeneous strains (due to process and cool down stress for example) also give rise to a linear-in-momentum SOI that can be much stronger than the present interactions \cite{Uriel22}. This strain-induced SOI is, nonetheless, itself superseded by strain-induced $g$-TMR under in-plane magnetic fields. The $g$-TMR mechanisms thus generally prevail over the linear Rashba and Dresselhaus interactions when the magnetic field lies in the plane of the heterostructure.

\subsection{$g$-factor corrections}
\label{sec:gfactorcorr}

Besides the Dresselhaus interaction, the interface Hamiltonian, Eq.~\eqref{eq:Hint}, also gives rise to $g$-factor corrections. This can be evidenced with the same Schrieffer-Wolff transformation as in Eq.~\eqref{eq:SW}, now using $H_\mathrm{c}=H_\mathrm{int}$, $H_\mathrm{c}^\prime=H_\mathrm{Z}$ or vice-versa. The resulting effective Hamiltonian is:
\begin{equation}
{\cal H}=\frac{s_\mathrm{int}c_\mathrm{int}}{\Delta_\mathrm{LH}}|\psi_0(z_0)|^2\kappa\mu_B\left(B_y\sigma_1-B_x\sigma_2\right)\,.
\end{equation}
Equation~\eqref{eq:HZeff} can then be generalized as:
\begin{equation}
\tilde{H}_\mathrm{Z}=-\frac{1}{2}\mu_B\boldsymbol{\sigma}\cdot \hat{g}\vec{B}\,,
\label{eq:HZeff2}
\end{equation}
where $\boldsymbol{\sigma}=(\sigma_1,\sigma_2,\sigma_3)$ and $\hat{g}$ is the $g$-matrix or tensor:
\begin{equation}
\hat{g}=\begin{pmatrix}
g_\parallel & -g_\otimes & 0 \\
g_\otimes & -g_\parallel & 0 \\
0 & 0 & g_\perp
\end{pmatrix}
\end{equation}
with
\begin{equation}
g_\otimes=2\kappa\frac{s_\mathrm{int}c_\mathrm{int}}{\Delta_\mathrm{LH}}|\psi_0(z_0)|^2\,.
\end{equation}
The $g$-matrix is actually diagonal in the primed Pauli matrices and coordinates introduced in section \ref{sec:SO}:
\begin{equation}
\tilde{H}_\mathrm{Z}=-\frac{1}{2}\mu_B(g_{x^\prime} B_{x^\prime}\sigma_1^\prime+g_{y^\prime} B_{y^\prime}\sigma_2^\prime+g_\perp B_z\sigma_3^\prime)\,,
\end{equation}
where:
\begin{subequations}
\begin{align}
g_{x^\prime}&=g_\otimes-g_\parallel \\
g_{y^\prime}&=g_\otimes+g_\parallel \,.
\end{align}
\end{subequations}
The principal magnetic axes of the dot are, therefore, $x^\prime$, $y^\prime$, and $z$, and the principal $g$-factors $g_{x^\prime}$ and $g_{y^\prime}$ differ in magnitude \cite{Venitucci18}. The $g$-factor correction $g_\otimes$ is proportional to the probability of presence of the hole at the interface. In a Ge well, the corrections at both interfaces add up if $s_\mathrm{int}$ is the same (even number of MLs), and cancel if $s_\mathrm{int}$ is different (odd number of MLs). This is, remarkably, the opposite trend as for the Dresselhaus coefficient $\alpha_\mathrm{D}$ originating from the same $H_\mathrm{int}$.

The sign and magnitude of the $g$-factor correction hence depends critically on the number (even/odd) of Ge MLs in the well, on the position of the well (the sign of $g_\otimes$ changes each time the whole Ge well is shifted by one ML), and on the degree of interdiffusion. We will further estimate the strength of $g_\otimes$ with numerical simulations in the next paragraph.

\subsection{Numerical calculations}

\begin{figure*}
\includegraphics[width=0.95\textwidth]{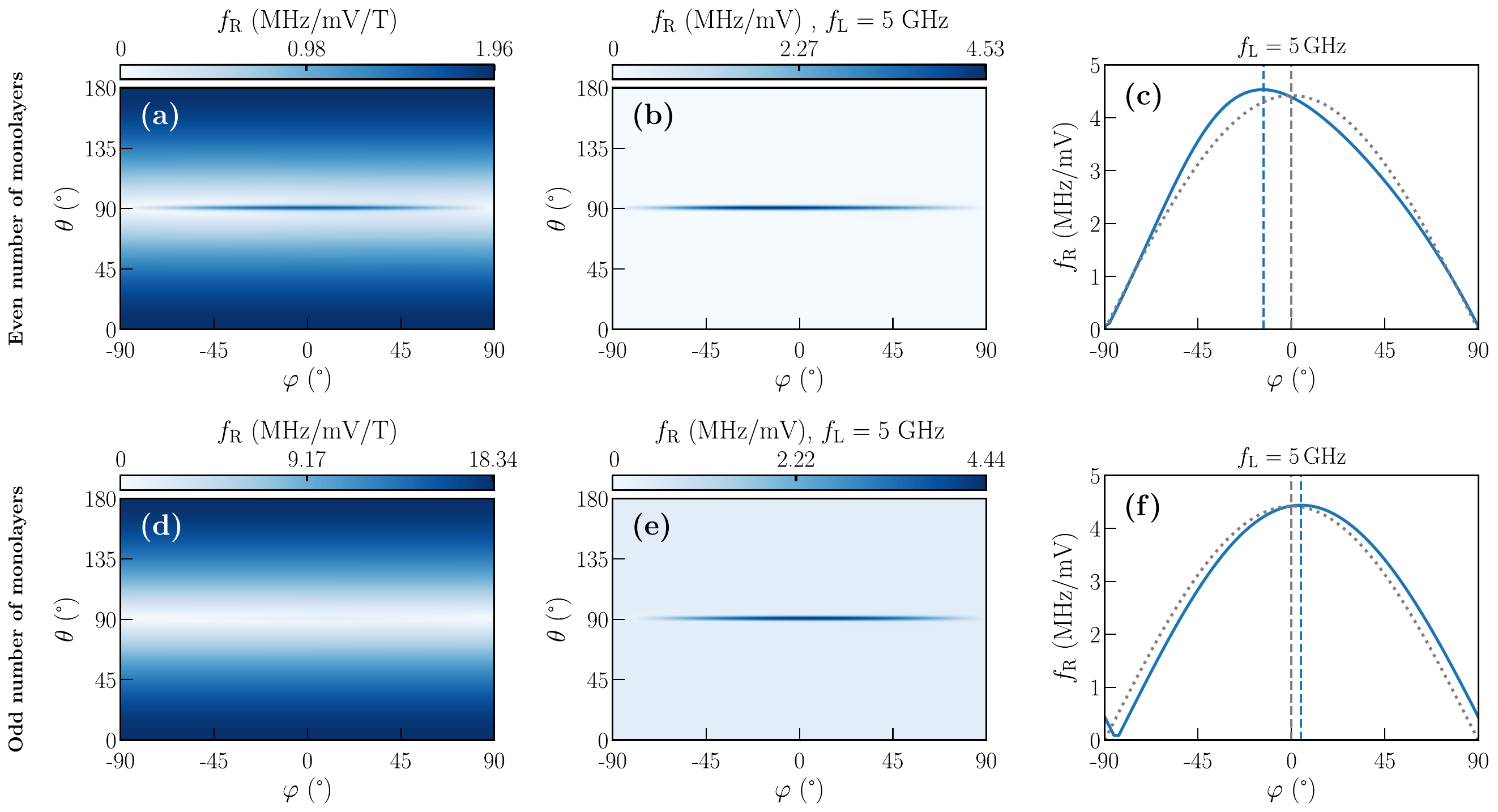}
\caption{(a) Rabi frequency $f_\mathrm{R}$ as a function of the angles $\theta$ and $\varphi$ of the magnetic field defined in Fig.~\ref{fig:device}, at constant magnetic field strength $B=1$ T and drive amplitude $V_\mathrm{ac}=1$\,mV ($V_\mathrm{C}=-40$\,meV). (b) Same at constant Larmor frequency $f_\mathrm{L}=5$\,GHz. (c) Line cut of panel (b) at $\theta=90^\circ$ (solid blue line). The dotted gray line is the Rabi frequency computed at $c_\mathrm{int}=0$, and the vertical dashed lines highlight the maximum Rabi frequency. We have assumed perfect interfaces and an even number of Ge MLs. (d, e, f) Same as (a, b, c) for an odd number of Ge MLs.}
\label{fig:numerics}
\end{figure*}

In order to make a more detailed assessment of the effects of the different SOIs on Ge spin qubits, we have performed comprehensive numerical simulations on the device of Fig.~\ref{fig:device}. For that purpose, we solve Poisson's equation for the potential with a finite volumes method, then compute the hole wave functions with a finite differences implementation of the LK model, and finally calculate the Rabi frequencies with a numerical $g$-matrix formalism \cite{Venitucci18}. We only account for the linear Dresselhaus SOI (with the interface Hamiltonian $H_\mathrm{int}$) and leave out the smaller linear Rashba interaction. We also assume for the sake of simplicity that the strains are homogeneous in the device (no additional SOIs due to inhomogeneous cool-down strains \cite{Uriel22}).

The Rabi frequencies calculated at $V_\mathrm{C}=-40$\,mV are displayed in Fig.~\ref{fig:numerics} in two relevant limiting cases: an ideal Ge well with abrupt interfaces and an even (first row) or odd number of MLs (second row). They are plotted as a function of the orientation of the magnetic field characterized by the polar and azimuthal angles $\theta$ and $\varphi$ defined in Fig.~\ref{fig:device}. The hole is driven by opposite modulations $\delta V_\mathrm{L}=-\delta V_\mathrm{R}=(V_\mathrm{ac}/2)\cos\omega_\mathrm{L}t$ on the L and R gates as in section \ref{sec:analytical}. In the linear response regime, $\delta x_\mathrm{ac}\propto V_\mathrm{ac}$ so that $f_\mathrm{R}$ is proportional to both $B$ and $V_\mathrm{ac}$ [see Eq.~\eqref{eq:fRabi}]. The Rabi frequency is thus normalized at constant magnetic field $B=1$\,T and drive amplitude $V_\mathrm{ac}=1$\,mV in panels (a, d). Practically, many experiments are however performed at constant Larmor frequency $f_\mathrm{L}$ rather than constant $B$. The Rabi frequency at $f_\mathrm{L}=5$\,GHz is, therefore, also plotted in panels (b, e), along with a line cut at $\theta=90^\circ$ (in-plane magnetic field) in panels (c, f). The Rabi frequency computed without interface correction ($c_\mathrm{int}=0$) is also reported as a dotted gray line in these two panels. The complete map of Rabi frequency in that reference case can be found in Ref.~\cite{martinez2022hole}.

In Fig.~\ref{fig:numerics}a, the sharp feature at $\theta=90^\circ$ is dominated by the $g$-TMR NS mechanism arising from the coupling between the in-plane and vertical motions of the hole in the non-separable confinement potential and drive field \cite{martinez2022hole}. The out-of plane, $\propto B_z$ background results from the linear Dresselhaus and cubic Rashba SOIs. Their contributions are cut off in-plane by the small $g$-factors $g_x$ and $g_y$ [see Eq.~\eqref{eq:fRabi}]. The balance between these two interactions can be assessed by comparing the Rabi frequencies along $z$ at finite $c_\mathrm{int}$ (linear Dresselhaus plus cubic Rashba SOIs) and at $c_\mathrm{int}=0$ (cubic Rashba SOI only). In the present case (even number of Ge MLs), $\alpha_\mathrm{D}$ is small because symmetry is broken only by the weak vertical electric field of the gates. Yet the linear Dresselhaus interaction already outweighs cubic Rashba SOI [$f_\mathrm{R}(\vec{B}\parallel\vec{z})=1.96$\,MHz/mV/T at $c_\mathrm{int}=55.9$\,meV\,\AA\ {\it vs} $f_\mathrm{R}(\vec{B}\parallel\vec{z})=0.61$\,MHz/mV/T at $c_\mathrm{int}=0$]. Using Eq.~\eqref{eq:fRabi} along $z$ and $y$ we estimate $\alpha_\mathrm{D}\approx-0.19$\,meV\,\AA, which would correspond to an average vertical electric field $E_z=0.29$\,mV/nm according to Fig.~\ref{fig:DresselhausKPEz}, in line with the estimates of section \ref{sec:analytical}. The prevalence of linear Dresselhaus SOI is even more striking for an odd number of Ge MLs (Fig.~\ref{fig:numerics}d), where $f_\mathrm{R}(\vec{B}\parallel\vec{z})=18.34$\,MHz/mV/T is $\approx 30$ times larger than for cubic Rashba SOI only ($\alpha_\mathrm{D}\approx-1.20$\,meV\,\AA\ as estimated in section \ref{sec:analytical}). The in-plane $g$-TMR feature is almost unchanged but becomes hardly visible on the scale of Fig.~\ref{fig:numerics}d. If the interfaces are not too much interdiffused, the linear Dresselhaus interaction shall therefore dominate over cubic Rashba SOI (although the LK model may underestimate the latter). 

Nonetheless, the linear Dresselhaus and cubic Rashba SOIs remain far less efficient than the in-plane $g$-TMR mechanisms at given Larmor frequency. Indeed, $|\vec{B}|$ must be rapidly decreased once the magnetic field goes out of plane since $g_\perp\gg g_\parallel$ (Fig.~\ref{fig:numerics}b,e). The line cuts of panels (c, f) highlight the role of $H_\mathrm{int}$ on the in-plane physics. For an odd number of Ge MLs, the whole plot is shifted (with respect to $c_\mathrm{int}=0$) by a small angle $\delta\varphi$ as a result of the interference between the NS mechanism (with a $\propto\cos\varphi$ dependence) and the linear Dresselhaus SOI (with a $\propto\sin\varphi$ dependence). As a consequence, the Rabi ``sweet spot'' ($f_\mathrm{R}=0$) and the Rabi ``hot spot'' ($f_\mathrm{R}$ maximum) are moved away, respectively, from $\vec{B}\parallel\vec{y}$ and $\vec{B}\parallel\vec{x}$ [$f_\mathrm{R}(\vec{B}\parallel\vec{y})=0.44$\,MHz/mV, close to the estimate of Eqs.~\eqref{eq:fRestimates}]. The shift $\delta\varphi$ is negligible for an even number of Ge MLs ($\alpha_\mathrm{D}$ being much smaller), yet the whole plot is significantly skewed and the  Rabi ``hot spot'' displaced by en even larger $\delta\varphi_\mathrm{max}$. This now results from the imbalance between the in-plane $g$-factors $|g_{x^\prime}|$ and $|g_{y^\prime}|$ discussed in section \ref{sec:gfactorcorr}. As a matter of fact, $|g_{x^\prime}|=0.131\approx |g_{y^\prime}|=0.128$ for an odd number of Ge MLs, but $g_{x^\prime}=-0.147$ and $g_{y^\prime}=0.113$ for an even number of Ge MLs. The magnetic field needed to reach $f_\mathrm{L}=5$\,GHz is therefore larger at $\varphi=+45^\circ$ than at $\varphi=-45^\circ$, and so is the Rabi frequency. The difference $|g_{x^\prime}|-|g_{y^\prime}|$, although small, can be a significant fraction of the in-plane $g$-factors and thus shift the Rabi hot spot by $|\delta\varphi_\mathrm{max}|>10^\circ$. The L/R/T/B gates may be rotated by $45^\circ$ in order to drive the dot along $x^\prime=[110]$ and best benefit from the smaller $|g_{x^\prime}|$. Yet $\alpha_\mathrm{D}$ changes sign each time the whole Ge well is shifted up or down by one ML, which would give rise to significant device-to-device variations of $f_\mathrm{R}(\vec{B}\parallel\vec{x}^\prime)$ (of the order of $(|g_{x^\prime}|-|g_{y^\prime}|)/(|g_{x^\prime}|+|g_{y^\prime}|)=\pm 13\%$. In the present set up (with the side gates laid down along the $\{100\}$ axes), the Rabi frequency along $x$ is independent on the status of the Ge/GeSi interfaces since the Dresselhaus interaction SOI does not give rise to Rabi oscillations for that orientation. It is, however, non optimal (the hot spot being either on the right or left of $\varphi=0$ depending on the number of Ge MLs and position of the interfaces).

\YMN{In the presence of interdiffusion and/or interface steps, the effective value of $c_\mathrm{int}$ will be renormalized down. This can give rise to a significant variability of the out-of-plane Rabi frequencies \cite{Martinez2022}, $f_\mathrm{R}(\vec{B}\parallel\vec{z})$ being expected to range from about $0.6$\,MHz/mV/T at $c_\mathrm{int}=0$ up to $18.34$\,MHz/mV/T at $c_\mathrm{int}=55.9$\,meV\,\AA\  (Fig.~\ref{fig:numerics}d). The in-plane Rabi frequencies will be far less affected, as evidenced by the solid blue and dotted gray lines of Figs.~\ref{fig:numerics}c and \ref{fig:numerics}f.}

\section{Conclusions}

To conclude, we have investigated the spin-orbit interactions arising in the valence band of planar Ge/GeSi heterostructures with atomistic tight-binding calculations, and we have discussed their impact on the operation of Ge/GeSi spin qubits. We have shown, in particular, that symmetry breaking by the Ge/GeSi interfaces gives rise to a linear-in-momentum Dresselhaus-type SOI for heavy holes. The strength $\alpha_\mathrm{D}$ of this interaction is strongly dependent on the parity of the number of Ge monolayers in the well, on its position ($\alpha_\mathrm{D}$ changes sign each time the Ge well is shifted up or down by one monolayer), and on the degree of interdiffusion of the Ge/GeSi interface(s). It is, indeed, almost completely suppressed when the interfaces are interdiffused over more than 5 monolayers. Furthermore, the tight-binding calculations also highlight the existence of a small linear Rashba-type SOI on top of the usual cubic Rashba interaction. This linear Rashba SOI is not related to the Ge/GeSi interfaces but arises from the mixing between the HH/LH manifold and the remote conduction bands allowed by the structural inversion asymmetry of the heterostructure.

These linear-in-momentum spin-orbit interactions may be leveraged to drive hole spin qubits. The Dresselhaus component can actually be stronger than cubic Rashba SOI and may dominate the physics of quasi-circular dots under out-of-plane magnetic fields (provided the strains are homogeneous \cite{Uriel22}). It is possibly a significant source of device-to-device variability given its sensitivity to the status of the interfaces. When the magnetic field lies in-plane (as is the case in most experiments), it however competes with $g$-tensor modulation resonance mechanisms that are usually more efficient (motion in non-separable and inhomogeneous confinement potentials and drive fields \cite{martinez2022hole}, motion in inhomogeneous strains \cite{Uriel22}). Nonetheless, the same interface-induced HH/LH mixings that give rise to the linear Dresselhaus SOI also shift the $g$-factors of the holes, possibly leaving visible fingerprints in the in-plane properties. Interdiffused interfaces shall, therefore, actually show less variability than sharp interfaces.

The Luttinger-Kohn, four bands $\vec{k}\cdot\vec{p}$ Hamiltonian can be corrected with interface terms that provide a reliable description of the linear Dresselhaus interaction and of the $g$-factor shifts. It still misses, nonetheless, the linear Rashba SOI and underestimates the cubic Rashba SOI (with respect to tight-binding). The Rashba interactions do not, however, appear to dominate the physics of spin qubit devices in most practical cases where the magnetic field lies in-plane. 

The spin-orbit interactions discussed in this work are not specific to Ge/GeSi interfaces and may, in particular, be relevant for (sharp enough) Si/SiO$_2$ interfaces. While the contributions of the Si/SiO$_2$ interfaces to the spin-orbit interactions of electrons \YMN{have been extensively investigated \cite{Golub04,Nestoklon08,Veldhorst15,Ruskov18,Ferdous18b,Tanttu19,Hosseinkhani21}}, the case of holes remains unexplored.

\section*{Acknowledgements}

This work is supported by the French National Research Agency under the programme ``France 2030'' (PEPR PRESQUILE - ANR-22-PETQ-0002). JCAU is supported by a fellowship from the Fundación General CSIC's ComFuturo programme which has received funding from the European Union's Horizon 2020 research and innovation programme under the Marie Skłodowska-Curie grant agreement No. 101034263.

\appendix

\section{Spin $\tfrac{3}{2}$ transformations}
\label{app:symmetries}

The $\ket{0,\pm\tfrac{3}{2}}$ states transform under the symmetry operations as the bulk $\ket{J=\tfrac{3}{2}, J_z=\pm\tfrac{3}{2}}$ Bloch functions \cite{Winkler03}:
\begin{subequations}
\label{eq:threehalf1}
\begin{align}
\ket{J=\tfrac{3}{2}, J_z=+\tfrac{3}{2}}&=-\frac{1}{\sqrt{2}}\left(\ket{X}+i\ket{Y}\right)\otimes\ket{\uparrow} \\
\ket{J=\tfrac{3}{2}, J_z=-\tfrac{3}{2}}&=+\frac{1}{\sqrt{2}}\left(\ket{X}-i\ket{Y}\right)\otimes\ket{\downarrow}\,,
\end{align}
\end{subequations}
where $\ket{X}$ and $\ket{Y}$ transform as the $p_x$ and $p_y$ orbitals (or as the $x$ and $y$ coordinates). The physical spin $\mathbf{S}$ is quantized along $z$, and the phase chosen so that $S_x=\tfrac{1}{2}\hbar\sigma_1$, $S_y=\tfrac{1}{2}\hbar\sigma_2$ and $S_z=\tfrac{1}{2}\hbar\sigma_3$ in the $\{\ket{\uparrow},\ket{\downarrow}\}$ basis set. The symmetry operations must be applied to both the orbital and spin parts \cite{Winkler03,Merzbacher97}, so that the $\tfrac{3}{2}$ spins do not necessarily transform like $\tfrac{1}{2}$ spins although they are routinely mapped onto an effective pseudospin.

We can alternatively work with the following 
$\ket{J=\tfrac{3}{2}, J_z=\pm\tfrac{3}{2}}^\prime$ Bloch functions:
\begin{subequations}
\label{eq:threehalf2}
\begin{align}
\ket{J=\tfrac{3}{2}, J_z=+\tfrac{3}{2}}^\prime&=-\frac{e^{-i\tfrac{\pi}{8}}}{\sqrt{2}}\left(\ket{X^\prime}+i\ket{Y^\prime}\right)\otimes\ket{\uparrow}  \nonumber  \\
&=e^{-i\tfrac{3\pi}{8}}\ket{J=\tfrac{3}{2}, J_z=+\tfrac{3}{2}} \\
\ket{J=\tfrac{3}{2}, J_z=-\tfrac{3}{2}}^\prime&=+\frac{e^{i\tfrac{\pi}{8}}}{\sqrt{2}}\left(\ket{X^\prime}-i\ket{Y^\prime}\right)\otimes\ket{\downarrow} \nonumber \\ 
&=e^{i\tfrac{3\pi}{8}}\ket{J=\tfrac{3}{2}, J_z=-\tfrac{3}{2}} \,,
\end{align}
\end{subequations}
where $\ket{X^\prime}$ and $\ket{Y^\prime}$ now transform as $p_{x^\prime}$ and $p_{y^\prime}$ orbitals. In the $\ket{J=\tfrac{3}{2}, J_z=\pm\tfrac{3}{2}}^\prime$ basis set, the Pauli matrices transform as in Table \ref{tab:symmetries}, which leaves Eqs.~\eqref{eq:H1} and \eqref{eq:H3} as possible invariants. In the $\ket{J=\tfrac{3}{2}, J_z=\pm\tfrac{3}{2}}$ basis set, the possible invariants are equivalently Eqs.~\eqref{eq:H1_cubic} and \eqref{eq:H3_cubic}.  

\section{Case of a Ge/Si superlattice}
\label{app:GeSi}

\begin{figure}
\includegraphics[width=0.9\columnwidth]{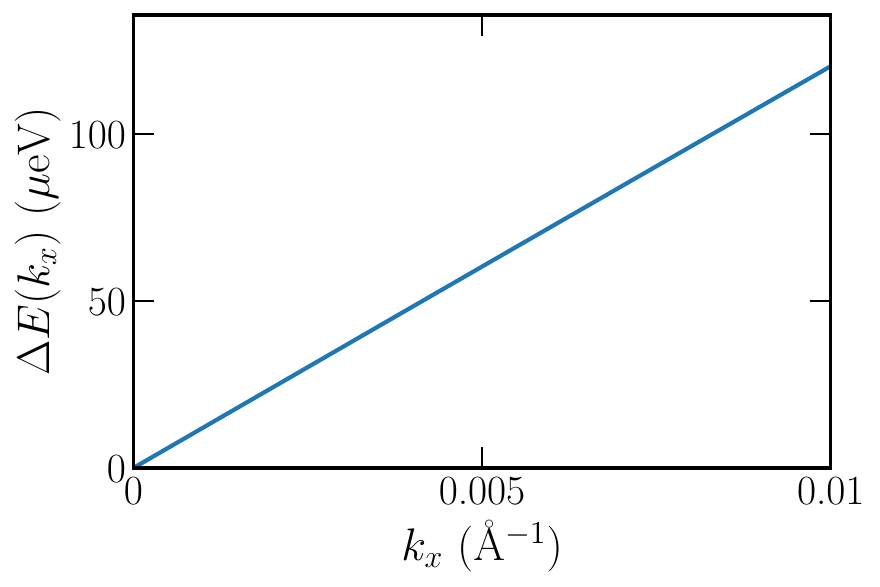}
\caption{TB spin splitting computed along $x=[100]$ in a (Ge)$_{40}$/(Si)$_{20}$ superlattice at vertical electric field $E_z=10$\,meV/nm.}
\label{fig:comparison}
\end{figure}

We have benchmarked TB against the pseudo-potential calculations of Ref.~\cite{Xiong21}. For that purpose, we have simulated the same (Ge)$_{40}$/(Si)$_{20}$ superlattice with pure Si barriers. The TB spin splitting computed at $E_z=10$\,meV/nm is plotted along $x=[100]$ in Fig.~\ref{fig:comparison}. We find as Ref.~\cite{Xiong21} a significant linear-in-$k$ splitting with no sizable cubic correction. The slope $\alpha=6.03$\,meV\,\AA\ is, however, twice larger than in Fig.~2b of Ref.~\cite{Xiong21}. Moreover, we can unambiguously deembed (with the methodology of section \ref{sec:TB}) a dominant Dresselhaus SOI with strength $\alpha_\mathrm{D}=-5.93$\,meV\,\AA\ along with a smaller Rashba SOI with strength $\alpha_\mathrm{R}=-1.12$\,meV\,\AA. 

\section{\textit{Ab initio} calculations}
\label{app:abinitio}

\begin{figure}
\includegraphics[width=0.8\columnwidth]{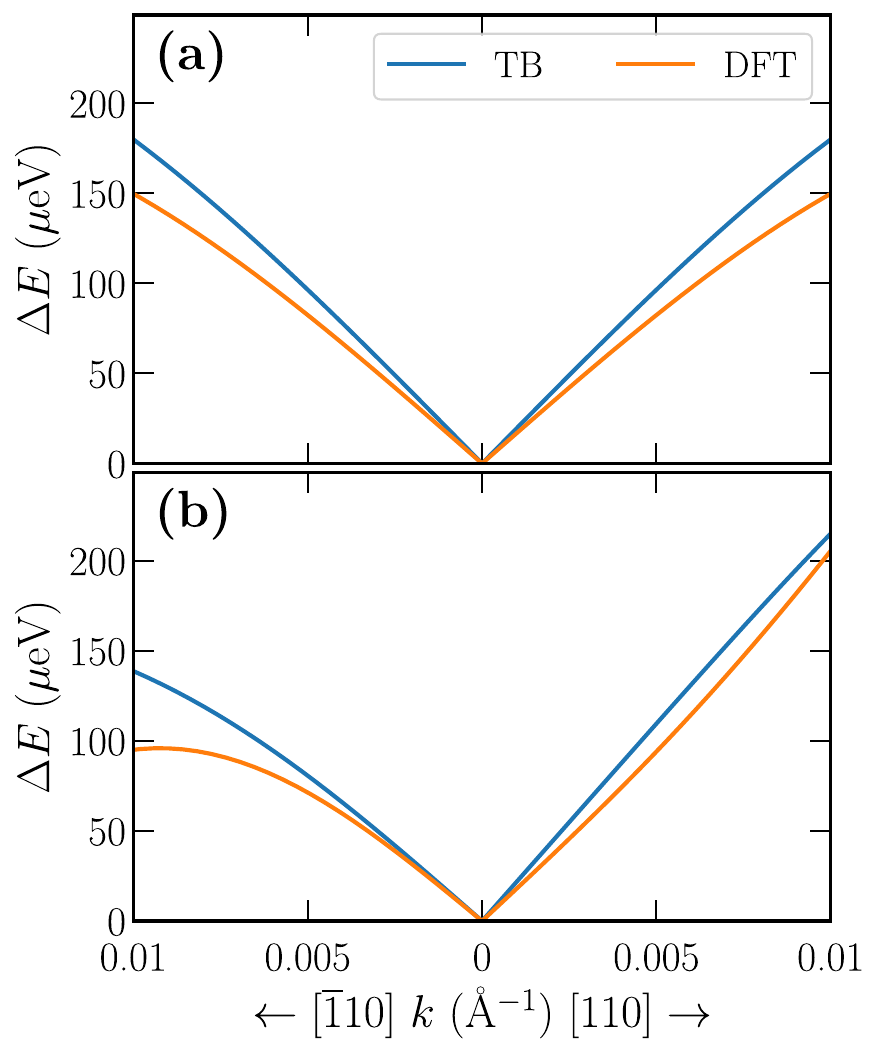}
\caption{(a) DFT and TB spin splittings in a (Si)$_{19}$/(Ge)$_{41}$/(Si)$_{19}$ slab, along a path from $k_{y^\prime}=0.1$\,\AA$^{-1}$ to $\Gamma$ then to $k_{x^\prime}=0.1$\,\AA$^{-1}$, at zero vertical electric field $E_z$. (b) Spin splitting along the same path at $E_z=6$ mV/nm.}
\label{fig:DFT_results}
\end{figure}

In order to consolidate the results, we have also benchmarked TB against {\it ab initio} density functional theory (DFT) on test Ge/Si structures. The DFT calculations are performed with the Vienna {\it ab initio} simulation package (VASP) \cite{Kresse93,Kresse94,Kresse96,Kresse96b}, the Perdew-Burke-Ernzerhof (PBE) exchange-correlation functional \cite{PBE96}, an energy cut-off of $350$\,eV and a $8\times 8\times 1$ $k$-points mesh.

We consider a (Si)$_{19}$/(Ge)$_{41}$/(Si)$_{19}$ slab instead of a superlattice as this eases the introduction of an electric field in the DFT calculations. The dangling bonds at the top and bottom surfaces of the slab are saturated with hydrogen atoms in both DFT and TB calculations. Also, the whole structure is hydrostatically strained to the lattice parameter $a=5.579$\,\AA\ in order to ensure a finite band gap (bulk, unstrained Ge being a semi-metal with PBE).

The DFT and TB spin splittings (computed with the same assumptions) are compared along the $k_{y^\prime}=0.01$\,\AA$^{-1}\to\Gamma\to k_{x^\prime}=0.01$\,\AA$^{-1}$ path in Fig.~\ref{fig:DFT_results}a. The vertical electric field is zero. Therefore, these splittings are expected to result from the linear and cubic Dresselhaus SOI, and shall be safe against the underestimation of the band gap by DFT. Fitting the spin splittings with
\begin{equation}
\Delta E(k)=2\alpha |k|+2\beta |k|^3\,,
\end{equation}
we find $\alpha_\mathrm{DFT}=8.17$ meV\,\AA\ and $\beta_\mathrm{DFT}=-10.25$ eV\,\AA$^3$, close to $\alpha_\mathrm{TB}=9.38$\,meV\,\AA\ and $\beta_\mathrm{TB}=-8.33$ eV\,\AA$^3$. We conclude, therefore, that the TB model provides a reliable description of the Dresselhaus SOI. 

We also plot in Fig.~\ref{fig:DFT_results}b the same splittings at finite electric field $E_z=6$ mV/nm. The agreement remains very good, although the DFT shows even stronger cubic corrections than TB. This may, however, result from the underestimation of the band gap by DFT, which shall enhance the mixing with the remote conduction bands.

\section{Boundary conditions at the Ge/GeSi interfaces.}
\label{app:BF}

\YMN{The Luttinger parameters $\gamma_i$ are discontinuous at the Ge/GeSi interface. The treatment of this discontinuity in finite-difference codes depends on the choices made to enforce hermiticity at the interfaces. The products $\gamma_i k_\alpha k_\beta$ may be replaced by $-\tfrac{1}{2}(\tfrac{\partial}{\partial\alpha}\gamma_i\tfrac{\partial}{\partial\beta}+\tfrac{\partial}{\partial\beta}\gamma_i\tfrac{\partial}{\partial\alpha})$ (symmetric operator ordering), or by the Burt-Foreman (BF) operator ordering scheme \cite{Burt92,Foreman93,Foreman97,Veprek07}.}

\YMN{We have computed the spin splittings with both the symmetric and the BF Luttinger-Kohn Hamiltonians. We have also compared them with a ``homogeneous'' solution where the Luttinger parameters in GeSi are the same as in Ge (in which case BF and symmetric ordering are equivalent). First of all, the topmost valence bands are almost indistinguishable on a few meV scale in the symmetric and BF cases. The spin splittings of the ground HH subband (on the few tens of $\mu$eV scale) are however different. On the one hand, the linear Dresselhaus coefficient is barely dependent on the ordering. On the other hand, the cubic Rashba coefficients almost double with BF ordering. The BF coefficients are thus more consistent with the TB data, but this agreement looks fortuitous. Indeed, the enhancement of the BF coefficients is an interface effect as {\it i}) the symmetric and BF Hamiltonians only differ at the interfaces, and {\it ii}) the symmetric coefficients are almost the same as the homogeneous ones (which highlights that they primarily arise from the bulk of the Ge well that hosts the hole). This is, however, not backed up by the TB calculations. The TB cubic Rashba coefficients are indeed little sensitive to interdiffusion (which suggests again that they mostly arise from the bulk of the materials). Moreover, this apparent agreement is spoiled once the split-off bands are added to the model. A comprehensive assessment of cubic Rashba interactions and boundary conditions at the interfaces calls, therefore, for an extended $\vec{k}\cdot\vec{p}$ model including at least the split-off and lowest conduction bands.}

\bibliography{biblio}

\begin{thebibliography}{94}%
\makeatletter
\providecommand \@ifxundefined [1]{%
 \@ifx{#1\undefined}
}%
\providecommand \@ifnum [1]{%
 \ifnum #1\expandafter \@firstoftwo
 \else \expandafter \@secondoftwo
 \fi
}%
\providecommand \@ifx [1]{%
 \ifx #1\expandafter \@firstoftwo
 \else \expandafter \@secondoftwo
 \fi
}%
\providecommand \natexlab [1]{#1}%
\providecommand \enquote  [1]{``#1''}%
\providecommand \bibnamefont  [1]{#1}%
\providecommand \bibfnamefont [1]{#1}%
\providecommand \citenamefont [1]{#1}%
\providecommand \href@noop [0]{\@secondoftwo}%
\providecommand \href [0]{\begingroup \@sanitize@url \@href}%
\providecommand \@href[1]{\@@startlink{#1}\@@href}%
\providecommand \@@href[1]{\endgroup#1\@@endlink}%
\providecommand \@sanitize@url [0]{\catcode `\\12\catcode `\$12\catcode
  `\&12\catcode `\#12\catcode `\^12\catcode `\_12\catcode `\%12\relax}%
\providecommand \@@startlink[1]{}%
\providecommand \@@endlink[0]{}%
\providecommand \url  [0]{\begingroup\@sanitize@url \@url }%
\providecommand \@url [1]{\endgroup\@href {#1}{\urlprefix }}%
\providecommand \urlprefix  [0]{URL }%
\providecommand \Eprint [0]{\href }%
\providecommand \doibase [0]{https://doi.org/}%
\providecommand \selectlanguage [0]{\@gobble}%
\providecommand \bibinfo  [0]{\@secondoftwo}%
\providecommand \bibfield  [0]{\@secondoftwo}%
\providecommand \translation [1]{[#1]}%
\providecommand \BibitemOpen [0]{}%
\providecommand \bibitemStop [0]{}%
\providecommand \bibitemNoStop [0]{.\EOS\space}%
\providecommand \EOS [0]{\spacefactor3000\relax}%
\providecommand \BibitemShut  [1]{\csname bibitem#1\endcsname}%
\let\auto@bib@innerbib\@empty
\bibitem [{\citenamefont {Burkard}\ \emph {et~al.}(2023)\citenamefont
  {Burkard}, \citenamefont {Ladd}, \citenamefont {Pan}, \citenamefont
  {Nichol},\ and\ \citenamefont {Petta}}]{Burkard23}%
  \BibitemOpen
  \bibfield  {author} {\bibinfo {author} {\bibfnamefont {G.}~\bibnamefont
  {Burkard}}, \bibinfo {author} {\bibfnamefont {T.~D.}\ \bibnamefont {Ladd}},
  \bibinfo {author} {\bibfnamefont {A.}~\bibnamefont {Pan}}, \bibinfo {author}
  {\bibfnamefont {J.~M.}\ \bibnamefont {Nichol}},\ and\ \bibinfo {author}
  {\bibfnamefont {J.~R.}\ \bibnamefont {Petta}},\ }\bibfield  {title} {\bibinfo
  {title} {Semiconductor spin qubits},\ }\href
  {https://doi.org/10.1103/RevModPhys.95.025003} {\bibfield  {journal}
  {\bibinfo  {journal} {Review of Modern Physics}\ }\textbf {\bibinfo {volume}
  {95}},\ \bibinfo {pages} {025003} (\bibinfo {year} {2023})}\BibitemShut
  {NoStop}%
\bibitem [{\citenamefont {Fang}\ \emph {et~al.}(2023)\citenamefont {Fang},
  \citenamefont {Philippopoulos}, \citenamefont {Culcer}, \citenamefont
  {Coish},\ and\ \citenamefont {Chesi}}]{Fang23}%
  \BibitemOpen
  \bibfield  {author} {\bibinfo {author} {\bibfnamefont {Y.}~\bibnamefont
  {Fang}}, \bibinfo {author} {\bibfnamefont {P.}~\bibnamefont
  {Philippopoulos}}, \bibinfo {author} {\bibfnamefont {D.}~\bibnamefont
  {Culcer}}, \bibinfo {author} {\bibfnamefont {W.~A.}\ \bibnamefont {Coish}},\
  and\ \bibinfo {author} {\bibfnamefont {S.}~\bibnamefont {Chesi}},\ }\bibfield
   {title} {\bibinfo {title} {Recent advances in hole-spin qubits},\ }\href
  {https://doi.org/10.1088/2633-4356/acb87e} {\bibfield  {journal} {\bibinfo
  {journal} {Materials for Quantum Technology}\ }\textbf {\bibinfo {volume}
  {3}},\ \bibinfo {pages} {012003} (\bibinfo {year} {2023})}\BibitemShut
  {NoStop}%
\bibitem [{\citenamefont {Winkler}(2003)}]{Winkler03}%
  \BibitemOpen
  \bibfield  {author} {\bibinfo {author} {\bibfnamefont {R.}~\bibnamefont
  {Winkler}},\ }\href {https://doi.org/10.1007/b13586} {\emph {\bibinfo {title}
  {Spin-orbit coupling in two-dimensional electron and hole systems}}}\
  (\bibinfo  {publisher} {Springer},\ \bibinfo {address} {Berlin},\ \bibinfo
  {year} {2003})\BibitemShut {NoStop}%
\bibitem [{\citenamefont {Kloeffel}\ \emph {et~al.}(2011)\citenamefont
  {Kloeffel}, \citenamefont {Trif},\ and\ \citenamefont {Loss}}]{Kloeffel11}%
  \BibitemOpen
  \bibfield  {author} {\bibinfo {author} {\bibfnamefont {C.}~\bibnamefont
  {Kloeffel}}, \bibinfo {author} {\bibfnamefont {M.}~\bibnamefont {Trif}},\
  and\ \bibinfo {author} {\bibfnamefont {D.}~\bibnamefont {Loss}},\ }\bibfield
  {title} {\bibinfo {title} {Strong spin-orbit interaction and helical hole
  states in {Ge/Si} nanowires},\ }\href
  {https://doi.org/10.1103/PhysRevB.84.195314} {\bibfield  {journal} {\bibinfo
  {journal} {Physical Review B}\ }\textbf {\bibinfo {volume} {84}},\ \bibinfo
  {pages} {195314} (\bibinfo {year} {2011})}\BibitemShut {NoStop}%
\bibitem [{\citenamefont {Kloeffel}\ \emph {et~al.}(2018)\citenamefont
  {Kloeffel}, \citenamefont {Ran\ifmmode \check{c}\else
  \v{c}\fi{}i\ifmmode~\acute{c}\else \'{c}\fi{}},\ and\ \citenamefont
  {Loss}}]{Kloeffel18}%
  \BibitemOpen
  \bibfield  {author} {\bibinfo {author} {\bibfnamefont {C.}~\bibnamefont
  {Kloeffel}}, \bibinfo {author} {\bibfnamefont {M.~J.}\ \bibnamefont
  {Ran\ifmmode \check{c}\else \v{c}\fi{}i\ifmmode~\acute{c}\else \'{c}\fi{}}},\
  and\ \bibinfo {author} {\bibfnamefont {D.}~\bibnamefont {Loss}},\ }\bibfield
  {title} {\bibinfo {title} {Direct {Rashba} spin-orbit interaction in {Si} and
  {Ge} nanowires with different growth directions},\ }\href
  {https://doi.org/10.1103/PhysRevB.97.235422} {\bibfield  {journal} {\bibinfo
  {journal} {Physical Review B}\ }\textbf {\bibinfo {volume} {97}},\ \bibinfo
  {pages} {235422} (\bibinfo {year} {2018})}\BibitemShut {NoStop}%
\bibitem [{\citenamefont {Maurand}\ \emph {et~al.}(2016)\citenamefont
  {Maurand}, \citenamefont {Jehl}, \citenamefont {Kotekar-Patil}, \citenamefont
  {Corna}, \citenamefont {Bohuslavskyi}, \citenamefont {Lavi\'{e}ville},
  \citenamefont {Hutin}, \citenamefont {Barraud}, \citenamefont {Vinet},
  \citenamefont {Sanquer},\ and\ \citenamefont {de~Franceschi}}]{Maurand16}%
  \BibitemOpen
  \bibfield  {author} {\bibinfo {author} {\bibfnamefont {R.}~\bibnamefont
  {Maurand}}, \bibinfo {author} {\bibfnamefont {X.}~\bibnamefont {Jehl}},
  \bibinfo {author} {\bibfnamefont {D.}~\bibnamefont {Kotekar-Patil}}, \bibinfo
  {author} {\bibfnamefont {A.}~\bibnamefont {Corna}}, \bibinfo {author}
  {\bibfnamefont {H.}~\bibnamefont {Bohuslavskyi}}, \bibinfo {author}
  {\bibfnamefont {R.}~\bibnamefont {Lavi\'{e}ville}}, \bibinfo {author}
  {\bibfnamefont {L.}~\bibnamefont {Hutin}}, \bibinfo {author} {\bibfnamefont
  {S.}~\bibnamefont {Barraud}}, \bibinfo {author} {\bibfnamefont
  {M.}~\bibnamefont {Vinet}}, \bibinfo {author} {\bibfnamefont
  {M.}~\bibnamefont {Sanquer}},\ and\ \bibinfo {author} {\bibfnamefont
  {S.}~\bibnamefont {de~Franceschi}},\ }\bibfield  {title} {\bibinfo {title} {A
  {CMOS} silicon spin qubit},\ }\href {https://doi.org/10.1038/ncomms13575}
  {\bibfield  {journal} {\bibinfo  {journal} {Nature Communications}\ }\textbf
  {\bibinfo {volume} {7}},\ \bibinfo {pages} {13575} (\bibinfo {year}
  {2016})}\BibitemShut {NoStop}%
\bibitem [{\citenamefont {Crippa}\ \emph {et~al.}(2018)\citenamefont {Crippa},
  \citenamefont {Maurand}, \citenamefont {Bourdet}, \citenamefont
  {Kotekar-Patil}, \citenamefont {Amisse}, \citenamefont {Jehl}, \citenamefont
  {Sanquer}, \citenamefont {Laviéville}, \citenamefont {Bohuslavskyi},
  \citenamefont {Hutin}, \citenamefont {Barraud}, \citenamefont {Vinet},
  \citenamefont {Niquet},\ and\ \citenamefont {{De Franceschi}}}]{Crippa18}%
  \BibitemOpen
  \bibfield  {author} {\bibinfo {author} {\bibfnamefont {A.}~\bibnamefont
  {Crippa}}, \bibinfo {author} {\bibfnamefont {R.}~\bibnamefont {Maurand}},
  \bibinfo {author} {\bibfnamefont {L.}~\bibnamefont {Bourdet}}, \bibinfo
  {author} {\bibfnamefont {D.}~\bibnamefont {Kotekar-Patil}}, \bibinfo {author}
  {\bibfnamefont {A.}~\bibnamefont {Amisse}}, \bibinfo {author} {\bibfnamefont
  {X.}~\bibnamefont {Jehl}}, \bibinfo {author} {\bibfnamefont {M.}~\bibnamefont
  {Sanquer}}, \bibinfo {author} {\bibfnamefont {R.}~\bibnamefont
  {Laviéville}}, \bibinfo {author} {\bibfnamefont {H.}~\bibnamefont
  {Bohuslavskyi}}, \bibinfo {author} {\bibfnamefont {L.}~\bibnamefont {Hutin}},
  \bibinfo {author} {\bibfnamefont {S.}~\bibnamefont {Barraud}}, \bibinfo
  {author} {\bibfnamefont {M.}~\bibnamefont {Vinet}}, \bibinfo {author}
  {\bibfnamefont {Y.-M.}\ \bibnamefont {Niquet}},\ and\ \bibinfo {author}
  {\bibfnamefont {S.}~\bibnamefont {{De Franceschi}}},\ }\bibfield  {title}
  {\bibinfo {title} {Electrical spin driving by $g$-matrix modulation in
  spin-orbit qubits},\ }\href {https://doi.org/10.1103/PhysRevLett.120.137702}
  {\bibfield  {journal} {\bibinfo  {journal} {Physical Review Letters}\
  }\textbf {\bibinfo {volume} {120}},\ \bibinfo {pages} {137702} (\bibinfo
  {year} {2018})}\BibitemShut {NoStop}%
\bibitem [{\citenamefont {Camenzind}\ \emph {et~al.}(2022)\citenamefont
  {Camenzind}, \citenamefont {Geyer}, \citenamefont {Fuhrer}, \citenamefont
  {Warburton}, \citenamefont {Zumbühl},\ and\ \citenamefont
  {Kuhlmann}}]{Camenzind22}%
  \BibitemOpen
  \bibfield  {author} {\bibinfo {author} {\bibfnamefont {L.~C.}\ \bibnamefont
  {Camenzind}}, \bibinfo {author} {\bibfnamefont {S.}~\bibnamefont {Geyer}},
  \bibinfo {author} {\bibfnamefont {A.}~\bibnamefont {Fuhrer}}, \bibinfo
  {author} {\bibfnamefont {R.~J.}\ \bibnamefont {Warburton}}, \bibinfo {author}
  {\bibfnamefont {D.~M.}\ \bibnamefont {Zumbühl}},\ and\ \bibinfo {author}
  {\bibfnamefont {A.~V.}\ \bibnamefont {Kuhlmann}},\ }\bibfield  {title}
  {\bibinfo {title} {A hole spin qubit in a fin field-effect transistor above 4
  kelvin},\ }\href {https://doi.org/10.1038/s41928-022-00722-0} {\bibfield
  {journal} {\bibinfo  {journal} {Nature Electronics}\ }\textbf {\bibinfo
  {volume} {5}},\ \bibinfo {pages} {178} (\bibinfo {year} {2022})}\BibitemShut
  {NoStop}%
\bibitem [{\citenamefont {Watzinger}\ \emph {et~al.}(2018)\citenamefont
  {Watzinger}, \citenamefont {Kuku\v{c}ka}, \citenamefont {Vuku\v{s}i\'c},
  \citenamefont {Gao}, \citenamefont {Wang}, \citenamefont {Sch\"affler},
  \citenamefont {Zhang},\ and\ \citenamefont {Katsaros}}]{Watzinger18}%
  \BibitemOpen
  \bibfield  {author} {\bibinfo {author} {\bibfnamefont {H.}~\bibnamefont
  {Watzinger}}, \bibinfo {author} {\bibfnamefont {J.}~\bibnamefont
  {Kuku\v{c}ka}}, \bibinfo {author} {\bibfnamefont {L.}~\bibnamefont
  {Vuku\v{s}i\'c}}, \bibinfo {author} {\bibfnamefont {F.}~\bibnamefont {Gao}},
  \bibinfo {author} {\bibfnamefont {T.}~\bibnamefont {Wang}}, \bibinfo {author}
  {\bibfnamefont {F.}~\bibnamefont {Sch\"affler}}, \bibinfo {author}
  {\bibfnamefont {J.-J.}\ \bibnamefont {Zhang}},\ and\ \bibinfo {author}
  {\bibfnamefont {G.}~\bibnamefont {Katsaros}},\ }\bibfield  {title} {\bibinfo
  {title} {A germanium hole spin qubit},\ }\href
  {https://doi.org/10.1038/s41467-018-06418-4} {\bibfield  {journal} {\bibinfo
  {journal} {Nature Communications}\ }\textbf {\bibinfo {volume} {9}},\
  \bibinfo {pages} {3902} (\bibinfo {year} {2018})}\BibitemShut {NoStop}%
\bibitem [{\citenamefont {Hendrickx}\ \emph
  {et~al.}(2020{\natexlab{a}})\citenamefont {Hendrickx}, \citenamefont
  {Lawrie}, \citenamefont {Petit}, \citenamefont {Sammak}, \citenamefont
  {Scappucci},\ and\ \citenamefont {Veldhorst}}]{Hendrickx20b}%
  \BibitemOpen
  \bibfield  {author} {\bibinfo {author} {\bibfnamefont {N.~W.}\ \bibnamefont
  {Hendrickx}}, \bibinfo {author} {\bibfnamefont {W.~I.~L.}\ \bibnamefont
  {Lawrie}}, \bibinfo {author} {\bibfnamefont {L.}~\bibnamefont {Petit}},
  \bibinfo {author} {\bibfnamefont {A.}~\bibnamefont {Sammak}}, \bibinfo
  {author} {\bibfnamefont {G.}~\bibnamefont {Scappucci}},\ and\ \bibinfo
  {author} {\bibfnamefont {M.}~\bibnamefont {Veldhorst}},\ }\bibfield  {title}
  {\bibinfo {title} {A single-hole spin qubit},\ }\href
  {https://doi.org/10.1038/s41467-020-17211-7} {\bibfield  {journal} {\bibinfo
  {journal} {Nature Communications}\ }\textbf {\bibinfo {volume} {11}},\
  \bibinfo {pages} {3478} (\bibinfo {year} {2020}{\natexlab{a}})}\BibitemShut
  {NoStop}%
\bibitem [{\citenamefont {Froning}\ \emph {et~al.}(2021)\citenamefont
  {Froning}, \citenamefont {Camenzind}, \citenamefont {van~der Molen},
  \citenamefont {Li}, \citenamefont {Bakkers}, \citenamefont {Zumbühl},\ and\
  \citenamefont {Braakman}}]{Froning21}%
  \BibitemOpen
  \bibfield  {author} {\bibinfo {author} {\bibfnamefont {F.~N.~M.}\
  \bibnamefont {Froning}}, \bibinfo {author} {\bibfnamefont {L.~C.}\
  \bibnamefont {Camenzind}}, \bibinfo {author} {\bibfnamefont {O.~A.~H.}\
  \bibnamefont {van~der Molen}}, \bibinfo {author} {\bibfnamefont
  {A.}~\bibnamefont {Li}}, \bibinfo {author} {\bibfnamefont {E.~P. A.~M.}\
  \bibnamefont {Bakkers}}, \bibinfo {author} {\bibfnamefont {D.~M.}\
  \bibnamefont {Zumbühl}},\ and\ \bibinfo {author} {\bibfnamefont {F.~R.}\
  \bibnamefont {Braakman}},\ }\bibfield  {title} {\bibinfo {title} {Ultrafast
  hole spin qubit with gate-tunable spin–orbit switch functionality},\ }\href
  {https://doi.org/10.1038/s41565-020-00828-6} {\bibfield  {journal} {\bibinfo
  {journal} {Nature Nanotechnology}\ }\textbf {\bibinfo {volume} {16}},\
  \bibinfo {pages} {308} (\bibinfo {year} {2021})}\BibitemShut {NoStop}%
\bibitem [{\citenamefont {Wang}\ \emph
  {et~al.}(2022{\natexlab{a}})\citenamefont {Wang}, \citenamefont {Xu},
  \citenamefont {Gao}, \citenamefont {Liu}, \citenamefont {Ma}, \citenamefont
  {Zhang}, \citenamefont {Wang}, \citenamefont {Cao}, \citenamefont {Wang},
  \citenamefont {Zhang}, \citenamefont {Culcer}, \citenamefont {Hu},
  \citenamefont {Jiang}, \citenamefont {Li}, \citenamefont {Guo},\ and\
  \citenamefont {Guo}}]{wang2022ultrafast}%
  \BibitemOpen
  \bibfield  {author} {\bibinfo {author} {\bibfnamefont {K.}~\bibnamefont
  {Wang}}, \bibinfo {author} {\bibfnamefont {G.}~\bibnamefont {Xu}}, \bibinfo
  {author} {\bibfnamefont {F.}~\bibnamefont {Gao}}, \bibinfo {author}
  {\bibfnamefont {H.}~\bibnamefont {Liu}}, \bibinfo {author} {\bibfnamefont
  {R.-L.}\ \bibnamefont {Ma}}, \bibinfo {author} {\bibfnamefont
  {X.}~\bibnamefont {Zhang}}, \bibinfo {author} {\bibfnamefont
  {Z.}~\bibnamefont {Wang}}, \bibinfo {author} {\bibfnamefont {G.}~\bibnamefont
  {Cao}}, \bibinfo {author} {\bibfnamefont {T.}~\bibnamefont {Wang}}, \bibinfo
  {author} {\bibfnamefont {J.-J.}\ \bibnamefont {Zhang}}, \bibinfo {author}
  {\bibfnamefont {D.}~\bibnamefont {Culcer}}, \bibinfo {author} {\bibfnamefont
  {X.}~\bibnamefont {Hu}}, \bibinfo {author} {\bibfnamefont {H.-W.}\
  \bibnamefont {Jiang}}, \bibinfo {author} {\bibfnamefont {H.-O.}\ \bibnamefont
  {Li}}, \bibinfo {author} {\bibfnamefont {G.-C.}\ \bibnamefont {Guo}},\ and\
  \bibinfo {author} {\bibfnamefont {G.-P.}\ \bibnamefont {Guo}},\ }\bibfield
  {title} {\bibinfo {title} {Ultrafast coherent control of a hole spin qubit in
  a germanium quantum dot},\ }\href
  {https://doi.org/10.1038/s41467-021-27880-7} {\bibfield  {journal} {\bibinfo
  {journal} {Nature Communications}\ }\textbf {\bibinfo {volume} {13}},\
  \bibinfo {pages} {206} (\bibinfo {year} {2022}{\natexlab{a}})}\BibitemShut
  {NoStop}%
\bibitem [{\citenamefont {Fischer}\ \emph {et~al.}(2008)\citenamefont
  {Fischer}, \citenamefont {Coish}, \citenamefont {Bulaev},\ and\ \citenamefont
  {Loss}}]{fischer2008spin}%
  \BibitemOpen
  \bibfield  {author} {\bibinfo {author} {\bibfnamefont {J.}~\bibnamefont
  {Fischer}}, \bibinfo {author} {\bibfnamefont {W.~A.}\ \bibnamefont {Coish}},
  \bibinfo {author} {\bibfnamefont {D.~V.}\ \bibnamefont {Bulaev}},\ and\
  \bibinfo {author} {\bibfnamefont {D.}~\bibnamefont {Loss}},\ }\bibfield
  {title} {\bibinfo {title} {Spin decoherence of a heavy hole coupled to
  nuclear spins in a quantum dot},\ }\href
  {https://doi.org/10.1103/PhysRevB.78.155329} {\bibfield  {journal} {\bibinfo
  {journal} {Physical Review B}\ }\textbf {\bibinfo {volume} {78}},\ \bibinfo
  {pages} {155329} (\bibinfo {year} {2008})}\BibitemShut {NoStop}%
\bibitem [{\citenamefont {Mazzocchi}\ \emph {et~al.}(2019)\citenamefont
  {Mazzocchi}, \citenamefont {Sennikov}, \citenamefont {Bulanov}, \citenamefont
  {Churbanov}, \citenamefont {Bertrand}, \citenamefont {Hutin}, \citenamefont
  {Barnes}, \citenamefont {Drozdov}, \citenamefont {Hartmann},\ and\
  \citenamefont {Sanquer}}]{Mazzocchi19}%
  \BibitemOpen
  \bibfield  {author} {\bibinfo {author} {\bibfnamefont {V.}~\bibnamefont
  {Mazzocchi}}, \bibinfo {author} {\bibfnamefont {P.}~\bibnamefont {Sennikov}},
  \bibinfo {author} {\bibfnamefont {A.}~\bibnamefont {Bulanov}}, \bibinfo
  {author} {\bibfnamefont {M.}~\bibnamefont {Churbanov}}, \bibinfo {author}
  {\bibfnamefont {B.}~\bibnamefont {Bertrand}}, \bibinfo {author}
  {\bibfnamefont {L.}~\bibnamefont {Hutin}}, \bibinfo {author} {\bibfnamefont
  {J.}~\bibnamefont {Barnes}}, \bibinfo {author} {\bibfnamefont
  {M.}~\bibnamefont {Drozdov}}, \bibinfo {author} {\bibfnamefont
  {J.}~\bibnamefont {Hartmann}},\ and\ \bibinfo {author} {\bibfnamefont
  {M.}~\bibnamefont {Sanquer}},\ }\bibfield  {title} {\bibinfo {title}
  {99.992\% $^{28}${Si} {CVD}-grown epilayer on 300 mm substrates for large
  scale integration of silicon spin qubits},\ }\href
  {https://doi.org/10.1016/j.jcrysgro.2018.12.010} {\bibfield  {journal}
  {\bibinfo  {journal} {Journal of Crystal Growth}\ }\textbf {\bibinfo {volume}
  {509}},\ \bibinfo {pages} {1} (\bibinfo {year} {2019})}\BibitemShut {NoStop}%
\bibitem [{\citenamefont {Moutanabbir}\ \emph {et~al.}(2023)\citenamefont
  {Moutanabbir}, \citenamefont {Assali}, \citenamefont {Attiaoui},
  \citenamefont {Daligou}, \citenamefont {Daoust}, \citenamefont {Vecchio},
  \citenamefont {Koelling}, \citenamefont {Luo},\ and\ \citenamefont
  {Rotaru}}]{moutanabbir23}%
  \BibitemOpen
  \bibfield  {author} {\bibinfo {author} {\bibfnamefont {O.}~\bibnamefont
  {Moutanabbir}}, \bibinfo {author} {\bibfnamefont {S.}~\bibnamefont {Assali}},
  \bibinfo {author} {\bibfnamefont {A.}~\bibnamefont {Attiaoui}}, \bibinfo
  {author} {\bibfnamefont {G.}~\bibnamefont {Daligou}}, \bibinfo {author}
  {\bibfnamefont {P.}~\bibnamefont {Daoust}}, \bibinfo {author} {\bibfnamefont
  {P.~D.}\ \bibnamefont {Vecchio}}, \bibinfo {author} {\bibfnamefont
  {S.}~\bibnamefont {Koelling}}, \bibinfo {author} {\bibfnamefont
  {L.}~\bibnamefont {Luo}},\ and\ \bibinfo {author} {\bibfnamefont
  {N.}~\bibnamefont {Rotaru}},\ }\bibfield  {title} {\bibinfo {title} {Nuclear
  spin-depleted, isotopically enriched $^{70}${Ge}/$^{28}${Si}$^{70}${Ge}
  quantum wells},\ }\href {https://arxiv.org/abs/2306.04052} {\bibfield
  {journal} {\bibinfo  {journal} {arXiv:2306.04052}\ } (\bibinfo {year}
  {2023})}\BibitemShut {NoStop}%
\bibitem [{\citenamefont {Bosco}\ \emph
  {et~al.}(2021{\natexlab{a}})\citenamefont {Bosco}, \citenamefont
  {Het\'enyi},\ and\ \citenamefont {Loss}}]{Bosco21}%
  \BibitemOpen
  \bibfield  {author} {\bibinfo {author} {\bibfnamefont {S.}~\bibnamefont
  {Bosco}}, \bibinfo {author} {\bibfnamefont {B.}~\bibnamefont {Het\'enyi}},\
  and\ \bibinfo {author} {\bibfnamefont {D.}~\bibnamefont {Loss}},\ }\bibfield
  {title} {\bibinfo {title} {Hole spin qubits in $\mathrm{Si}$ {FinFETs} with
  fully tunable spin-orbit coupling and sweet spots for charge noise},\ }\href
  {https://doi.org/10.1103/PRXQuantum.2.010348} {\bibfield  {journal} {\bibinfo
   {journal} {PRX Quantum}\ }\textbf {\bibinfo {volume} {2}},\ \bibinfo {pages}
  {010348} (\bibinfo {year} {2021}{\natexlab{a}})}\BibitemShut {NoStop}%
\bibitem [{\citenamefont {Wang}\ \emph {et~al.}(2021)\citenamefont {Wang},
  \citenamefont {Marcellina}, \citenamefont {Hamilton}, \citenamefont {Cullen},
  \citenamefont {Rogge}, \citenamefont {Salfi},\ and\ \citenamefont
  {Culcer}}]{Wang21}%
  \BibitemOpen
  \bibfield  {author} {\bibinfo {author} {\bibfnamefont {Z.}~\bibnamefont
  {Wang}}, \bibinfo {author} {\bibfnamefont {E.}~\bibnamefont {Marcellina}},
  \bibinfo {author} {\bibfnamefont {A.~R.}\ \bibnamefont {Hamilton}}, \bibinfo
  {author} {\bibfnamefont {J.~H.}\ \bibnamefont {Cullen}}, \bibinfo {author}
  {\bibfnamefont {S.}~\bibnamefont {Rogge}}, \bibinfo {author} {\bibfnamefont
  {J.}~\bibnamefont {Salfi}},\ and\ \bibinfo {author} {\bibfnamefont
  {D.}~\bibnamefont {Culcer}},\ }\bibfield  {title} {\bibinfo {title} {Optimal
  operation points for ultrafast, highly coherent {Ge} hole spin-orbit
  qubits},\ }\href {https://doi.org/10.1038/s41534-021-00386-2} {\bibfield
  {journal} {\bibinfo  {journal} {npj Quantum Information}\ }\textbf {\bibinfo
  {volume} {7}},\ \bibinfo {pages} {54} (\bibinfo {year} {2021})}\BibitemShut
  {NoStop}%
\bibitem [{\citenamefont {Piot}\ \emph {et~al.}(2022)\citenamefont {Piot},
  \citenamefont {Brun}, \citenamefont {Schmitt}, \citenamefont {Zihlmann},
  \citenamefont {Michal}, \citenamefont {Apra}, \citenamefont {Abadillo-Uriel},
  \citenamefont {Jehl}, \citenamefont {Bertrand}, \citenamefont {Niebojewski},
  \citenamefont {Hutin}, \citenamefont {Vinet}, \citenamefont {Urdampilleta},
  \citenamefont {Meunier}, \citenamefont {Niquet}, \citenamefont {Maurand},\
  and\ \citenamefont {De~Franceschi}}]{Piot22}%
  \BibitemOpen
  \bibfield  {author} {\bibinfo {author} {\bibfnamefont {N.}~\bibnamefont
  {Piot}}, \bibinfo {author} {\bibfnamefont {B.}~\bibnamefont {Brun}}, \bibinfo
  {author} {\bibfnamefont {V.}~\bibnamefont {Schmitt}}, \bibinfo {author}
  {\bibfnamefont {S.}~\bibnamefont {Zihlmann}}, \bibinfo {author}
  {\bibfnamefont {V.~P.}\ \bibnamefont {Michal}}, \bibinfo {author}
  {\bibfnamefont {A.}~\bibnamefont {Apra}}, \bibinfo {author} {\bibfnamefont
  {J.~C.}\ \bibnamefont {Abadillo-Uriel}}, \bibinfo {author} {\bibfnamefont
  {X.}~\bibnamefont {Jehl}}, \bibinfo {author} {\bibfnamefont {B.}~\bibnamefont
  {Bertrand}}, \bibinfo {author} {\bibfnamefont {H.}~\bibnamefont
  {Niebojewski}}, \bibinfo {author} {\bibfnamefont {L.}~\bibnamefont {Hutin}},
  \bibinfo {author} {\bibfnamefont {M.}~\bibnamefont {Vinet}}, \bibinfo
  {author} {\bibfnamefont {M.}~\bibnamefont {Urdampilleta}}, \bibinfo {author}
  {\bibfnamefont {T.}~\bibnamefont {Meunier}}, \bibinfo {author} {\bibfnamefont
  {Y.-M.}\ \bibnamefont {Niquet}}, \bibinfo {author} {\bibfnamefont
  {R.}~\bibnamefont {Maurand}},\ and\ \bibinfo {author} {\bibfnamefont
  {S.}~\bibnamefont {De~Franceschi}},\ }\bibfield  {title} {\bibinfo {title} {A
  single hole spin with enhanced coherence in natural silicon},\ }\href
  {https://doi.org/10.1038/s41565-022-01196-z} {\bibfield  {journal} {\bibinfo
  {journal} {Nature Nanotechnology}\ }\textbf {\bibinfo {volume} {17}},\
  \bibinfo {pages} {1072} (\bibinfo {year} {2022})}\BibitemShut {NoStop}%
\bibitem [{\citenamefont {Hendrickx}\ \emph {et~al.}(2023)\citenamefont
  {Hendrickx}, \citenamefont {Massai}, \citenamefont {Mergenthaler},
  \citenamefont {Schupp}, \citenamefont {Paredes}, \citenamefont {Bedell},
  \citenamefont {Salis},\ and\ \citenamefont {Fuhrer}}]{Hendrickx23}%
  \BibitemOpen
  \bibfield  {author} {\bibinfo {author} {\bibfnamefont {N.~W.}\ \bibnamefont
  {Hendrickx}}, \bibinfo {author} {\bibfnamefont {L.}~\bibnamefont {Massai}},
  \bibinfo {author} {\bibfnamefont {M.}~\bibnamefont {Mergenthaler}}, \bibinfo
  {author} {\bibfnamefont {F.}~\bibnamefont {Schupp}}, \bibinfo {author}
  {\bibfnamefont {S.}~\bibnamefont {Paredes}}, \bibinfo {author} {\bibfnamefont
  {S.~W.}\ \bibnamefont {Bedell}}, \bibinfo {author} {\bibfnamefont
  {G.}~\bibnamefont {Salis}},\ and\ \bibinfo {author} {\bibfnamefont
  {A.}~\bibnamefont {Fuhrer}},\ }\bibfield  {title} {\bibinfo {title}
  {Sweet-spot operation of a germanium hole spin qubit with highly anisotropic
  noise sensitivity},\ }\href {https://arxiv.org/abs/2305.13150} {\bibfield
  {journal} {\bibinfo  {journal} {arXiv:2305.13150}\ } (\bibinfo {year}
  {2023})}\BibitemShut {NoStop}%
\bibitem [{\citenamefont {Michal}\ \emph {et~al.}(2023)\citenamefont {Michal},
  \citenamefont {Abadillo-Uriel}, \citenamefont {Zihlmann}, \citenamefont
  {Maurand}, \citenamefont {Niquet},\ and\ \citenamefont
  {Filippone}}]{michal2022tunable}%
  \BibitemOpen
  \bibfield  {author} {\bibinfo {author} {\bibfnamefont {V.~P.}\ \bibnamefont
  {Michal}}, \bibinfo {author} {\bibfnamefont {J.~C.}\ \bibnamefont
  {Abadillo-Uriel}}, \bibinfo {author} {\bibfnamefont {S.}~\bibnamefont
  {Zihlmann}}, \bibinfo {author} {\bibfnamefont {R.}~\bibnamefont {Maurand}},
  \bibinfo {author} {\bibfnamefont {Y.-M.}\ \bibnamefont {Niquet}},\ and\
  \bibinfo {author} {\bibfnamefont {M.}~\bibnamefont {Filippone}},\ }\bibfield
  {title} {\bibinfo {title} {Tunable hole spin-photon interaction based on
  $\mathtt{g}$-matrix modulation},\ }\href
  {https://doi.org/10.1103/PhysRevB.107.L041303} {\bibfield  {journal}
  {\bibinfo  {journal} {Physical Review B}\ }\textbf {\bibinfo {volume}
  {107}},\ \bibinfo {pages} {L041303} (\bibinfo {year} {2023})}\BibitemShut
  {NoStop}%
\bibitem [{\citenamefont {Yoneda}\ \emph {et~al.}(2018)\citenamefont {Yoneda},
  \citenamefont {Takeda}, \citenamefont {Otsuka}, \citenamefont {Nakajima},
  \citenamefont {Delbecq}, \citenamefont {Allison}, \citenamefont {Honda},
  \citenamefont {Kodera}, \citenamefont {Oda}, \citenamefont {Hoshi},
  \citenamefont {Usami}, \citenamefont {Itoh},\ and\ \citenamefont
  {Tarucha}}]{Yoneda18}%
  \BibitemOpen
  \bibfield  {author} {\bibinfo {author} {\bibfnamefont {J.}~\bibnamefont
  {Yoneda}}, \bibinfo {author} {\bibfnamefont {K.}~\bibnamefont {Takeda}},
  \bibinfo {author} {\bibfnamefont {T.}~\bibnamefont {Otsuka}}, \bibinfo
  {author} {\bibfnamefont {T.}~\bibnamefont {Nakajima}}, \bibinfo {author}
  {\bibfnamefont {M.~R.}\ \bibnamefont {Delbecq}}, \bibinfo {author}
  {\bibfnamefont {G.}~\bibnamefont {Allison}}, \bibinfo {author} {\bibfnamefont
  {T.}~\bibnamefont {Honda}}, \bibinfo {author} {\bibfnamefont
  {T.}~\bibnamefont {Kodera}}, \bibinfo {author} {\bibfnamefont
  {S.}~\bibnamefont {Oda}}, \bibinfo {author} {\bibfnamefont {Y.}~\bibnamefont
  {Hoshi}}, \bibinfo {author} {\bibfnamefont {N.}~\bibnamefont {Usami}},
  \bibinfo {author} {\bibfnamefont {K.~M.}\ \bibnamefont {Itoh}},\ and\
  \bibinfo {author} {\bibfnamefont {S.}~\bibnamefont {Tarucha}},\ }\bibfield
  {title} {\bibinfo {title} {A quantum-dot spin qubit with coherence limited by
  charge noise and fidelity higher than 99.9\%},\ }\href
  {https://doi.org/https://doi.org/10.1038/s41565-017-0014-x} {\bibfield
  {journal} {\bibinfo  {journal} {Nature Nanotechnology}\ }\textbf {\bibinfo
  {volume} {13}},\ \bibinfo {pages} {102} (\bibinfo {year} {2018})}\BibitemShut
  {NoStop}%
\bibitem [{\citenamefont {Kloeffel}\ \emph {et~al.}(2013)\citenamefont
  {Kloeffel}, \citenamefont {Trif}, \citenamefont {Stano},\ and\ \citenamefont
  {Loss}}]{Kloeffel13}%
  \BibitemOpen
  \bibfield  {author} {\bibinfo {author} {\bibfnamefont {C.}~\bibnamefont
  {Kloeffel}}, \bibinfo {author} {\bibfnamefont {M.}~\bibnamefont {Trif}},
  \bibinfo {author} {\bibfnamefont {P.}~\bibnamefont {Stano}},\ and\ \bibinfo
  {author} {\bibfnamefont {D.}~\bibnamefont {Loss}},\ }\bibfield  {title}
  {\bibinfo {title} {{Circuit QED with hole-spin qubits in Ge/Si nanowire
  quantum dots}},\ }\href {https://doi.org/10.1103/PhysRevB.88.241405}
  {\bibfield  {journal} {\bibinfo  {journal} {Physical Review B}\ }\textbf
  {\bibinfo {volume} {88}},\ \bibinfo {pages} {241405} (\bibinfo {year}
  {2013})}\BibitemShut {NoStop}%
\bibitem [{\citenamefont {Bosco}\ \emph {et~al.}(2022)\citenamefont {Bosco},
  \citenamefont {Scarlino}, \citenamefont {Klinovaja},\ and\ \citenamefont
  {Loss}}]{Bosco22}%
  \BibitemOpen
  \bibfield  {author} {\bibinfo {author} {\bibfnamefont {S.}~\bibnamefont
  {Bosco}}, \bibinfo {author} {\bibfnamefont {P.}~\bibnamefont {Scarlino}},
  \bibinfo {author} {\bibfnamefont {J.}~\bibnamefont {Klinovaja}},\ and\
  \bibinfo {author} {\bibfnamefont {D.}~\bibnamefont {Loss}},\ }\bibfield
  {title} {\bibinfo {title} {Fully tunable longitudinal spin-photon
  interactions in {Si} and {Ge} quantum dots},\ }\href
  {https://doi.org/10.1103/PhysRevLett.129.066801} {\bibfield  {journal}
  {\bibinfo  {journal} {Physical Review Letters}\ }\textbf {\bibinfo {volume}
  {129}},\ \bibinfo {pages} {066801} (\bibinfo {year} {2022})}\BibitemShut
  {NoStop}%
\bibitem [{\citenamefont {Yu}\ \emph {et~al.}(2023)\citenamefont {Yu},
  \citenamefont {Zihlmann}, \citenamefont {Abadillo-Uriel}, \citenamefont
  {Michal}, \citenamefont {Rambal}, \citenamefont {Niebojewski}, \citenamefont
  {Bedecarrats}, \citenamefont {Vinet}, \citenamefont {Dumur}, \citenamefont
  {Filippone}, \citenamefont {Bertrand}, \citenamefont {De~Franceschi},
  \citenamefont {Niquet},\ and\ \citenamefont {Maurand}}]{yu2022strong}%
  \BibitemOpen
  \bibfield  {author} {\bibinfo {author} {\bibfnamefont {C.~X.}\ \bibnamefont
  {Yu}}, \bibinfo {author} {\bibfnamefont {S.}~\bibnamefont {Zihlmann}},
  \bibinfo {author} {\bibfnamefont {J.~C.}\ \bibnamefont {Abadillo-Uriel}},
  \bibinfo {author} {\bibfnamefont {V.~P.}\ \bibnamefont {Michal}}, \bibinfo
  {author} {\bibfnamefont {N.}~\bibnamefont {Rambal}}, \bibinfo {author}
  {\bibfnamefont {H.}~\bibnamefont {Niebojewski}}, \bibinfo {author}
  {\bibfnamefont {T.}~\bibnamefont {Bedecarrats}}, \bibinfo {author}
  {\bibfnamefont {M.}~\bibnamefont {Vinet}}, \bibinfo {author} {\bibfnamefont
  {{\'E}.}~\bibnamefont {Dumur}}, \bibinfo {author} {\bibfnamefont
  {M.}~\bibnamefont {Filippone}}, \bibinfo {author} {\bibfnamefont
  {B.}~\bibnamefont {Bertrand}}, \bibinfo {author} {\bibfnamefont
  {S.}~\bibnamefont {De~Franceschi}}, \bibinfo {author} {\bibfnamefont {Y.-M.}\
  \bibnamefont {Niquet}},\ and\ \bibinfo {author} {\bibfnamefont
  {R.}~\bibnamefont {Maurand}},\ }\bibfield  {title} {\bibinfo {title} {Strong
  coupling between a photon and a hole spin in silicon},\ }\href
  {https://doi.org/10.1038/s41565-023-01332-3} {\bibfield  {journal} {\bibinfo
  {journal} {Nature Nanotechnology}\ }\textbf {\bibinfo {volume} {18}},\
  \bibinfo {pages} {741} (\bibinfo {year} {2023})}\BibitemShut {NoStop}%
\bibitem [{\citenamefont {Scappucci}\ \emph {et~al.}(2020)\citenamefont
  {Scappucci}, \citenamefont {Kloeffel}, \citenamefont {Zwanenburg},
  \citenamefont {Loss}, \citenamefont {Myronov}, \citenamefont {Zhang},
  \citenamefont {De~Franceschi}, \citenamefont {Katsaros},\ and\ \citenamefont
  {Veldhorst}}]{Scappucci20}%
  \BibitemOpen
  \bibfield  {author} {\bibinfo {author} {\bibfnamefont {G.}~\bibnamefont
  {Scappucci}}, \bibinfo {author} {\bibfnamefont {C.}~\bibnamefont {Kloeffel}},
  \bibinfo {author} {\bibfnamefont {F.~A.}\ \bibnamefont {Zwanenburg}},
  \bibinfo {author} {\bibfnamefont {D.}~\bibnamefont {Loss}}, \bibinfo {author}
  {\bibfnamefont {M.}~\bibnamefont {Myronov}}, \bibinfo {author} {\bibfnamefont
  {J.-J.}\ \bibnamefont {Zhang}}, \bibinfo {author} {\bibfnamefont
  {S.}~\bibnamefont {De~Franceschi}}, \bibinfo {author} {\bibfnamefont
  {G.}~\bibnamefont {Katsaros}},\ and\ \bibinfo {author} {\bibfnamefont
  {M.}~\bibnamefont {Veldhorst}},\ }\bibfield  {title} {\bibinfo {title} {The
  germanium quantum information route},\ }\bibfield  {journal} {\bibinfo
  {journal} {Nature Reviews Materials}\ }\href
  {https://doi.org/10.1038/s41578-020-00262-z} {10.1038/s41578-020-00262-z}
  (\bibinfo {year} {2020})\BibitemShut {NoStop}%
\bibitem [{\citenamefont {Hendrickx}\ \emph
  {et~al.}(2020{\natexlab{b}})\citenamefont {Hendrickx}, \citenamefont
  {Franke}, \citenamefont {Sammak}, \citenamefont {Scappucci},\ and\
  \citenamefont {Veldhorst}}]{Hendrickx20}%
  \BibitemOpen
  \bibfield  {author} {\bibinfo {author} {\bibfnamefont {N.~W.}\ \bibnamefont
  {Hendrickx}}, \bibinfo {author} {\bibfnamefont {D.~P.}\ \bibnamefont
  {Franke}}, \bibinfo {author} {\bibfnamefont {A.}~\bibnamefont {Sammak}},
  \bibinfo {author} {\bibfnamefont {G.}~\bibnamefont {Scappucci}},\ and\
  \bibinfo {author} {\bibfnamefont {M.}~\bibnamefont {Veldhorst}},\ }\bibfield
  {title} {\bibinfo {title} {{Fast two-qubit logic with holes in germanium}},\
  }\href {https://doi.org/10.1038/s41586-019-1919-3} {\bibfield  {journal}
  {\bibinfo  {journal} {Nature}\ }\textbf {\bibinfo {volume} {577}},\ \bibinfo
  {pages} {487} (\bibinfo {year} {2020}{\natexlab{b}})}\BibitemShut {NoStop}%
\bibitem [{\citenamefont {Hendrickx}\ \emph {et~al.}(2021)\citenamefont
  {Hendrickx}, \citenamefont {Lawrie~William}, \citenamefont {Russ},
  \citenamefont {van Riggelen}, \citenamefont {de~Snoo}, \citenamefont
  {Schouten}, \citenamefont {Sammak}, \citenamefont {Scappucci},\ and\
  \citenamefont {Veldhorst}}]{Hendrickx21}%
  \BibitemOpen
  \bibfield  {author} {\bibinfo {author} {\bibfnamefont {N.~W.}\ \bibnamefont
  {Hendrickx}}, \bibinfo {author} {\bibfnamefont {I.~L.}\ \bibnamefont
  {Lawrie~William}}, \bibinfo {author} {\bibfnamefont {M.}~\bibnamefont
  {Russ}}, \bibinfo {author} {\bibfnamefont {F.}~\bibnamefont {van Riggelen}},
  \bibinfo {author} {\bibfnamefont {S.~L.}\ \bibnamefont {de~Snoo}}, \bibinfo
  {author} {\bibfnamefont {R.~N.}\ \bibnamefont {Schouten}}, \bibinfo {author}
  {\bibfnamefont {A.}~\bibnamefont {Sammak}}, \bibinfo {author} {\bibfnamefont
  {G.}~\bibnamefont {Scappucci}},\ and\ \bibinfo {author} {\bibfnamefont
  {M.}~\bibnamefont {Veldhorst}},\ }\bibfield  {title} {\bibinfo {title} {A
  four-qubit germanium quantum processor},\ }\href
  {https://doi.org/10.1038/s41586-021-03332-6} {\bibfield  {journal} {\bibinfo
  {journal} {Nature}\ }\textbf {\bibinfo {volume} {591}},\ \bibinfo {pages}
  {580} (\bibinfo {year} {2021})}\BibitemShut {NoStop}%
\bibitem [{\citenamefont {Borsoi}\ \emph {et~al.}(2023)\citenamefont {Borsoi},
  \citenamefont {Hendrickx}, \citenamefont {John}, \citenamefont {Meyer},
  \citenamefont {Motz}, \citenamefont {van Riggelen}, \citenamefont {Sammak},
  \citenamefont {de~Snoo}, \citenamefont {Scappucci},\ and\ \citenamefont
  {Veldhorst}}]{Borsoi22}%
  \BibitemOpen
  \bibfield  {author} {\bibinfo {author} {\bibfnamefont {F.}~\bibnamefont
  {Borsoi}}, \bibinfo {author} {\bibfnamefont {N.~W.}\ \bibnamefont
  {Hendrickx}}, \bibinfo {author} {\bibfnamefont {V.}~\bibnamefont {John}},
  \bibinfo {author} {\bibfnamefont {M.}~\bibnamefont {Meyer}}, \bibinfo
  {author} {\bibfnamefont {S.}~\bibnamefont {Motz}}, \bibinfo {author}
  {\bibfnamefont {F.}~\bibnamefont {van Riggelen}}, \bibinfo {author}
  {\bibfnamefont {A.}~\bibnamefont {Sammak}}, \bibinfo {author} {\bibfnamefont
  {S.~L.}\ \bibnamefont {de~Snoo}}, \bibinfo {author} {\bibfnamefont
  {G.}~\bibnamefont {Scappucci}},\ and\ \bibinfo {author} {\bibfnamefont
  {M.}~\bibnamefont {Veldhorst}},\ }\bibfield  {title} {\bibinfo {title}
  {Shared control of a 16 semiconductor quantum dot crossbar array},\
  }\bibfield  {journal} {\bibinfo  {journal} {Nature Nanotechnology}\ }\href
  {https://doi.org/10.1038/s41565-023-01491-3} {10.1038/s41565-023-01491-3}
  (\bibinfo {year} {2023})\BibitemShut {NoStop}%
\bibitem [{\citenamefont {Martinez}\ and\ \citenamefont
  {Niquet}(2022)}]{Martinez2022}%
  \BibitemOpen
  \bibfield  {author} {\bibinfo {author} {\bibfnamefont {B.}~\bibnamefont
  {Martinez}}\ and\ \bibinfo {author} {\bibfnamefont {Y.-M.}\ \bibnamefont
  {Niquet}},\ }\bibfield  {title} {\bibinfo {title} {Variability of electron
  and hole spin qubits due to interface roughness and charge traps},\ }\href
  {https://doi.org/10.1103/PhysRevApplied.17.024022} {\bibfield  {journal}
  {\bibinfo  {journal} {Physical Review Applied}\ }\textbf {\bibinfo {volume}
  {17}},\ \bibinfo {pages} {024022} (\bibinfo {year} {2022})}\BibitemShut
  {NoStop}%
\bibitem [{\citenamefont {Luttinger}(1956)}]{Luttinger56}%
  \BibitemOpen
  \bibfield  {author} {\bibinfo {author} {\bibfnamefont {J.~M.}\ \bibnamefont
  {Luttinger}},\ }\bibfield  {title} {\bibinfo {title} {Quantum theory of
  cyclotron resonance in semiconductors: General theory},\ }\href
  {https://doi.org/10.1103/PhysRev.102.1030} {\bibfield  {journal} {\bibinfo
  {journal} {Physical Review}\ }\textbf {\bibinfo {volume} {102}},\ \bibinfo
  {pages} {1030} (\bibinfo {year} {1956})}\BibitemShut {NoStop}%
\bibitem [{\citenamefont {Lew Yan~Voon}\ and\ \citenamefont
  {Willatzen}(2009)}]{KP09}%
  \BibitemOpen
  \bibfield  {author} {\bibinfo {author} {\bibfnamefont {L.~C.}\ \bibnamefont
  {Lew Yan~Voon}}\ and\ \bibinfo {author} {\bibfnamefont {M.}~\bibnamefont
  {Willatzen}},\ }\href {https://doi.org/10.1007/978-3-540-92872-0} {\emph
  {\bibinfo {title} {The k p Method}}}\ (\bibinfo  {publisher} {Springer},\
  \bibinfo {address} {Berlin},\ \bibinfo {year} {2009})\BibitemShut {NoStop}%
\bibitem [{\citenamefont {Rashba}\ and\ \citenamefont
  {Sherman}(1988)}]{Rashba88}%
  \BibitemOpen
  \bibfield  {author} {\bibinfo {author} {\bibfnamefont {E.~I.}\ \bibnamefont
  {Rashba}}\ and\ \bibinfo {author} {\bibfnamefont {E.~Y.}\ \bibnamefont
  {Sherman}},\ }\bibfield  {title} {\bibinfo {title} {Spin-orbital band
  splitting in symmetric quantum wells},\ }\href
  {https://doi.org/https://doi.org/10.1016/0375-9601(88)90140-5} {\bibfield
  {journal} {\bibinfo  {journal} {Physics Letters A}\ }\textbf {\bibinfo
  {volume} {129}},\ \bibinfo {pages} {175} (\bibinfo {year}
  {1988})}\BibitemShut {NoStop}%
\bibitem [{\citenamefont {Marcellina}\ \emph {et~al.}(2017)\citenamefont
  {Marcellina}, \citenamefont {Hamilton}, \citenamefont {Winkler},\ and\
  \citenamefont {Culcer}}]{Marcellina17}%
  \BibitemOpen
  \bibfield  {author} {\bibinfo {author} {\bibfnamefont {E.}~\bibnamefont
  {Marcellina}}, \bibinfo {author} {\bibfnamefont {A.~R.}\ \bibnamefont
  {Hamilton}}, \bibinfo {author} {\bibfnamefont {R.}~\bibnamefont {Winkler}},\
  and\ \bibinfo {author} {\bibfnamefont {D.}~\bibnamefont {Culcer}},\
  }\bibfield  {title} {\bibinfo {title} {Spin-orbit interactions in
  inversion-asymmetric two-dimensional hole systems: A variational analysis},\
  }\href {https://doi.org/10.1103/PhysRevB.95.075305} {\bibfield  {journal}
  {\bibinfo  {journal} {Physical Review B}\ }\textbf {\bibinfo {volume} {95}},\
  \bibinfo {pages} {075305} (\bibinfo {year} {2017})}\BibitemShut {NoStop}%
\bibitem [{\citenamefont {Terrazos}\ \emph {et~al.}(2021)\citenamefont
  {Terrazos}, \citenamefont {Marcellina}, \citenamefont {Wang}, \citenamefont
  {Coppersmith}, \citenamefont {Friesen}, \citenamefont {Hamilton},
  \citenamefont {Hu}, \citenamefont {Koiller}, \citenamefont {Saraiva},
  \citenamefont {Culcer},\ and\ \citenamefont {Capaz}}]{Terrazos21}%
  \BibitemOpen
  \bibfield  {author} {\bibinfo {author} {\bibfnamefont {L.~A.}\ \bibnamefont
  {Terrazos}}, \bibinfo {author} {\bibfnamefont {E.}~\bibnamefont
  {Marcellina}}, \bibinfo {author} {\bibfnamefont {Z.}~\bibnamefont {Wang}},
  \bibinfo {author} {\bibfnamefont {S.~N.}\ \bibnamefont {Coppersmith}},
  \bibinfo {author} {\bibfnamefont {M.}~\bibnamefont {Friesen}}, \bibinfo
  {author} {\bibfnamefont {A.~R.}\ \bibnamefont {Hamilton}}, \bibinfo {author}
  {\bibfnamefont {X.}~\bibnamefont {Hu}}, \bibinfo {author} {\bibfnamefont
  {B.}~\bibnamefont {Koiller}}, \bibinfo {author} {\bibfnamefont {A.~L.}\
  \bibnamefont {Saraiva}}, \bibinfo {author} {\bibfnamefont {D.}~\bibnamefont
  {Culcer}},\ and\ \bibinfo {author} {\bibfnamefont {R.~B.}\ \bibnamefont
  {Capaz}},\ }\bibfield  {title} {\bibinfo {title} {Theory of hole-spin qubits
  in strained germanium quantum dots},\ }\href
  {https://doi.org/10.1103/PhysRevB.103.125201} {\bibfield  {journal} {\bibinfo
   {journal} {Physical Review B}\ }\textbf {\bibinfo {volume} {103}},\ \bibinfo
  {pages} {125201} (\bibinfo {year} {2021})}\BibitemShut {NoStop}%
\bibitem [{\citenamefont {Bosco}\ \emph
  {et~al.}(2021{\natexlab{b}})\citenamefont {Bosco}, \citenamefont {Benito},
  \citenamefont {Adelsberger},\ and\ \citenamefont {Loss}}]{Bosco21b}%
  \BibitemOpen
  \bibfield  {author} {\bibinfo {author} {\bibfnamefont {S.}~\bibnamefont
  {Bosco}}, \bibinfo {author} {\bibfnamefont {M.}~\bibnamefont {Benito}},
  \bibinfo {author} {\bibfnamefont {C.}~\bibnamefont {Adelsberger}},\ and\
  \bibinfo {author} {\bibfnamefont {D.}~\bibnamefont {Loss}},\ }\bibfield
  {title} {\bibinfo {title} {Squeezed hole spin qubits in {Ge} quantum dots
  with ultrafast gates at low power},\ }\href
  {https://doi.org/10.1103/PhysRevB.104.115425} {\bibfield  {journal} {\bibinfo
   {journal} {Physical Review B}\ }\textbf {\bibinfo {volume} {104}},\ \bibinfo
  {pages} {115425} (\bibinfo {year} {2021}{\natexlab{b}})}\BibitemShut
  {NoStop}%
\bibitem [{\citenamefont {Ares}\ \emph {et~al.}(2013)\citenamefont {Ares},
  \citenamefont {Golovach}, \citenamefont {Katsaros}, \citenamefont {Stoffel},
  \citenamefont {Fournel}, \citenamefont {Glazman}, \citenamefont {Schmidt},\
  and\ \citenamefont {De~Franceschi}}]{Ares13}%
  \BibitemOpen
  \bibfield  {author} {\bibinfo {author} {\bibfnamefont {N.}~\bibnamefont
  {Ares}}, \bibinfo {author} {\bibfnamefont {V.~N.}\ \bibnamefont {Golovach}},
  \bibinfo {author} {\bibfnamefont {G.}~\bibnamefont {Katsaros}}, \bibinfo
  {author} {\bibfnamefont {M.}~\bibnamefont {Stoffel}}, \bibinfo {author}
  {\bibfnamefont {F.}~\bibnamefont {Fournel}}, \bibinfo {author} {\bibfnamefont
  {L.~I.}\ \bibnamefont {Glazman}}, \bibinfo {author} {\bibfnamefont {O.~G.}\
  \bibnamefont {Schmidt}},\ and\ \bibinfo {author} {\bibfnamefont
  {S.}~\bibnamefont {De~Franceschi}},\ }\bibfield  {title} {\bibinfo {title}
  {Nature of tunable hole $g$ factors in quantum dots},\ }\href
  {https://doi.org/10.1103/PhysRevLett.110.046602} {\bibfield  {journal}
  {\bibinfo  {journal} {Physical Review Letters}\ }\textbf {\bibinfo {volume}
  {110}},\ \bibinfo {pages} {046602} (\bibinfo {year} {2013})}\BibitemShut
  {NoStop}%
\bibitem [{\citenamefont {Michal}\ \emph {et~al.}(2021)\citenamefont {Michal},
  \citenamefont {Venitucci},\ and\ \citenamefont {Niquet}}]{Michal21}%
  \BibitemOpen
  \bibfield  {author} {\bibinfo {author} {\bibfnamefont {V.~P.}\ \bibnamefont
  {Michal}}, \bibinfo {author} {\bibfnamefont {B.}~\bibnamefont {Venitucci}},\
  and\ \bibinfo {author} {\bibfnamefont {Y.-M.}\ \bibnamefont {Niquet}},\
  }\bibfield  {title} {\bibinfo {title} {Longitudinal and transverse electric
  field manipulation of hole spin-orbit qubits in one-dimensional channels},\
  }\href {https://doi.org/10.1103/PhysRevB.103.045305} {\bibfield  {journal}
  {\bibinfo  {journal} {Physical Review B}\ }\textbf {\bibinfo {volume}
  {103}},\ \bibinfo {pages} {045305} (\bibinfo {year} {2021})}\BibitemShut
  {NoStop}%
\bibitem [{\citenamefont {Abadillo-Uriel}\ \emph {et~al.}(2023)\citenamefont
  {Abadillo-Uriel}, \citenamefont {Rodr\'{\i}guez-Mena}, \citenamefont
  {Martinez},\ and\ \citenamefont {Niquet}}]{Uriel22}%
  \BibitemOpen
  \bibfield  {author} {\bibinfo {author} {\bibfnamefont {J.~C.}\ \bibnamefont
  {Abadillo-Uriel}}, \bibinfo {author} {\bibfnamefont {E.~A.}\ \bibnamefont
  {Rodr\'{\i}guez-Mena}}, \bibinfo {author} {\bibfnamefont {B.}~\bibnamefont
  {Martinez}},\ and\ \bibinfo {author} {\bibfnamefont {Y.-M.}\ \bibnamefont
  {Niquet}},\ }\bibfield  {title} {\bibinfo {title} {Hole-spin driving by
  strain-induced spin-orbit interactions},\ }\href
  {https://doi.org/10.1103/PhysRevLett.131.097002} {\bibfield  {journal}
  {\bibinfo  {journal} {Physical Review Letters}\ }\textbf {\bibinfo {volume}
  {131}},\ \bibinfo {pages} {097002} (\bibinfo {year} {2023})}\BibitemShut
  {NoStop}%
\bibitem [{\citenamefont {Rashba}\ and\ \citenamefont
  {Efros}(2003)}]{Rashba03}%
  \BibitemOpen
  \bibfield  {author} {\bibinfo {author} {\bibfnamefont {E.~I.}\ \bibnamefont
  {Rashba}}\ and\ \bibinfo {author} {\bibfnamefont {A.~L.}\ \bibnamefont
  {Efros}},\ }\bibfield  {title} {\bibinfo {title} {Orbital mechanisms of
  electron-spin manipulation by an electric field},\ }\href
  {https://doi.org/10.1103/PhysRevLett.91.126405} {\bibfield  {journal}
  {\bibinfo  {journal} {Physical Review Letters}\ }\textbf {\bibinfo {volume}
  {91}},\ \bibinfo {pages} {126405} (\bibinfo {year} {2003})}\BibitemShut
  {NoStop}%
\bibitem [{\citenamefont {Golovach}\ \emph {et~al.}(2006)\citenamefont
  {Golovach}, \citenamefont {Borhani},\ and\ \citenamefont
  {Loss}}]{Golovach06}%
  \BibitemOpen
  \bibfield  {author} {\bibinfo {author} {\bibfnamefont {V.~N.}\ \bibnamefont
  {Golovach}}, \bibinfo {author} {\bibfnamefont {M.}~\bibnamefont {Borhani}},\
  and\ \bibinfo {author} {\bibfnamefont {D.}~\bibnamefont {Loss}},\ }\bibfield
  {title} {\bibinfo {title} {Electric-dipole-induced spin resonance in quantum
  dots},\ }\href {https://doi.org/10.1103/PhysRevB.74.165319} {\bibfield
  {journal} {\bibinfo  {journal} {Physical Review B}\ }\textbf {\bibinfo
  {volume} {74}},\ \bibinfo {pages} {165319} (\bibinfo {year}
  {2006})}\BibitemShut {NoStop}%
\bibitem [{\citenamefont {Rashba}(2008)}]{Rashba08}%
  \BibitemOpen
  \bibfield  {author} {\bibinfo {author} {\bibfnamefont {E.~I.}\ \bibnamefont
  {Rashba}},\ }\bibfield  {title} {\bibinfo {title} {Theory of electric dipole
  spin resonance in quantum dots: Mean field theory with gaussian fluctuations
  and beyond},\ }\href {https://doi.org/10.1103/PhysRevB.78.195302} {\bibfield
  {journal} {\bibinfo  {journal} {Physical Review B}\ }\textbf {\bibinfo
  {volume} {78}},\ \bibinfo {pages} {195302} (\bibinfo {year}
  {2008})}\BibitemShut {NoStop}%
\bibitem [{\citenamefont {Kato}\ \emph {et~al.}(2003)\citenamefont {Kato},
  \citenamefont {Myers}, \citenamefont {Driscoll}, \citenamefont {Gossard},
  \citenamefont {Levy},\ and\ \citenamefont {Awschalom}}]{Kato03}%
  \BibitemOpen
  \bibfield  {author} {\bibinfo {author} {\bibfnamefont {Y.}~\bibnamefont
  {Kato}}, \bibinfo {author} {\bibfnamefont {R.~C.}\ \bibnamefont {Myers}},
  \bibinfo {author} {\bibfnamefont {D.~C.}\ \bibnamefont {Driscoll}}, \bibinfo
  {author} {\bibfnamefont {A.~C.}\ \bibnamefont {Gossard}}, \bibinfo {author}
  {\bibfnamefont {J.}~\bibnamefont {Levy}},\ and\ \bibinfo {author}
  {\bibfnamefont {D.~D.}\ \bibnamefont {Awschalom}},\ }\bibfield  {title}
  {\bibinfo {title} {Gigahertz electron spin manipulation using
  voltage-controlled $g$-tensor modulation},\ }\href
  {https://doi.org/10.1126/science.1080880} {\bibfield  {journal} {\bibinfo
  {journal} {Science}\ }\textbf {\bibinfo {volume} {299}},\ \bibinfo {pages}
  {1201} (\bibinfo {year} {2003})}\BibitemShut {NoStop}%
\bibitem [{\citenamefont {Venitucci}\ \emph {et~al.}(2018)\citenamefont
  {Venitucci}, \citenamefont {Bourdet}, \citenamefont {Pouzada},\ and\
  \citenamefont {Niquet}}]{Venitucci18}%
  \BibitemOpen
  \bibfield  {author} {\bibinfo {author} {\bibfnamefont {B.}~\bibnamefont
  {Venitucci}}, \bibinfo {author} {\bibfnamefont {L.}~\bibnamefont {Bourdet}},
  \bibinfo {author} {\bibfnamefont {D.}~\bibnamefont {Pouzada}},\ and\ \bibinfo
  {author} {\bibfnamefont {Y.-M.}\ \bibnamefont {Niquet}},\ }\bibfield  {title}
  {\bibinfo {title} {Electrical manipulation of semiconductor spin qubits
  within the $g$-matrix formalism},\ }\href
  {https://doi.org/10.1103/PhysRevB.98.155319} {\bibfield  {journal} {\bibinfo
  {journal} {Physical Review B}\ }\textbf {\bibinfo {volume} {98}},\ \bibinfo
  {pages} {155319} (\bibinfo {year} {2018})}\BibitemShut {NoStop}%
\bibitem [{\citenamefont {Martinez}\ \emph {et~al.}(2022)\citenamefont
  {Martinez}, \citenamefont {Abadillo-Uriel}, \citenamefont
  {Rodr\'{\i}guez-Mena},\ and\ \citenamefont {Niquet}}]{martinez2022hole}%
  \BibitemOpen
  \bibfield  {author} {\bibinfo {author} {\bibfnamefont {B.}~\bibnamefont
  {Martinez}}, \bibinfo {author} {\bibfnamefont {J.~C.}\ \bibnamefont
  {Abadillo-Uriel}}, \bibinfo {author} {\bibfnamefont {E.~A.}\ \bibnamefont
  {Rodr\'{\i}guez-Mena}},\ and\ \bibinfo {author} {\bibfnamefont {Y.-M.}\
  \bibnamefont {Niquet}},\ }\bibfield  {title} {\bibinfo {title} {Hole spin
  manipulation in inhomogeneous and nonseparable electric fields},\ }\href
  {https://doi.org/10.1103/PhysRevB.106.235426} {\bibfield  {journal} {\bibinfo
   {journal} {Physical Review B}\ }\textbf {\bibinfo {volume} {106}},\ \bibinfo
  {pages} {235426} (\bibinfo {year} {2022})}\BibitemShut {NoStop}%
\bibitem [{\citenamefont {Corley-Wiciak}\ \emph {et~al.}(2023)\citenamefont
  {Corley-Wiciak}, \citenamefont {Richter}, \citenamefont {Zoellner},
  \citenamefont {Zaitsev}, \citenamefont {Manganelli}, \citenamefont
  {Zatterin}, \citenamefont {Sch{\"u}lli}, \citenamefont {Corley-Wiciak},
  \citenamefont {Katzer}, \citenamefont {Reichmann}, \citenamefont {Klesse},
  \citenamefont {Hendrickx}, \citenamefont {Sammak}, \citenamefont {Veldhorst},
  \citenamefont {Scappucci}, \citenamefont {Virgilio},\ and\ \citenamefont
  {Capellini}}]{Corley2023}%
  \BibitemOpen
  \bibfield  {author} {\bibinfo {author} {\bibfnamefont {C.}~\bibnamefont
  {Corley-Wiciak}}, \bibinfo {author} {\bibfnamefont {C.}~\bibnamefont
  {Richter}}, \bibinfo {author} {\bibfnamefont {M.~H.}\ \bibnamefont
  {Zoellner}}, \bibinfo {author} {\bibfnamefont {I.}~\bibnamefont {Zaitsev}},
  \bibinfo {author} {\bibfnamefont {C.~L.}\ \bibnamefont {Manganelli}},
  \bibinfo {author} {\bibfnamefont {E.}~\bibnamefont {Zatterin}}, \bibinfo
  {author} {\bibfnamefont {T.~U.}\ \bibnamefont {Sch{\"u}lli}}, \bibinfo
  {author} {\bibfnamefont {A.~A.}\ \bibnamefont {Corley-Wiciak}}, \bibinfo
  {author} {\bibfnamefont {J.}~\bibnamefont {Katzer}}, \bibinfo {author}
  {\bibfnamefont {F.}~\bibnamefont {Reichmann}}, \bibinfo {author}
  {\bibfnamefont {W.~M.}\ \bibnamefont {Klesse}}, \bibinfo {author}
  {\bibfnamefont {N.~W.}\ \bibnamefont {Hendrickx}}, \bibinfo {author}
  {\bibfnamefont {A.}~\bibnamefont {Sammak}}, \bibinfo {author} {\bibfnamefont
  {M.}~\bibnamefont {Veldhorst}}, \bibinfo {author} {\bibfnamefont
  {G.}~\bibnamefont {Scappucci}}, \bibinfo {author} {\bibfnamefont
  {M.}~\bibnamefont {Virgilio}},\ and\ \bibinfo {author} {\bibfnamefont
  {G.}~\bibnamefont {Capellini}},\ }\bibfield  {title} {\bibinfo {title}
  {Nanoscale mapping of the {3D} strain tensor in a germanium quantum well
  hosting a functional spin qubit device},\ }\href
  {https://doi.org/10.1021/acsami.2c17395} {\bibfield  {journal} {\bibinfo
  {journal} {ACS Applied Materials {\&} Interfaces}\ }\textbf {\bibinfo
  {volume} {15}},\ \bibinfo {pages} {3119} (\bibinfo {year}
  {2023})}\BibitemShut {NoStop}%
\bibitem [{\citenamefont {Ciocoiu}\ \emph {et~al.}(2022)\citenamefont
  {Ciocoiu}, \citenamefont {Khalifa},\ and\ \citenamefont {Salfi}}]{Salfi22}%
  \BibitemOpen
  \bibfield  {author} {\bibinfo {author} {\bibfnamefont {A.}~\bibnamefont
  {Ciocoiu}}, \bibinfo {author} {\bibfnamefont {M.}~\bibnamefont {Khalifa}},\
  and\ \bibinfo {author} {\bibfnamefont {J.}~\bibnamefont {Salfi}},\ }\bibfield
   {title} {\bibinfo {title} {Towards computer-assisted design of hole spin
  qubits in quantum dot devices},\ }\href {https://arxiv.org/abs/2209.12026}
  {\bibfield  {journal} {\bibinfo  {journal} {arXiv:2209.12026}\ } (\bibinfo
  {year} {2022})}\BibitemShut {NoStop}%
\bibitem [{\citenamefont {Malkoc}\ \emph {et~al.}(2022)\citenamefont {Malkoc},
  \citenamefont {Stano},\ and\ \citenamefont {Loss}}]{Malkoc22}%
  \BibitemOpen
  \bibfield  {author} {\bibinfo {author} {\bibfnamefont {O.}~\bibnamefont
  {Malkoc}}, \bibinfo {author} {\bibfnamefont {P.}~\bibnamefont {Stano}},\ and\
  \bibinfo {author} {\bibfnamefont {D.}~\bibnamefont {Loss}},\ }\bibfield
  {title} {\bibinfo {title} {Charge-noise-induced dephasing in silicon
  hole-spin qubits},\ }\href {https://doi.org/10.1103/PhysRevLett.129.247701}
  {\bibfield  {journal} {\bibinfo  {journal} {Physical Review Letters}\
  }\textbf {\bibinfo {volume} {129}},\ \bibinfo {pages} {247701} (\bibinfo
  {year} {2022})}\BibitemShut {NoStop}%
\bibitem [{\citenamefont {Wang}\ \emph
  {et~al.}(2022{\natexlab{b}})\citenamefont {Wang}, \citenamefont {Scappucci},
  \citenamefont {Veldhorst},\ and\ \citenamefont {Russ}}]{Wang22}%
  \BibitemOpen
  \bibfield  {author} {\bibinfo {author} {\bibfnamefont {C.-A.}\ \bibnamefont
  {Wang}}, \bibinfo {author} {\bibfnamefont {G.}~\bibnamefont {Scappucci}},
  \bibinfo {author} {\bibfnamefont {M.}~\bibnamefont {Veldhorst}},\ and\
  \bibinfo {author} {\bibfnamefont {M.}~\bibnamefont {Russ}},\ }\bibfield
  {title} {\bibinfo {title} {Modelling of planar germanium hole qubits in
  electric and magnetic fields},\ }\href
  {https://arxiv.org/abs/arXiv:2208.04795} {\bibfield  {journal} {\bibinfo
  {journal} {arXiv:2208.04795}\ } (\bibinfo {year}
  {2022}{\natexlab{b}})}\BibitemShut {NoStop}%
\bibitem [{\citenamefont {Sarkar}\ \emph {et~al.}(2023)\citenamefont {Sarkar},
  \citenamefont {Wang}, \citenamefont {Rendell}, \citenamefont {Hendrickx},
  \citenamefont {Veldhorst}, \citenamefont {Scappucci}, \citenamefont
  {Khalifa}, \citenamefont {Salfi}, \citenamefont {Saraiva}, \citenamefont
  {Dzurak}, \citenamefont {Hamilton},\ and\ \citenamefont
  {Culcer}}]{sarkar2023}%
  \BibitemOpen
  \bibfield  {author} {\bibinfo {author} {\bibfnamefont {A.}~\bibnamefont
  {Sarkar}}, \bibinfo {author} {\bibfnamefont {Z.}~\bibnamefont {Wang}},
  \bibinfo {author} {\bibfnamefont {M.}~\bibnamefont {Rendell}}, \bibinfo
  {author} {\bibfnamefont {N.~W.}\ \bibnamefont {Hendrickx}}, \bibinfo {author}
  {\bibfnamefont {M.}~\bibnamefont {Veldhorst}}, \bibinfo {author}
  {\bibfnamefont {G.}~\bibnamefont {Scappucci}}, \bibinfo {author}
  {\bibfnamefont {M.}~\bibnamefont {Khalifa}}, \bibinfo {author} {\bibfnamefont
  {J.}~\bibnamefont {Salfi}}, \bibinfo {author} {\bibfnamefont
  {A.}~\bibnamefont {Saraiva}}, \bibinfo {author} {\bibfnamefont {A.~S.}\
  \bibnamefont {Dzurak}}, \bibinfo {author} {\bibfnamefont {A.~R.}\
  \bibnamefont {Hamilton}},\ and\ \bibinfo {author} {\bibfnamefont
  {D.}~\bibnamefont {Culcer}},\ }\bibfield  {title} {\bibinfo {title}
  {Electrical operation of planar {Ge} hole spin qubits in an in-plane magnetic
  field},\ }\href {https://arxiv.org/abs/2307.01451} {\bibfield  {journal}
  {\bibinfo  {journal} {arXiv:2307.01451}\ } (\bibinfo {year}
  {2023})}\BibitemShut {NoStop}%
\bibitem [{\citenamefont {Xiong}\ \emph {et~al.}(2021)\citenamefont {Xiong},
  \citenamefont {Guan}, \citenamefont {Luo},\ and\ \citenamefont
  {Li}}]{Xiong21}%
  \BibitemOpen
  \bibfield  {author} {\bibinfo {author} {\bibfnamefont {J.-X.}\ \bibnamefont
  {Xiong}}, \bibinfo {author} {\bibfnamefont {S.}~\bibnamefont {Guan}},
  \bibinfo {author} {\bibfnamefont {J.-W.}\ \bibnamefont {Luo}},\ and\ \bibinfo
  {author} {\bibfnamefont {S.-S.}\ \bibnamefont {Li}},\ }\bibfield  {title}
  {\bibinfo {title} {Emergence of strong tunable linear {Rashba} spin-orbit
  coupling in two-dimensional hole gases in semiconductor quantum wells},\
  }\href {https://doi.org/10.1103/PhysRevB.103.085309} {\bibfield  {journal}
  {\bibinfo  {journal} {Physical Review B}\ }\textbf {\bibinfo {volume}
  {103}},\ \bibinfo {pages} {085309} (\bibinfo {year} {2021})}\BibitemShut
  {NoStop}%
\bibitem [{\citenamefont {Xiong}\ \emph
  {et~al.}(2022{\natexlab{a}})\citenamefont {Xiong}, \citenamefont {Guan},
  \citenamefont {Luo},\ and\ \citenamefont {Li}}]{Xiong22b}%
  \BibitemOpen
  \bibfield  {author} {\bibinfo {author} {\bibfnamefont {J.-X.}\ \bibnamefont
  {Xiong}}, \bibinfo {author} {\bibfnamefont {S.}~\bibnamefont {Guan}},
  \bibinfo {author} {\bibfnamefont {J.-W.}\ \bibnamefont {Luo}},\ and\ \bibinfo
  {author} {\bibfnamefont {S.-S.}\ \bibnamefont {Li}},\ }\bibfield  {title}
  {\bibinfo {title} {Orientation-dependent rashba spin-orbit coupling of
  two-dimensional hole gases in semiconductor quantum wells: Linear or cubic},\
  }\href {https://doi.org/10.1103/PhysRevB.105.115303} {\bibfield  {journal}
  {\bibinfo  {journal} {Physical Review B}\ }\textbf {\bibinfo {volume}
  {105}},\ \bibinfo {pages} {115303} (\bibinfo {year}
  {2022}{\natexlab{a}})}\BibitemShut {NoStop}%
\bibitem [{\citenamefont {Liu}\ \emph {et~al.}(2022)\citenamefont {Liu},
  \citenamefont {Xiong}, \citenamefont {Wang}, \citenamefont {Ma},
  \citenamefont {Guan}, \citenamefont {Luo},\ and\ \citenamefont {Li}}]{Liu22}%
  \BibitemOpen
  \bibfield  {author} {\bibinfo {author} {\bibfnamefont {Y.}~\bibnamefont
  {Liu}}, \bibinfo {author} {\bibfnamefont {J.-X.}\ \bibnamefont {Xiong}},
  \bibinfo {author} {\bibfnamefont {Z.}~\bibnamefont {Wang}}, \bibinfo {author}
  {\bibfnamefont {W.-L.}\ \bibnamefont {Ma}}, \bibinfo {author} {\bibfnamefont
  {S.}~\bibnamefont {Guan}}, \bibinfo {author} {\bibfnamefont {J.-W.}\
  \bibnamefont {Luo}},\ and\ \bibinfo {author} {\bibfnamefont {S.-S.}\
  \bibnamefont {Li}},\ }\bibfield  {title} {\bibinfo {title} {Emergent linear
  {Rashba} spin-orbit coupling offers fast manipulation of hole-spin qubits in
  germanium},\ }\href {https://doi.org/10.1103/PhysRevB.105.075313} {\bibfield
  {journal} {\bibinfo  {journal} {Physical Review B}\ }\textbf {\bibinfo
  {volume} {105}},\ \bibinfo {pages} {075313} (\bibinfo {year}
  {2022})}\BibitemShut {NoStop}%
\bibitem [{\citenamefont {Vervoort}\ \emph {et~al.}(1997)\citenamefont
  {Vervoort}, \citenamefont {Ferreira},\ and\ \citenamefont
  {Voisin}}]{Vervoort97}%
  \BibitemOpen
  \bibfield  {author} {\bibinfo {author} {\bibfnamefont {L.}~\bibnamefont
  {Vervoort}}, \bibinfo {author} {\bibfnamefont {R.}~\bibnamefont {Ferreira}},\
  and\ \bibinfo {author} {\bibfnamefont {P.}~\bibnamefont {Voisin}},\
  }\bibfield  {title} {\bibinfo {title} {Effects of interface asymmetry on hole
  subband degeneracies and spin-relaxation rates in quantum wells},\ }\href
  {https://doi.org/10.1103/PhysRevB.56.R12744} {\bibfield  {journal} {\bibinfo
  {journal} {Physical Review B}\ }\textbf {\bibinfo {volume} {56}},\ \bibinfo
  {pages} {R12744} (\bibinfo {year} {1997})}\BibitemShut {NoStop}%
\bibitem [{\citenamefont {Vervoort}\ \emph {et~al.}(1999)\citenamefont
  {Vervoort}, \citenamefont {Ferreira},\ and\ \citenamefont
  {Voisin}}]{Vervoort99}%
  \BibitemOpen
  \bibfield  {author} {\bibinfo {author} {\bibfnamefont {L.}~\bibnamefont
  {Vervoort}}, \bibinfo {author} {\bibfnamefont {R.}~\bibnamefont {Ferreira}},\
  and\ \bibinfo {author} {\bibfnamefont {P.}~\bibnamefont {Voisin}},\
  }\bibfield  {title} {\bibinfo {title} {Spin-splitting of the subbands of
  ingaas-inp and other ‘no common atom’ quantum wells},\ }\href
  {https://doi.org/10.1088/0268-1242/14/3/004} {\bibfield  {journal} {\bibinfo
  {journal} {Semiconductor Science and Technology}\ }\textbf {\bibinfo {volume}
  {14}},\ \bibinfo {pages} {227} (\bibinfo {year} {1999})}\BibitemShut
  {NoStop}%
\bibitem [{\citenamefont {Luo}\ \emph {et~al.}(2010)\citenamefont {Luo},
  \citenamefont {Chantis}, \citenamefont {van Schilfgaarde}, \citenamefont
  {Bester},\ and\ \citenamefont {Zunger}}]{Luo10}%
  \BibitemOpen
  \bibfield  {author} {\bibinfo {author} {\bibfnamefont {J.-W.}\ \bibnamefont
  {Luo}}, \bibinfo {author} {\bibfnamefont {A.~N.}\ \bibnamefont {Chantis}},
  \bibinfo {author} {\bibfnamefont {M.}~\bibnamefont {van Schilfgaarde}},
  \bibinfo {author} {\bibfnamefont {G.}~\bibnamefont {Bester}},\ and\ \bibinfo
  {author} {\bibfnamefont {A.}~\bibnamefont {Zunger}},\ }\bibfield  {title}
  {\bibinfo {title} {Discovery of a novel linear-in-$k$ spin splitting for
  holes in the 2d $\mathrm{GaAs}/\mathrm{AlAs}$ system},\ }\href
  {https://doi.org/10.1103/PhysRevLett.104.066405} {\bibfield  {journal}
  {\bibinfo  {journal} {Physical Review Letters}\ }\textbf {\bibinfo {volume}
  {104}},\ \bibinfo {pages} {066405} (\bibinfo {year} {2010})}\BibitemShut
  {NoStop}%
\bibitem [{\citenamefont {Aleiner}\ and\ \citenamefont
  {Ivchenko}(1992)}]{Aleiner92}%
  \BibitemOpen
  \bibfield  {author} {\bibinfo {author} {\bibfnamefont {I.~L.}\ \bibnamefont
  {Aleiner}}\ and\ \bibinfo {author} {\bibfnamefont {E.~L.}\ \bibnamefont
  {Ivchenko}},\ }\bibfield  {title} {\bibinfo {title} {Anisotropic exchange
  splitting in type-{II} {GaAs}-{AlAs} superlattices},\ }\href
  {http://jetpletters.ru/ps/1278/article_19336.shtml} {\bibfield  {journal}
  {\bibinfo  {journal} {JETP}\ }\textbf {\bibinfo {volume} {55}},\ \bibinfo
  {pages} {692} (\bibinfo {year} {1992})}\BibitemShut {NoStop}%
\bibitem [{\citenamefont {Ivchenko}\ \emph {et~al.}(1996)\citenamefont
  {Ivchenko}, \citenamefont {Kaminski},\ and\ \citenamefont
  {R\"ossler}}]{Ivchenko96}%
  \BibitemOpen
  \bibfield  {author} {\bibinfo {author} {\bibfnamefont {E.~L.}\ \bibnamefont
  {Ivchenko}}, \bibinfo {author} {\bibfnamefont {A.~Y.}\ \bibnamefont
  {Kaminski}},\ and\ \bibinfo {author} {\bibfnamefont {U.}~\bibnamefont
  {R\"ossler}},\ }\bibfield  {title} {\bibinfo {title} {Heavy-light hole mixing
  at zinc-blende (001) interfaces under normal incidence},\ }\href
  {https://doi.org/10.1103/PhysRevB.54.5852} {\bibfield  {journal} {\bibinfo
  {journal} {Physical Review B}\ }\textbf {\bibinfo {volume} {54}},\ \bibinfo
  {pages} {5852} (\bibinfo {year} {1996})}\BibitemShut {NoStop}%
\bibitem [{\citenamefont {Golub}\ and\ \citenamefont
  {Ivchenko}(2004)}]{Golub04}%
  \BibitemOpen
  \bibfield  {author} {\bibinfo {author} {\bibfnamefont {L.~E.}\ \bibnamefont
  {Golub}}\ and\ \bibinfo {author} {\bibfnamefont {E.~L.}\ \bibnamefont
  {Ivchenko}},\ }\bibfield  {title} {\bibinfo {title} {Spin splitting in
  symmetrical {SiGe} quantum wells},\ }\href
  {https://doi.org/10.1103/PhysRevB.69.115333} {\bibfield  {journal} {\bibinfo
  {journal} {Physical Review B}\ }\textbf {\bibinfo {volume} {69}},\ \bibinfo
  {pages} {115333} (\bibinfo {year} {2004})}\BibitemShut {NoStop}%
\bibitem [{\citenamefont {Nestoklon}\ \emph {et~al.}(2008)\citenamefont
  {Nestoklon}, \citenamefont {Ivchenko}, \citenamefont {Jancu},\ and\
  \citenamefont {Voisin}}]{Nestoklon08}%
  \BibitemOpen
  \bibfield  {author} {\bibinfo {author} {\bibfnamefont {M.~O.}\ \bibnamefont
  {Nestoklon}}, \bibinfo {author} {\bibfnamefont {E.~L.}\ \bibnamefont
  {Ivchenko}}, \bibinfo {author} {\bibfnamefont {J.-M.}\ \bibnamefont
  {Jancu}},\ and\ \bibinfo {author} {\bibfnamefont {P.}~\bibnamefont
  {Voisin}},\ }\bibfield  {title} {\bibinfo {title} {Electric field effect on
  electron spin splitting in $\mathrm{Si}\mathrm{Ge}/\mathrm{Si}$ quantum
  wells},\ }\href {https://doi.org/10.1103/PhysRevB.77.155328} {\bibfield
  {journal} {\bibinfo  {journal} {Physical Review B}\ }\textbf {\bibinfo
  {volume} {77}},\ \bibinfo {pages} {155328} (\bibinfo {year}
  {2008})}\BibitemShut {NoStop}%
\bibitem [{Note1()}]{Note1}%
  \BibitemOpen
  \bibinfo {note} {The Rashba and Dresselhaus Hamiltonians for holes slightly
  differ from those for electrons, $H_\protect \mathrm {R}\propto k_x\sigma
  _2-k_y\sigma _1$ and $H_\protect \mathrm {D}\propto k_x\sigma _1-k_y\sigma
  _2$, because spins $\protect \genfrac {}{}{}1{3}{2}$ do not transform as
  spins $\protect \genfrac {}{}{}1{1}{2}$ under the symmetry operations of
  Table~\ref {tab:symmetries} (also see Appendix \ref
  {app:symmetries}).}\BibitemShut {Stop}%
\bibitem [{\citenamefont {{Di Carlo}}(2002)}]{DiCarlo03}%
  \BibitemOpen
  \bibfield  {author} {\bibinfo {author} {\bibfnamefont {A.}~\bibnamefont {{Di
  Carlo}}},\ }\bibfield  {title} {\bibinfo {title} {Microscopic theory of
  nanostructured semiconductor devices: beyond the envelope-function
  approximation},\ }\href {https://doi.org/10.1088/0268-1242/18/1/201}
  {\bibfield  {journal} {\bibinfo  {journal} {Semiconductor Science and
  Technology}\ }\textbf {\bibinfo {volume} {18}},\ \bibinfo {pages} {R1}
  (\bibinfo {year} {2002})}\BibitemShut {NoStop}%
\bibitem [{\citenamefont {Niquet}\ \emph {et~al.}(2009)\citenamefont {Niquet},
  \citenamefont {Rideau}, \citenamefont {Tavernier}, \citenamefont {Jaouen},\
  and\ \citenamefont {Blase}}]{Niquet09}%
  \BibitemOpen
  \bibfield  {author} {\bibinfo {author} {\bibfnamefont {Y.~M.}\ \bibnamefont
  {Niquet}}, \bibinfo {author} {\bibfnamefont {D.}~\bibnamefont {Rideau}},
  \bibinfo {author} {\bibfnamefont {C.}~\bibnamefont {Tavernier}}, \bibinfo
  {author} {\bibfnamefont {H.}~\bibnamefont {Jaouen}},\ and\ \bibinfo {author}
  {\bibfnamefont {X.}~\bibnamefont {Blase}},\ }\bibfield  {title} {\bibinfo
  {title} {{Onsite matrix elements of the tight-binding Hamiltonian of a
  strained crystal: Application to silicon, germanium, and their alloys}},\
  }\href {https://doi.org/10.1103/PhysRevB.79.245201} {\bibfield  {journal}
  {\bibinfo  {journal} {Physical Review B}\ }\textbf {\bibinfo {volume} {79}},\
  \bibinfo {pages} {245201} (\bibinfo {year} {2009})}\BibitemShut {NoStop}%
\bibitem [{Note2()}]{Note2}%
  \BibitemOpen
  \bibinfo {note} {In a Ge$_{1-x}$Si$_x$ alloy, the atomic energies $E_\alpha $
  of orbital $\alpha \in \{s,p,d,s^*\}$ (as well as the other on-site
  parameters) are averaged according to the probability of finding a Si or Ge
  atom in the alloy: \begin {equation*} E_\alpha (x)=xE_\alpha (\protect
  \mathrm {Si})+(1-x)E_\alpha (\protect \mathrm {Ge})\protect \tmspace
  +\thinmuskip {.1667em}. \end {equation*} The nearest-neighbor parameters
  $V_{\alpha \beta }$ are averaged according to the probability of finding a
  Si$-$Si, Ge$-$Ge, Si$-$Ge or Ge$-$Si bond: \begin {align*} V_{\alpha \beta
  }(x)&=x^2V_{\alpha \beta }(\protect \mathrm {Si}-\protect \mathrm
  {Si})+(1-x)^2V_{\alpha \beta }(\protect \mathrm {Ge}-\protect \mathrm {Ge})
  \\ &+x(1-x)[V_{\alpha \beta }(\protect \mathrm {Si}-\protect \mathrm
  {Ge})+V_{\alpha \beta }(\protect \mathrm {Ge}-\protect \mathrm {Si})]\protect
  \tmspace +\thinmuskip {.1667em}, \end {align*} where $V_{\alpha \beta
  }(\protect \mathrm {A}-\protect \mathrm {B})$ is the matrix element between
  orbital $\alpha $ on atom A and orbital $\beta $ on atom B.}\BibitemShut
  {Stop}%
\bibitem [{\citenamefont {Sammak}\ \emph {et~al.}(2019)\citenamefont {Sammak},
  \citenamefont {Sabbagh}, \citenamefont {Hendrickx}, \citenamefont {Lodari},
  \citenamefont {Paquelet~Wuetz}, \citenamefont {Tosato}, \citenamefont {Yeoh},
  \citenamefont {Bollani}, \citenamefont {Virgilio}, \citenamefont {Schubert},
  \citenamefont {Zaumseil}, \citenamefont {Capellini}, \citenamefont
  {Veldhorst},\ and\ \citenamefont {Scappucci}}]{Sammak19}%
  \BibitemOpen
  \bibfield  {author} {\bibinfo {author} {\bibfnamefont {A.}~\bibnamefont
  {Sammak}}, \bibinfo {author} {\bibfnamefont {D.}~\bibnamefont {Sabbagh}},
  \bibinfo {author} {\bibfnamefont {N.~W.}\ \bibnamefont {Hendrickx}}, \bibinfo
  {author} {\bibfnamefont {M.}~\bibnamefont {Lodari}}, \bibinfo {author}
  {\bibfnamefont {B.}~\bibnamefont {Paquelet~Wuetz}}, \bibinfo {author}
  {\bibfnamefont {A.}~\bibnamefont {Tosato}}, \bibinfo {author} {\bibfnamefont
  {L.}~\bibnamefont {Yeoh}}, \bibinfo {author} {\bibfnamefont {M.}~\bibnamefont
  {Bollani}}, \bibinfo {author} {\bibfnamefont {M.}~\bibnamefont {Virgilio}},
  \bibinfo {author} {\bibfnamefont {M.~A.}\ \bibnamefont {Schubert}}, \bibinfo
  {author} {\bibfnamefont {P.}~\bibnamefont {Zaumseil}}, \bibinfo {author}
  {\bibfnamefont {G.}~\bibnamefont {Capellini}}, \bibinfo {author}
  {\bibfnamefont {M.}~\bibnamefont {Veldhorst}},\ and\ \bibinfo {author}
  {\bibfnamefont {G.}~\bibnamefont {Scappucci}},\ }\bibfield  {title} {\bibinfo
  {title} {Shallow and undoped germanium quantum wells: A playground for spin
  and hybrid quantum technology},\ }\href
  {https://doi.org/10.1002/adfm.201807613} {\bibfield  {journal} {\bibinfo
  {journal} {Advanced Functional Materials}\ }\textbf {\bibinfo {volume}
  {29}},\ \bibinfo {pages} {1807613} (\bibinfo {year} {2019})}\BibitemShut
  {NoStop}%
\bibitem [{\citenamefont {Keating}(1966)}]{Keating66}%
  \BibitemOpen
  \bibfield  {author} {\bibinfo {author} {\bibfnamefont {P.~N.}\ \bibnamefont
  {Keating}},\ }\bibfield  {title} {\bibinfo {title} {Effect of invariance
  requirements on the elastic strain energy of crystals with application to the
  diamond structure},\ }\href {https://doi.org/10.1103/PhysRev.145.637}
  {\bibfield  {journal} {\bibinfo  {journal} {Physical Review}\ }\textbf
  {\bibinfo {volume} {145}},\ \bibinfo {pages} {637} (\bibinfo {year}
  {1966})}\BibitemShut {NoStop}%
\bibitem [{Note3()}]{Note3}%
  \BibitemOpen
  \bibinfo {note} {More precisely, the $\protect \ket {0,+\protect \genfrac
  {}{}{}1{3}{2}}$ and $\protect \ket {0,-\protect \genfrac {}{}{}1{3}{2}}$
  states are first computed at zero magnetic field, then the matrix $B_zS_z$
  (with $S_z$ the physical spin) is diagonalized in the subspace $\{\protect
  \ket {0,+\protect \genfrac {}{}{}1{3}{2}}$,$\protect \ket {0,-\protect
  \genfrac {}{}{}1{3}{2}}\}$. The amplitude of $B_z$ is, therefore,
  irrelevant.}\BibitemShut {Stop}%
\bibitem [{Note4()}]{Note4}%
  \BibitemOpen
  \bibinfo {note} {In Ge$_{0.8}$Si$_{0.2}$, $\gamma _1=11.56$, $\gamma
  _2=3.46$, $\gamma _3=4.84$, $\kappa =2.64$, $q=0.05$, and $b_v=-2.19$\protect
  \tmspace +\thinmuskip {.1667em}eV. The unstrained valence band offset is
  $\Delta _\protect \mathrm {VBO}=0.138$\protect \tmspace +\thinmuskip
  {.1667em}eV, and the hydrostatic valence band deformation potentials are
  $a_v=2$\protect \tmspace +\thinmuskip {.1667em}eV in Ge and
  $a_v=2.02$\protect \tmspace +\thinmuskip {.1667em}eV in
  Ge$_{0.8}$Si$_{0.2}$}\BibitemShut {NoStop}%
\bibitem [{\citenamefont {Luttinger}\ and\ \citenamefont
  {Kohn}(1955)}]{Luttinger55}%
  \BibitemOpen
  \bibfield  {author} {\bibinfo {author} {\bibfnamefont {J.~M.}\ \bibnamefont
  {Luttinger}}\ and\ \bibinfo {author} {\bibfnamefont {W.}~\bibnamefont
  {Kohn}},\ }\bibfield  {title} {\bibinfo {title} {Motion of electrons and
  holes in perturbed periodic fields},\ }\href
  {https://doi.org/10.1103/PhysRev.97.869} {\bibfield  {journal} {\bibinfo
  {journal} {Physical Review}\ }\textbf {\bibinfo {volume} {97}},\ \bibinfo
  {pages} {869} (\bibinfo {year} {1955})}\BibitemShut {NoStop}%
\bibitem [{Note5()}]{Note5}%
  \BibitemOpen
  \bibinfo {note} {In Figs.~\ref {fig:DresselhausKPEz} to \ref
  {fig:alpha3beta3Ez}, we use the Luttinger parameters, deformation potentials
  and strains of the TB and valence force field models \cite {Niquet09}, in
  order to achieve a meaningful comparison between the TB and LK data. In
  section \ref {sec:qubits}, we use the parameters given in the main text for
  consistency with our previous works \cite {martinez2022hole,Uriel22}. There
  are no qualitative differences between the two sets of parameters,
  though.}\BibitemShut {Stop}%
\bibitem [{\citenamefont {Burt}(1992)}]{Burt92}%
  \BibitemOpen
  \bibfield  {author} {\bibinfo {author} {\bibfnamefont {M.~G.}\ \bibnamefont
  {Burt}},\ }\bibfield  {title} {\bibinfo {title} {The justification for
  applying the effective-mass approximation to microstructures},\ }\href
  {https://doi.org/10.1088/0953-8984/4/32/003} {\bibfield  {journal} {\bibinfo
  {journal} {Journal of Physics: Condensed Matter}\ }\textbf {\bibinfo {volume}
  {4}},\ \bibinfo {pages} {6651} (\bibinfo {year} {1992})}\BibitemShut
  {NoStop}%
\bibitem [{\citenamefont {Foreman}(1993)}]{Foreman93}%
  \BibitemOpen
  \bibfield  {author} {\bibinfo {author} {\bibfnamefont {B.~A.}\ \bibnamefont
  {Foreman}},\ }\bibfield  {title} {\bibinfo {title} {Effective-mass
  hamiltonian and boundary conditions for the valence bands of semiconductor
  microstructures},\ }\href {https://doi.org/10.1103/PhysRevB.48.4964}
  {\bibfield  {journal} {\bibinfo  {journal} {Physical Review B}\ }\textbf
  {\bibinfo {volume} {48}},\ \bibinfo {pages} {4964} (\bibinfo {year}
  {1993})}\BibitemShut {NoStop}%
\bibitem [{\citenamefont {Foreman}(1997)}]{Foreman97}%
  \BibitemOpen
  \bibfield  {author} {\bibinfo {author} {\bibfnamefont {B.~A.}\ \bibnamefont
  {Foreman}},\ }\bibfield  {title} {\bibinfo {title} {Elimination of spurious
  solutions from eight-band $\mathbf{k}\ensuremath{\cdot}\mathbf{p}$ theory},\
  }\href {https://doi.org/10.1103/PhysRevB.56.R12748} {\bibfield  {journal}
  {\bibinfo  {journal} {Physical Review B}\ }\textbf {\bibinfo {volume} {56}},\
  \bibinfo {pages} {R12748} (\bibinfo {year} {1997})}\BibitemShut {NoStop}%
\bibitem [{\citenamefont {Veprek}\ \emph {et~al.}(2007)\citenamefont {Veprek},
  \citenamefont {Steiger},\ and\ \citenamefont {Witzigmann}}]{Veprek07}%
  \BibitemOpen
  \bibfield  {author} {\bibinfo {author} {\bibfnamefont {R.~G.}\ \bibnamefont
  {Veprek}}, \bibinfo {author} {\bibfnamefont {S.}~\bibnamefont {Steiger}},\
  and\ \bibinfo {author} {\bibfnamefont {B.}~\bibnamefont {Witzigmann}},\
  }\bibfield  {title} {\bibinfo {title} {Ellipticity and the spurious solution
  problem of $\mathbf{k}\cdot\mathbf{p}$ envelope equations},\ }\href
  {https://doi.org/10.1103/PhysRevB.76.165320} {\bibfield  {journal} {\bibinfo
  {journal} {Physical Review B}\ }\textbf {\bibinfo {volume} {76}},\ \bibinfo
  {pages} {165320} (\bibinfo {year} {2007})}\BibitemShut {NoStop}%
\bibitem [{Note6()}]{Note6}%
  \BibitemOpen
  \bibinfo {note} {In the limit $\ell _\protect \mathrm {E}\ll L_\protect
  \mathrm {W}$, the potential in Ge can be approximated by a triangular well.
  It can be shown from the eigensolutions of this triangular well (the Airy
  function) or from a simple dimensional analysis that both $\Delta _\protect
  \mathrm {LH}\propto 1/\ell _\protect \mathrm {E}^2\propto E_z^{2/3}$ and
  $d|\Psi (z)|^2/dz\propto 1/\ell _\protect \mathrm {E}^2\propto E_z^{2/3}$ so
  that $\alpha _\protect \mathrm {D}$ is expected to saturate. This saturation
  is however delayed by the biaxial strain contribution to $\Delta _\protect
  \mathrm {LH}$ [second term of Eq.~\protect \textup {\hbox {\mathsurround \z@
  \protect \normalfont (\ignorespaces \ref {eq:deltaLH}\unskip \@@italiccorr
  )}}] and by the finite band offset between Ge and GeSi.}\BibitemShut {Stop}%
\bibitem [{\citenamefont {Richard}\ \emph {et~al.}(2004)\citenamefont
  {Richard}, \citenamefont {Aniel},\ and\ \citenamefont {Fishman}}]{Richard04}%
  \BibitemOpen
  \bibfield  {author} {\bibinfo {author} {\bibfnamefont {S.}~\bibnamefont
  {Richard}}, \bibinfo {author} {\bibfnamefont {F.}~\bibnamefont {Aniel}},\
  and\ \bibinfo {author} {\bibfnamefont {G.}~\bibnamefont {Fishman}},\
  }\bibfield  {title} {\bibinfo {title} {Energy-band structure of {Ge}, {Si},
  and {GaAs}: A thirty-band $\mathbf{k}\cdot\mathbf{p}$ method},\ }\href
  {https://doi.org/10.1103/PhysRevB.70.235204} {\bibfield  {journal} {\bibinfo
  {journal} {Physical Review B}\ }\textbf {\bibinfo {volume} {70}},\ \bibinfo
  {pages} {235204} (\bibinfo {year} {2004})}\BibitemShut {NoStop}%
\bibitem [{\citenamefont {Durnev}\ \emph {et~al.}(2014)\citenamefont {Durnev},
  \citenamefont {Glazov},\ and\ \citenamefont {Ivchenko}}]{Durnev14}%
  \BibitemOpen
  \bibfield  {author} {\bibinfo {author} {\bibfnamefont {M.~V.}\ \bibnamefont
  {Durnev}}, \bibinfo {author} {\bibfnamefont {M.~M.}\ \bibnamefont {Glazov}},\
  and\ \bibinfo {author} {\bibfnamefont {E.~L.}\ \bibnamefont {Ivchenko}},\
  }\bibfield  {title} {\bibinfo {title} {Spin-orbit splitting of valence
  subbands in semiconductor nanostructures},\ }\href
  {https://doi.org/10.1103/PhysRevB.89.075430} {\bibfield  {journal} {\bibinfo
  {journal} {Physical Review B}\ }\textbf {\bibinfo {volume} {89}},\ \bibinfo
  {pages} {075430} (\bibinfo {year} {2014})}\BibitemShut {NoStop}%
\bibitem [{\citenamefont {Dresselhaus}\ \emph {et~al.}(1955)\citenamefont
  {Dresselhaus}, \citenamefont {Kip},\ and\ \citenamefont
  {Kittel}}]{Dresselhaus55}%
  \BibitemOpen
  \bibfield  {author} {\bibinfo {author} {\bibfnamefont {G.}~\bibnamefont
  {Dresselhaus}}, \bibinfo {author} {\bibfnamefont {A.~F.}\ \bibnamefont
  {Kip}},\ and\ \bibinfo {author} {\bibfnamefont {C.}~\bibnamefont {Kittel}},\
  }\bibfield  {title} {\bibinfo {title} {Cyclotron resonance of electrons and
  holes in silicon and germanium crystals},\ }\href
  {https://doi.org/10.1103/PhysRev.98.368} {\bibfield  {journal} {\bibinfo
  {journal} {Physical Review}\ }\textbf {\bibinfo {volume} {98}},\ \bibinfo
  {pages} {368} (\bibinfo {year} {1955})}\BibitemShut {NoStop}%
\bibitem [{\citenamefont {Moriya}\ \emph {et~al.}(2014)\citenamefont {Moriya},
  \citenamefont {Sawano}, \citenamefont {Hoshi}, \citenamefont {Masubuchi},
  \citenamefont {Shiraki}, \citenamefont {Wild}, \citenamefont {Neumann},
  \citenamefont {Abstreiter}, \citenamefont {Bougeard}, \citenamefont {Koga},\
  and\ \citenamefont {Machida}}]{Moriya14}%
  \BibitemOpen
  \bibfield  {author} {\bibinfo {author} {\bibfnamefont {R.}~\bibnamefont
  {Moriya}}, \bibinfo {author} {\bibfnamefont {K.}~\bibnamefont {Sawano}},
  \bibinfo {author} {\bibfnamefont {Y.}~\bibnamefont {Hoshi}}, \bibinfo
  {author} {\bibfnamefont {S.}~\bibnamefont {Masubuchi}}, \bibinfo {author}
  {\bibfnamefont {Y.}~\bibnamefont {Shiraki}}, \bibinfo {author} {\bibfnamefont
  {A.}~\bibnamefont {Wild}}, \bibinfo {author} {\bibfnamefont {C.}~\bibnamefont
  {Neumann}}, \bibinfo {author} {\bibfnamefont {G.}~\bibnamefont {Abstreiter}},
  \bibinfo {author} {\bibfnamefont {D.}~\bibnamefont {Bougeard}}, \bibinfo
  {author} {\bibfnamefont {T.}~\bibnamefont {Koga}},\ and\ \bibinfo {author}
  {\bibfnamefont {T.}~\bibnamefont {Machida}},\ }\bibfield  {title} {\bibinfo
  {title} {Cubic rashba spin-orbit interaction of a two-dimensional hole gas in
  a strained-$\mathrm{Ge}/\mathrm{SiGe}$ quantum well},\ }\href
  {https://doi.org/10.1103/PhysRevLett.113.086601} {\bibfield  {journal}
  {\bibinfo  {journal} {Physical Review Letters}\ }\textbf {\bibinfo {volume}
  {113}},\ \bibinfo {pages} {086601} (\bibinfo {year} {2014})}\BibitemShut
  {NoStop}%
\bibitem [{\citenamefont {Morrison}\ \emph {et~al.}(2016)\citenamefont
  {Morrison}, \citenamefont {Foronda}, \citenamefont {Wisniewski},
  \citenamefont {Rhead}, \citenamefont {Leadley},\ and\ \citenamefont
  {Myronov}}]{Morisson16}%
  \BibitemOpen
  \bibfield  {author} {\bibinfo {author} {\bibfnamefont {C.}~\bibnamefont
  {Morrison}}, \bibinfo {author} {\bibfnamefont {J.}~\bibnamefont {Foronda}},
  \bibinfo {author} {\bibfnamefont {P.}~\bibnamefont {Wisniewski}}, \bibinfo
  {author} {\bibfnamefont {S.}~\bibnamefont {Rhead}}, \bibinfo {author}
  {\bibfnamefont {D.}~\bibnamefont {Leadley}},\ and\ \bibinfo {author}
  {\bibfnamefont {M.}~\bibnamefont {Myronov}},\ }\bibfield  {title} {\bibinfo
  {title} {Evidence of strong spin–orbit interaction in strained epitaxial
  germanium},\ }\href {https://doi.org/10.1016/j.tsf.2015.09.063} {\bibfield
  {journal} {\bibinfo  {journal} {Thin Solid Films}\ }\textbf {\bibinfo
  {volume} {602}},\ \bibinfo {pages} {84} (\bibinfo {year} {2016})}\BibitemShut
  {NoStop}%
\bibitem [{\citenamefont {Mizokuchi}\ \emph {et~al.}(2017)\citenamefont
  {Mizokuchi}, \citenamefont {Torresani}, \citenamefont {Maurand},
  \citenamefont {Zeng}, \citenamefont {Niquet}, \citenamefont {Myronov},\ and\
  \citenamefont {De~Franceschi}}]{Mizokuchi17}%
  \BibitemOpen
  \bibfield  {author} {\bibinfo {author} {\bibfnamefont {R.}~\bibnamefont
  {Mizokuchi}}, \bibinfo {author} {\bibfnamefont {P.}~\bibnamefont
  {Torresani}}, \bibinfo {author} {\bibfnamefont {R.}~\bibnamefont {Maurand}},
  \bibinfo {author} {\bibfnamefont {Z.}~\bibnamefont {Zeng}}, \bibinfo {author}
  {\bibfnamefont {Y.-M.}\ \bibnamefont {Niquet}}, \bibinfo {author}
  {\bibfnamefont {M.}~\bibnamefont {Myronov}},\ and\ \bibinfo {author}
  {\bibfnamefont {S.}~\bibnamefont {De~Franceschi}},\ }\bibfield  {title}
  {\bibinfo {title} {{Hole weak anti-localization in a strained-Ge surface
  quantum well}},\ }\href {https://doi.org/10.1063/1.4997411} {\bibfield
  {journal} {\bibinfo  {journal} {Applied Physics Letters}\ }\textbf {\bibinfo
  {volume} {111}},\ \bibinfo {pages} {063102} (\bibinfo {year}
  {2017})}\BibitemShut {NoStop}%
\bibitem [{\citenamefont {Xiong}\ \emph
  {et~al.}(2022{\natexlab{b}})\citenamefont {Xiong}, \citenamefont {Liu},
  \citenamefont {Guan}, \citenamefont {Luo},\ and\ \citenamefont
  {Li}}]{Xiong22}%
  \BibitemOpen
  \bibfield  {author} {\bibinfo {author} {\bibfnamefont {J.-X.}\ \bibnamefont
  {Xiong}}, \bibinfo {author} {\bibfnamefont {Y.}~\bibnamefont {Liu}}, \bibinfo
  {author} {\bibfnamefont {S.}~\bibnamefont {Guan}}, \bibinfo {author}
  {\bibfnamefont {J.-W.}\ \bibnamefont {Luo}},\ and\ \bibinfo {author}
  {\bibfnamefont {S.-S.}\ \bibnamefont {Li}},\ }\bibfield  {title} {\bibinfo
  {title} {Why experiments fail to detect the finite linear rashba spin-orbit
  coupling of two-dimensional holes in semiconductor quantum wells: The case of
  ge/sige quantum wells},\ }\href {https://doi.org/10.1103/PhysRevB.106.155421}
  {\bibfield  {journal} {\bibinfo  {journal} {Physical Review B}\ }\textbf
  {\bibinfo {volume} {106}},\ \bibinfo {pages} {155421} (\bibinfo {year}
  {2022}{\natexlab{b}})}\BibitemShut {NoStop}%
\bibitem [{\citenamefont {Peña}\ \emph {et~al.}(2023)\citenamefont {Peña},
  \citenamefont {Koepke}, \citenamefont {Dycus}, \citenamefont {Mounce},
  \citenamefont {Baczewski}, \citenamefont {Jacobson},\ and\ \citenamefont
  {Bussmann}}]{Pena23}%
  \BibitemOpen
  \bibfield  {author} {\bibinfo {author} {\bibfnamefont {L.~F.}\ \bibnamefont
  {Peña}}, \bibinfo {author} {\bibfnamefont {J.~C.}\ \bibnamefont {Koepke}},
  \bibinfo {author} {\bibfnamefont {J.~H.}\ \bibnamefont {Dycus}}, \bibinfo
  {author} {\bibfnamefont {A.}~\bibnamefont {Mounce}}, \bibinfo {author}
  {\bibfnamefont {A.~D.}\ \bibnamefont {Baczewski}}, \bibinfo {author}
  {\bibfnamefont {N.~T.}\ \bibnamefont {Jacobson}},\ and\ \bibinfo {author}
  {\bibfnamefont {E.}~\bibnamefont {Bussmann}},\ }\bibfield  {title} {\bibinfo
  {title} {Utilizing multimodal microscopy to reconstruct {Si/SiGe} interfacial
  atomic disorder and infer its impacts on qubit variability},\ }\href
  {https://arxiv.org/abs/2306.15646} {\bibfield  {journal} {\bibinfo  {journal}
  {arXiv:2306.15646}\ } (\bibinfo {year} {2023})}\BibitemShut {NoStop}%
\bibitem [{\citenamefont {Su}\ \emph {et~al.}(2017)\citenamefont {Su},
  \citenamefont {Chuang}, \citenamefont {Liu}, \citenamefont {Li},\ and\
  \citenamefont {Lu}}]{Su17}%
  \BibitemOpen
  \bibfield  {author} {\bibinfo {author} {\bibfnamefont {Y.-H.}\ \bibnamefont
  {Su}}, \bibinfo {author} {\bibfnamefont {Y.}~\bibnamefont {Chuang}}, \bibinfo
  {author} {\bibfnamefont {C.-Y.}\ \bibnamefont {Liu}}, \bibinfo {author}
  {\bibfnamefont {J.-Y.}\ \bibnamefont {Li}},\ and\ \bibinfo {author}
  {\bibfnamefont {T.-M.}\ \bibnamefont {Lu}},\ }\bibfield  {title} {\bibinfo
  {title} {Effects of surface tunneling of two-dimensional hole gases in
  undoped {Ge/GeSi} heterostructures},\ }\href
  {https://doi.org/10.1103/PhysRevMaterials.1.044601} {\bibfield  {journal}
  {\bibinfo  {journal} {Physical Review Materials}\ }\textbf {\bibinfo {volume}
  {1}},\ \bibinfo {pages} {044601} (\bibinfo {year} {2017})}\BibitemShut
  {NoStop}%
\bibitem [{\citenamefont {Veldhorst}\ \emph {et~al.}(2015)\citenamefont
  {Veldhorst}, \citenamefont {Ruskov}, \citenamefont {Yang}, \citenamefont
  {Hwang}, \citenamefont {Hudson}, \citenamefont {Flatté}, \citenamefont
  {Tahan}, \citenamefont {Itoh}, \citenamefont {Morello},\ and\ \citenamefont
  {Dzurak}}]{Veldhorst15}%
  \BibitemOpen
  \bibfield  {author} {\bibinfo {author} {\bibfnamefont {M.}~\bibnamefont
  {Veldhorst}}, \bibinfo {author} {\bibfnamefont {R.}~\bibnamefont {Ruskov}},
  \bibinfo {author} {\bibfnamefont {C.~H.}\ \bibnamefont {Yang}}, \bibinfo
  {author} {\bibfnamefont {J.~C.~C.}\ \bibnamefont {Hwang}}, \bibinfo {author}
  {\bibfnamefont {F.~E.}\ \bibnamefont {Hudson}}, \bibinfo {author}
  {\bibfnamefont {M.~E.}\ \bibnamefont {Flatté}}, \bibinfo {author}
  {\bibfnamefont {C.}~\bibnamefont {Tahan}}, \bibinfo {author} {\bibfnamefont
  {K.~M.}\ \bibnamefont {Itoh}}, \bibinfo {author} {\bibfnamefont
  {A.}~\bibnamefont {Morello}},\ and\ \bibinfo {author} {\bibfnamefont {A.~S.}\
  \bibnamefont {Dzurak}},\ }\bibfield  {title} {\bibinfo {title} {{Spin-orbit
  coupling and operation of multivalley spin qubits}},\ }\href
  {https://doi.org/10.1103/PhysRevB.92.201401} {\bibfield  {journal} {\bibinfo
  {journal} {Physical Review B}\ }\textbf {\bibinfo {volume} {92}},\ \bibinfo
  {pages} {201401} (\bibinfo {year} {2015})}\BibitemShut {NoStop}%
\bibitem [{\citenamefont {Ruskov}\ \emph {et~al.}(2018)\citenamefont {Ruskov},
  \citenamefont {Veldhorst}, \citenamefont {Dzurak},\ and\ \citenamefont
  {Tahan}}]{Ruskov18}%
  \BibitemOpen
  \bibfield  {author} {\bibinfo {author} {\bibfnamefont {R.}~\bibnamefont
  {Ruskov}}, \bibinfo {author} {\bibfnamefont {M.}~\bibnamefont {Veldhorst}},
  \bibinfo {author} {\bibfnamefont {A.~S.}\ \bibnamefont {Dzurak}},\ and\
  \bibinfo {author} {\bibfnamefont {C.}~\bibnamefont {Tahan}},\ }\bibfield
  {title} {\bibinfo {title} {Electron $g$-factor of valley states in realistic
  silicon quantum dots},\ }\href {https://doi.org/10.1103/PhysRevB.98.245424}
  {\bibfield  {journal} {\bibinfo  {journal} {Physical Review B}\ }\textbf
  {\bibinfo {volume} {98}},\ \bibinfo {pages} {245424} (\bibinfo {year}
  {2018})}\BibitemShut {NoStop}%
\bibitem [{\citenamefont {Ferdous}\ \emph {et~al.}(2018)\citenamefont
  {Ferdous}, \citenamefont {Chan}, \citenamefont {Veldhorst}, \citenamefont
  {Hwang}, \citenamefont {Yang}, \citenamefont {Sahasrabudhe}, \citenamefont
  {Klimeck}, \citenamefont {Morello}, \citenamefont {Dzurak},\ and\
  \citenamefont {Rahman}}]{Ferdous18b}%
  \BibitemOpen
  \bibfield  {author} {\bibinfo {author} {\bibfnamefont {R.}~\bibnamefont
  {Ferdous}}, \bibinfo {author} {\bibfnamefont {K.~W.}\ \bibnamefont {Chan}},
  \bibinfo {author} {\bibfnamefont {M.}~\bibnamefont {Veldhorst}}, \bibinfo
  {author} {\bibfnamefont {J.~C.~C.}\ \bibnamefont {Hwang}}, \bibinfo {author}
  {\bibfnamefont {C.~H.}\ \bibnamefont {Yang}}, \bibinfo {author}
  {\bibfnamefont {H.}~\bibnamefont {Sahasrabudhe}}, \bibinfo {author}
  {\bibfnamefont {G.}~\bibnamefont {Klimeck}}, \bibinfo {author} {\bibfnamefont
  {A.}~\bibnamefont {Morello}}, \bibinfo {author} {\bibfnamefont {A.~S.}\
  \bibnamefont {Dzurak}},\ and\ \bibinfo {author} {\bibfnamefont
  {R.}~\bibnamefont {Rahman}},\ }\bibfield  {title} {\bibinfo {title}
  {Interface-induced spin-orbit interaction in silicon quantum dots and
  prospects for scalability},\ }\href
  {https://doi.org/10.1103/PhysRevB.97.241401} {\bibfield  {journal} {\bibinfo
  {journal} {Physical Review B}\ }\textbf {\bibinfo {volume} {97}},\ \bibinfo
  {pages} {241401} (\bibinfo {year} {2018})}\BibitemShut {NoStop}%
\bibitem [{\citenamefont {Tanttu}\ \emph {et~al.}(2019)\citenamefont {Tanttu},
  \citenamefont {Hensen}, \citenamefont {Chan}, \citenamefont {Yang},
  \citenamefont {Huang}, \citenamefont {Fogarty}, \citenamefont {Hudson},
  \citenamefont {Itoh}, \citenamefont {Culcer}, \citenamefont {Laucht},
  \citenamefont {Morello},\ and\ \citenamefont {Dzurak}}]{Tanttu19}%
  \BibitemOpen
  \bibfield  {author} {\bibinfo {author} {\bibfnamefont {T.}~\bibnamefont
  {Tanttu}}, \bibinfo {author} {\bibfnamefont {B.}~\bibnamefont {Hensen}},
  \bibinfo {author} {\bibfnamefont {K.~W.}\ \bibnamefont {Chan}}, \bibinfo
  {author} {\bibfnamefont {C.~H.}\ \bibnamefont {Yang}}, \bibinfo {author}
  {\bibfnamefont {W.~W.}\ \bibnamefont {Huang}}, \bibinfo {author}
  {\bibfnamefont {M.}~\bibnamefont {Fogarty}}, \bibinfo {author} {\bibfnamefont
  {F.}~\bibnamefont {Hudson}}, \bibinfo {author} {\bibfnamefont
  {K.}~\bibnamefont {Itoh}}, \bibinfo {author} {\bibfnamefont {D.}~\bibnamefont
  {Culcer}}, \bibinfo {author} {\bibfnamefont {A.}~\bibnamefont {Laucht}},
  \bibinfo {author} {\bibfnamefont {A.}~\bibnamefont {Morello}},\ and\ \bibinfo
  {author} {\bibfnamefont {A.}~\bibnamefont {Dzurak}},\ }\bibfield  {title}
  {\bibinfo {title} {Controlling spin-orbit interactions in silicon quantum
  dots using magnetic field direction},\ }\href
  {https://doi.org/10.1103/PhysRevX.9.021028} {\bibfield  {journal} {\bibinfo
  {journal} {Physical Review X}\ }\textbf {\bibinfo {volume} {9}},\ \bibinfo
  {pages} {021028} (\bibinfo {year} {2019})}\BibitemShut {NoStop}%
\bibitem [{\citenamefont {Hosseinkhani}\ and\ \citenamefont
  {Burkard}(2021)}]{Hosseinkhani21}%
  \BibitemOpen
  \bibfield  {author} {\bibinfo {author} {\bibfnamefont {A.}~\bibnamefont
  {Hosseinkhani}}\ and\ \bibinfo {author} {\bibfnamefont {G.}~\bibnamefont
  {Burkard}},\ }\bibfield  {title} {\bibinfo {title} {Relaxation of
  single-electron spin qubits in silicon in the presence of interface steps},\
  }\href {https://doi.org/10.1103/PhysRevB.104.085309} {\bibfield  {journal}
  {\bibinfo  {journal} {Physical Review B}\ }\textbf {\bibinfo {volume}
  {104}},\ \bibinfo {pages} {085309} (\bibinfo {year} {2021})}\BibitemShut
  {NoStop}%
\bibitem [{\citenamefont {Merzbacher}(1997)}]{Merzbacher97}%
  \BibitemOpen
  \bibfield  {author} {\bibinfo {author} {\bibfnamefont {E.}~\bibnamefont
  {Merzbacher}},\ }\href@noop {} {\emph {\bibinfo {title} {Quantum Mechanics,
  3$^{rd}$ edition}}}\ (\bibinfo  {publisher} {Wiley},\ \bibinfo {address}
  {New-York},\ \bibinfo {year} {1997})\BibitemShut {NoStop}%
\bibitem [{\citenamefont {Kresse}\ and\ \citenamefont
  {Hafner}(1993)}]{Kresse93}%
  \BibitemOpen
  \bibfield  {author} {\bibinfo {author} {\bibfnamefont {G.}~\bibnamefont
  {Kresse}}\ and\ \bibinfo {author} {\bibfnamefont {J.}~\bibnamefont
  {Hafner}},\ }\bibfield  {title} {\bibinfo {title} {Ab initio molecular
  dynamics for liquid metals},\ }\href
  {https://doi.org/10.1103/PhysRevB.47.558} {\bibfield  {journal} {\bibinfo
  {journal} {Physical Review B}\ }\textbf {\bibinfo {volume} {47}},\ \bibinfo
  {pages} {558} (\bibinfo {year} {1993})}\BibitemShut {NoStop}%
\bibitem [{\citenamefont {Kresse}\ and\ \citenamefont
  {Hafner}(1994)}]{Kresse94}%
  \BibitemOpen
  \bibfield  {author} {\bibinfo {author} {\bibfnamefont {G.}~\bibnamefont
  {Kresse}}\ and\ \bibinfo {author} {\bibfnamefont {J.}~\bibnamefont
  {Hafner}},\ }\bibfield  {title} {\bibinfo {title} {Ab initio
  molecular-dynamics simulation of the liquid-metal--amorphous-semiconductor
  transition in germanium},\ }\href {https://doi.org/10.1103/PhysRevB.49.14251}
  {\bibfield  {journal} {\bibinfo  {journal} {Physical Review B}\ }\textbf
  {\bibinfo {volume} {49}},\ \bibinfo {pages} {14251} (\bibinfo {year}
  {1994})}\BibitemShut {NoStop}%
\bibitem [{\citenamefont {Kresse}\ and\ \citenamefont
  {Furthmüller}(1996)}]{Kresse96}%
  \BibitemOpen
  \bibfield  {author} {\bibinfo {author} {\bibfnamefont {G.}~\bibnamefont
  {Kresse}}\ and\ \bibinfo {author} {\bibfnamefont {J.}~\bibnamefont
  {Furthmüller}},\ }\bibfield  {title} {\bibinfo {title} {Efficiency of
  ab-initio total energy calculations for metals and semiconductors using a
  plane-wave basis set},\ }\href {https://doi.org/10.1016/0927-0256(96)00008-0}
  {\bibfield  {journal} {\bibinfo  {journal} {Computational Materials Science}\
  }\textbf {\bibinfo {volume} {6}},\ \bibinfo {pages} {15} (\bibinfo {year}
  {1996})}\BibitemShut {NoStop}%
\bibitem [{\citenamefont {Kresse}\ and\ \citenamefont
  {Furthm\"uller}(1996)}]{Kresse96b}%
  \BibitemOpen
  \bibfield  {author} {\bibinfo {author} {\bibfnamefont {G.}~\bibnamefont
  {Kresse}}\ and\ \bibinfo {author} {\bibfnamefont {J.}~\bibnamefont
  {Furthm\"uller}},\ }\bibfield  {title} {\bibinfo {title} {Efficient iterative
  schemes for ab initio total-energy calculations using a plane-wave basis
  set},\ }\href {https://doi.org/10.1103/PhysRevB.54.11169} {\bibfield
  {journal} {\bibinfo  {journal} {Physical Review B}\ }\textbf {\bibinfo
  {volume} {54}},\ \bibinfo {pages} {11169} (\bibinfo {year}
  {1996})}\BibitemShut {NoStop}%
\bibitem [{\citenamefont {Perdew}\ \emph {et~al.}(1996)\citenamefont {Perdew},
  \citenamefont {Burke},\ and\ \citenamefont {Ernzerhof}}]{PBE96}%
  \BibitemOpen
  \bibfield  {author} {\bibinfo {author} {\bibfnamefont {J.~P.}\ \bibnamefont
  {Perdew}}, \bibinfo {author} {\bibfnamefont {K.}~\bibnamefont {Burke}},\ and\
  \bibinfo {author} {\bibfnamefont {M.}~\bibnamefont {Ernzerhof}},\ }\bibfield
  {title} {\bibinfo {title} {Generalized gradient approximation made simple},\
  }\href {https://doi.org/10.1103/PhysRevLett.77.3865} {\bibfield  {journal}
  {\bibinfo  {journal} {Physical Review Letters}\ }\textbf {\bibinfo {volume}
  {77}},\ \bibinfo {pages} {3865} (\bibinfo {year} {1996})}\BibitemShut
  {NoStop}%
\end{thebibliography}%

\end{document}